\pdftrailerid{}
\documentclass[11pt,a4paper]{article}
\usepackage[T1]{fontenc}
\usepackage[utf8]{inputenc}
\usepackage{lmodern}
\usepackage[margin=27mm]{geometry}
\usepackage{amsmath,amssymb,amsthm,mathtools,booktabs,array}
\usepackage{microtype}
\usepackage{enumitem}
\usepackage{url}
\usepackage{xurl}
\usepackage{tikz,graphicx}
\usepackage{flafter,placeins}
\usepackage{needspace}
\usepackage{float}
\usepackage[font=small,labelfont=bf]{caption}
\usepackage{algorithm,algpseudocode}
\usepackage[hidelinks,hypertexnames=false]{hyperref}
\usepackage{orcidlink}
\makeatletter
\newcommand{\includegroupedpdf}[2][]{\begingroup
  \let\Gread@transgrouptrue\Gread@transgroupfalse
  \includegraphics[#1]{#2}\endgroup}
\makeatother
\hypersetup{pdftitle={Finding most likely Delaunay triangulations of imprecise points is NP-hard},pdfauthor={Pedro M. M. de Castro},pdfsubject={Computational geometry, modal structures and approximation}}
\usetikzlibrary{arrows.meta}
\allowdisplaybreaks

\newtheorem{theorem}{Theorem}
\newtheorem{lemma}[theorem]{Lemma}
\newtheorem{proposition}[theorem]{Proposition}

\theoremstyle{definition}\newtheorem{definition}[theorem]{Definition}
\newcommand{\R}{\mathbb R}
\newcommand{\Q}{\mathbb Q}
\newcommand{\E}{\mathbb E}
\newcommand{\Prob}{\mathbb P}
\newcommand{\DT}{\operatorname{DT}}
\newcommand{\sgn}{\operatorname{sgn}}
\newcommand{\A}{\mathcal A}

\newcommand{\TV}{d_{\mathrm{TV}}}
\title{Finding most likely Delaunay triangulations\\of imprecise points is NP-hard}
\author{Pedro M. M. de Castro\,\orcidlink{0000-0002-9470-7458}\\
\small Centro de Inform\'atica, Universidade Federal de Pernambuco\\
\small Recife, Pernambuco, Brazil\\
\small \texttt{pmmc@cin.ufpe.br}}
\date{December 2, 2026}
\begin{document}
\maketitle
\begin{abstract}
Given independent Gaussian sites, which labelled triangulation is most likely to be Delaunay?
With common isotropic variance, six competitors can beat the triangulation of the reported centres on four sites, and at least \((n-1)!/2\) can do so for all sufficiently large \(n\).
On the line, under common variance, sorting the means finds an unoriented Delaunay path of probability at least half the maximum in \(O(n\log n)\) comparisons.
In the plane, even this approximation is NP-hard.
We encode a planar cubic graph in the diagonals using rational generic centres and standard Gaussian displacements. Every qualifying answer to one query yields a maximum independent set, even at relative factor \(2^{-n^{1-\epsilon}}\) for fixed \(0<\epsilon<1\).
A boundary-cap construction proves NP-hardness for the most probable convex hull when sites have individual isotropic Gaussian variances.
A logarithmic moment condition covers the uniform law on a disk, the isotropic Cauchy law and some heavier-tailed laws. Delaunay hardness extends to their fixed invertible linear images. Polynomial-time sampling gives an additive approximation for Gaussian sites. All numbered results are formally verified in Lean~4.
\end{abstract}

\medskip\noindent\textbf{AI-use disclosure.}
The author used OpenAI's ChatGPT and Codex, together with Anthropic's Claude, as research tools for literature retrieval, mathematical exploration, symbolic and numerical checks, illustrations, and manuscript preparation and revision. Their contributions included proposed geometric constructions, probability estimates, and drafts of the reductions and proofs. Under the author's direction, Codex and Claude wrote the formalization, which was checked by Lean~4. Google's Antigravity contributed to editorial review and helped draft the exposition of the direct Gaussian reduction. The author formulated the questions, directed the search for weaker hypotheses, audited the proofs, and revised the text and figures throughout. Responsibility for the mathematical claims and their presentation rests with the author.



\medskip\noindent\textbf{Supplementary material.} \href{https://github.com/tashimir/most-likely-delaunay-triangulations}{\textcolor{blue!60!black}{\underline{Code, data, figures and Lean proofs on GitHub}}}.

\section{Introduction}\label{sec:introduction}
\begin{figure}[tb]
\centering\begingroup\makeatletter\let\Gread@transgrouptrue\Gread@transgroupfalse\includegraphics[width=\linewidth]{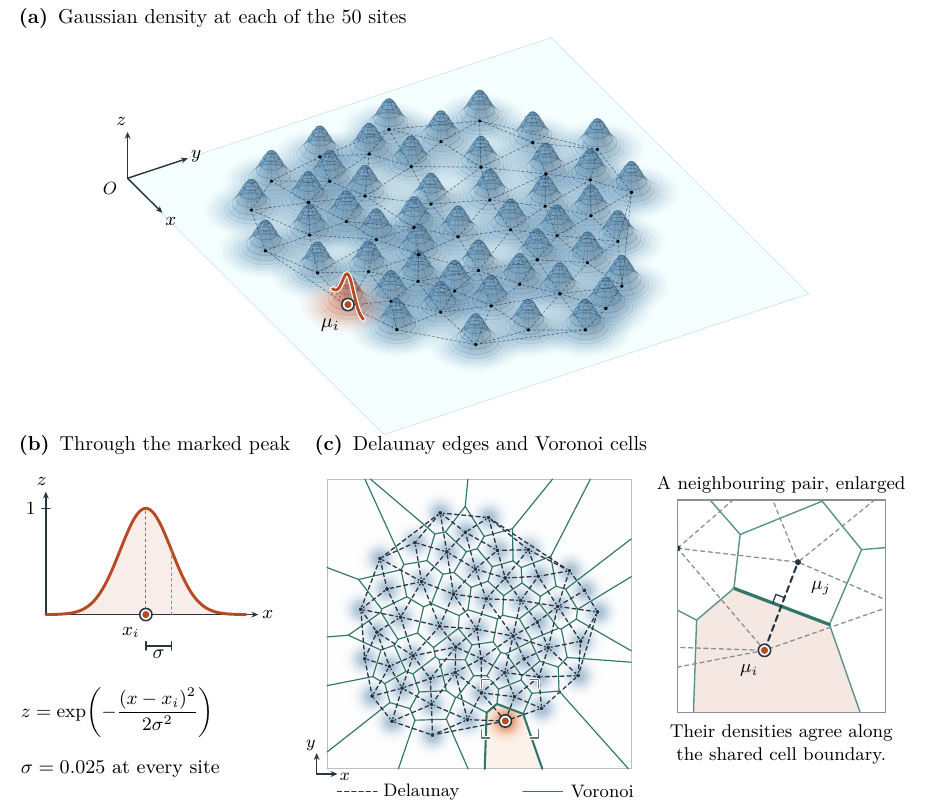}\endgroup
\caption{Gaussian neighbours. In (a), the 50 marginal densities are scaled to unit peak. In (b), the cut crosses the marked peak. In (c), the Delaunay edge crosses the equal-density boundary.}\label{fig:density-scene}
\end{figure}

\begin{figure}[tb]
\centering\includegraphics[width=\linewidth]{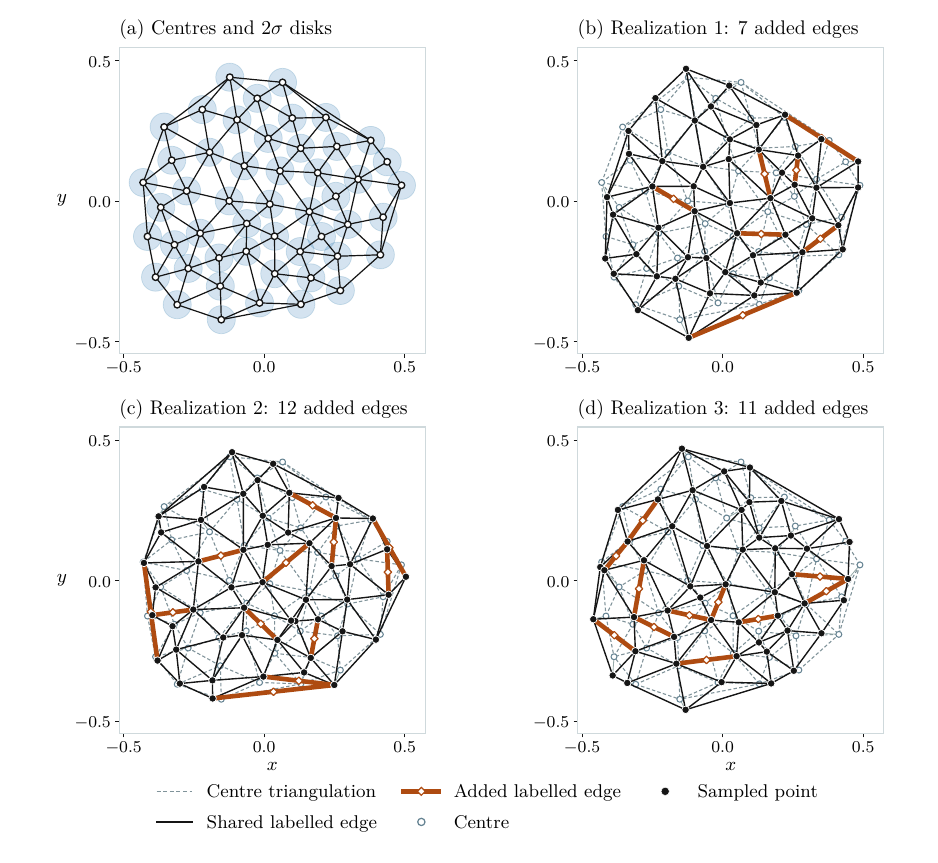}
\caption{New Delaunay neighbours. In (a), disks of radius \(2\sigma\) surround the centres of Fig.~\ref{fig:density-scene}(c). In (b)--(d), the centre triangulation is dashed. Bold orange edges with diamonds mark new neighbours.}\label{fig:vis-intro}
\end{figure}

A Delaunay triangulation divides the convex hull of planar sites into triangles with empty open circumdisks. Under independent Gaussian noise, the joint density peaks at the reported centres, whose Voronoi cells define the centre triangulation (Fig.~\ref{fig:density-scene}). Yet any single configuration occurs with probability zero: to find how likely a triangulation is, we integrate the density over all realizations with the same labelled triangles, and modest noise can flip many edges in one draw (Fig.~\ref{fig:vis-intro}). Almost surely, a sample is generic (no three sites collinear, no four cocircular). We record each \emph{type} by the label triples of its triangles, and its hull edges are the pairs in only one triangle. A \emph{mode} is a type of maximum probability \(p_*\).

For fixed generic centres, their triangulation has probability tending to one as the noise scale vanishes. At larger scale with common isotropic Gaussian variance, putting one centre inside the triangle of three others lets all six quadrilateral types beat the centre type (Proposition~\ref{prop:four-centres}). For all sufficiently large \(n\), placing centres on a tiny parabolic arc puts them in convex position near a point. Relabelling a mode at coincident centres yields at least \((n-1)!/2\) competitors that keep their lead over the centre type on the parabola, and convex position is so rare under noise that the centre type has probability at most \(\exp(-n\log n+O(n))\) times the maximum. One can choose arbitrarily close rational generic centres off the tie sets, keeping both bounds with a unique mode below \(3/(n-1)!\) (Proposition~\ref{prop:centre-rank}) that polynomially many independent draws miss with probability tending to one.

On the line, sorting the means of independent Gaussians of common variance maximizes the probability of the left-to-right order~\cite[Theorem~1]{palmer}. Each labelled Delaunay path corresponds to an order and its reverse, so sorting attains at least half the maximum path probability in \(O(n\log n)\) comparisons (Proposition~\ref{prop:line}). In the plane, even this factor is NP-hard for independent standard Gaussian displacements. Theorem~\ref{thm:socg-gaussian} proves hardness down to the relative factor \(2^{-n^{1-\epsilon}}\) for every fixed \(0<\epsilon<1\). The reduction makes one query on rational generic centres, and every qualifying answer recovers a maximum independent set.

The convex hull retains only the outer boundary. When the Gaussian sites have individual isotropic variances, finding a most probable set of extreme labels or their cyclic order up to reversal and rotation is NP-hard under either convention (Theorem~\ref{thm:gaussian-hull}). Here too, a single query suffices, and the reduction can use any mode that the query returns. The common-variance case for hulls remains open.

How much does Delaunay hardness depend on Gaussian noise? If the displacements are independent draws from a fixed bounded radial density, a finite positive moment of \(\log(1+\|Z\|)\) suffices for exact mode hardness (Theorem~\ref{thm:main}). The one-query reduction also gives hardness at the decaying approximation factor (Section~\ref{sec:amp-replication}). Uniform disks and isotropic Cauchy displacements both meet the condition. Directional normalization extends the construction to fixed invertible linear images of rotationally invariant laws satisfying the same logarithmic moment condition and having bounded projection densities, including every fixed positive definite Gaussian covariance (Theorem~\ref{thm:elliptical}). With additive error the situation changes. For Gaussian sites with rational centres and invertible rational covariance factors, sampling gives a triangulation of probability at least \(p_*-\varepsilon\) with confidence \(1-\delta\), in time polynomial in the input bit length, \(1/\varepsilon\) and \(\log(1/\delta)\) (Theorem~\ref{thm:additive}).

Lean~4 verifies every numbered result, the radial reduction, and the Gaussian sampling algorithm. The Delaunay and hull hardness theorems use the classical NP-hardness of planar cubic \textsc{Independent Set}~\cite{mohar_face} and \textsc{Subset Sum}~\cite{karp}, respectively (Section~\ref{sec:lean}).

Suri, Verbeek and Y{\i}ld{\i}z found the most probable hull in cubic time for independently present fixed sites, proved NP-hardness for the hull polygon of sites with finite support while amplifying the gap with copies~\cite[Theorems~1, 3 and 4]{suri}, and asked about Delaunay triangulations~\cite[Section~5]{suri}. Because Gaussian sites can land between repeated copies, we triangulate those gaps and bound flips there. Khanimov and Sharir linearized incircle determinants on perturbed regular hexagons and heptagons~\cite[Sections~2, 3.1 and 3.2]{khanimov}, and Leipold, Lazar, Brakke and Srolovitz studied correlated diagonals in perturbed lattices~\cite[Section~II]{leipold}. We encode each vertex as a cycle of tangent empty circles, so shared sites favour equal diagonals around a cycle and opposite choices at adjacent vertices. Other work preprocesses uncertainty regions to triangulate observed sites~\cite{loffler_snoeyink,buchin,devillers}, estimates a tetrahedron's Delaunay probability~\cite[Section~5]{almost}, finds a most probable nearest neighbour~\cite{voronoi}, computes probabilities that query points lie inside the hull~\cite{agarwal_hull}, or builds regions containing in expectation at least a specified fraction of Gaussian sites~\cite[Theorem~V.2]{atalay-hull}.

\FloatBarrier
\section{How misleading can the centres be?}\label{sec:centres}\label{sec:intro-centres}

\begin{figure}[!hb]
\centering
\includegraphics[width=\linewidth]{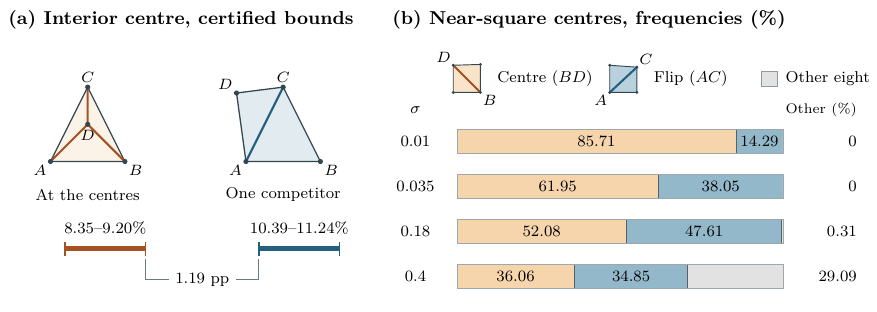}
\caption{Two four-site comparisons. In (a), at $\sigma=20$ each quadrilateral type beats the centre type by at least 1.19 percentage points (Proposition~\ref{prop:four-centres}). In (b), four million independent draws per scale add $N(0,\sigma^2 I_2)$ noise to unit-square centres with $C$ raised by $0.03$ (supplement, \texttt{gaussian\_sites/sample.py}).}
\label{fig:four-centres}
\end{figure}

Generic centres have a positive stability radius~\cite[pp.~221, 223--224]{abellanas}, so their triangulation leads at small scales. Near a square, the diagonal can flip while the hull cycle persists (Fig.~\ref{fig:four-centres}(b)). In Fig.~\ref{fig:four-centres}(a), all six quadrilateral types beat the centre type at scale \(20\).

\subsection{Four sites and ten possible outcomes}\label{app:four-proof}

\begin{proposition}\label{prop:four-centres}
Let \(A,B,C,D\) be independent Gaussian sites with centres \((-1,-1)\), \((1,-1)\), \((0,1)\), and \((0,0)\), respectively, and covariance \(20^2I_2\). Each labelled Delaunay triangulation with a triangular hull has probability in \([0.0835,0.0920]\), and each with a quadrilateral hull has probability in \([0.1039,0.1124]\). All six of the latter outrank the triangulation at the centres.
\end{proposition}

\begin{proof}
At the centres, \(D\) lies inside \(ABC\) with barycentric coordinates \((1/4,1/4,1/2)\). When all four means coincide at the origin, Maehara's formula~\cite{maehara} and relabelling give each quadrilateral type probability \(q=\arcsin(1/3)/\pi\) and each triangular type \(r=(1-6q)/4\). Blatter gives a geometric proof of Maehara's formula~\cite[p.~837]{blatter}, and Frick, Newman and Pegden give one using Gaussian order statistics~\cite{frick_newman_pegden}. Kabluchko and Panzo treat \(d+2\) Gaussian points in \(\mathbb R^d\)~\cite[Theorem~3.2]{sylvester_types}.

Subtracting the centroid of the four centres and dividing by \(20\) gives joint mean \(v\in\mathbb R^8\) with \(a=\|v\|^2=19/1600\). Because a half-turn preserves every face list, an equal mixture of the Gaussian laws with means \(v\) and \(-v\) preserves each probability while cancelling the linear term. Relative to a standard Gaussian \(Z\), this mixture has density ratio \(e^{-a/2}\cosh(v\cdot Z)\), with mean one and second moment \(\cosh a\). Cauchy--Schwarz bounds the change in each probability by \(\tfrac12\sqrt{\cosh a-1}\).

To enclose \(q=\arcsin(1/3)/\pi\), sum the first five terms of the arcsine series. Each successive term is at most \(1/9\) of its predecessor, so the omitted tail is at most \(9/8\) of the first omitted term. With \(3.14159<\pi<3.14160\), this gives \(0.10817<q<0.10818\). For \(a=19/1600\), the ratio of successive terms in \(\cosh a-1\), starting with the quadratic term, is at most \(a^2/12\), so
\[
 \tfrac14(\cosh a-1)\le \frac{a^2}{8(1-a^2/12)}<(0.0042)^2.
\]
Applying this bound on the probability shift to \(q\) and to \(r=(1-6q)/4\) gives \([0.1039,0.1124]\) for each quadrilateral type and \([0.0835,0.0920]\) for each triangular type. The accompanying \texttt{gaussian\_sites/certificate.py} verifies both enclosures in rational arithmetic.
\end{proof}

\hypertarget{gaussian-translation}{For later use}, note a direct Gaussian translation bound: for any nonzero shift \(u\in\mathbb R^d\) at scale \(\sigma\), the two densities agree along the perpendicular bisector of their means, so projecting onto \(u\) gives total variation \(2\Phi(\|u\|/(2\sigma))-1\le\|u\|/(\sigma\sqrt{2\pi})\le\|u\|/(2\sigma)\), where \(\Phi\) is the standard normal cumulative distribution function. In Proposition~\ref{prop:four-centres}, moving the joint mean in \(\mathbb R^8\) by less than \(0.23\) while keeping \(D\) inside \(ABC\) shifts each probability by under \(0.0058\), preserving all six comparisons since \(2\cdot 0.0058<0.1039-0.0920\).

\subsection{How many types can outrank the centre triangulation?}\label{app:rank-proof}

We start at coincident centres, where relabellings of a mode are equiprobable and yield at least \((n-1)!/2\) competitors. Shifting slightly into convex position keeps their lead over the centre triangulation, since matching the centres requires all sampled sites on the hull.

\begin{proposition}\label{prop:centre-rank}
For all sufficiently large \(n\), let \(\mu_i=(i,i^2)/(n!)^3\), \(1\le i\le n\), with independent standard Gaussian displacements. At least \((n-1)!/2\) labelled triangulations outrank the centre triangulation, whose probability is at most \(\exp(-n\log n+O(n))\) times the maximum. Arbitrarily close rational generic centres retain both bounds and have a unique mode of probability less than \(3/(n-1)!\).
\end{proposition}

\begin{proof}
\emph{Modal orbit at the origin.} At coincident centres, relabellings in one orbit have equal probability. Fixing a directed hull edge fixes the incident bounded face and, by edge propagation, the whole labelled triangulation, so its automorphism group has size at most \(2n\). Every modal orbit thus has at least \((n-1)!/2\) members, each with probability at most \(2/(n-1)!\) and at least \(1/(n!C^n)\) for an absolute \(C\). For the lower bound, root each planar triangulation at a directed hull edge with the exterior face on its right. The rooted map and its vertex labels determine the bounded face list. Tutte's formula bounds the number of rooted planar maps with \(e\) edges by \(12^e\)~\cite[(5.1)]{tutte-maps}. Since \(e\le3n-6\), there are at most \(n!\sum_{e=1}^{3n-6}12^e<n!(12^3)^n\) labelled types.

\emph{Rarity of convex position.} Sort the sample by \(x\)-coordinates, leaving the reordered \(y\)-coordinates independent standard normals. In convex position, the upper hull rises then falls and the lower falls then rises~\cite[Section~2]{tiwary}. Assigning each point to one of the four runs (breaking ties at shared endpoints and extrema) gives at most \(4^n\) partitions. A partition with run sizes \(m_j\) satisfies all four orders with probability \(1/\prod_j m_j!\le 4^n/n!\) by the multinomial theorem. Convex position thus has probability at most \(16^n/n!\), and since the full orbit of any type with all \(n\) labels on the hull lies in that event, each has probability at most \(2\cdot 16^n/(n!(n-1)!)\).

\emph{Shifting to the parabola and breaking ties.} The points \((i,i^2)\) lie in strictly convex position, and no four are cocircular since a circle meets \(y=x^2\) in a monic quartic with zero cubic term, impossible for four positive roots. Scaling preserves the centre triangulation. The \hyperlink{gaussian-translation}{Gaussian translation bound following Proposition~\ref*{prop:four-centres}} at \(\sigma=1\) limits every shift to \(\|\mu\|/2\). Each of the \(2n\) coordinates is at most \(n^2/(n!)^3\), so \(\|\mu\|/2\le\|\mu\|_1/2\le n^3/(n!)^3=o(1/(n!(n-1)!))\). Each modal-orbit member keeps probability at least \(1/(2n!C^n)\), while the centre type stays below \(3\cdot 16^n/(n!(n-1)!)\). The probability of the centre triangulation is at most \(6(16C)^n/(n-1)!\) times that of any orbit member, and the orbit bound yields \(p_*<3/(n-1)!\) at \(\mu\).

For the perturbation, fix a realizable face list. Its probability is real analytic in the joint mean. Indeed, the Gaussian integral over the realization region extends to complex means. On each compact set of means the integrand is bounded by a constant times \(e^{-\|x\|^2/4}\). Distinct realizable lists give distinct analytic functions. Indeed, dilate a generic realization of one list by a factor \(R\). After rescaling by \(1/R\), the Gaussian displacements converge to zero in probability. The list is constant in a neighbourhood of the original realization, so its probability tends to one and that of every other list tends to zero.

For each pair of lists the zero set of the difference is closed and has empty interior. The complement of the finitely many tie sets is therefore open and dense. The strict comparisons above and the centre triangulation persist in a neighbourhood of the stated means. Every smaller open neighbourhood there contains rational generic centres outside all the tie sets, giving a unique mode with the stated bounds.
\end{proof}

On the nearby unique-mode instances of Proposition~\ref{prop:centre-rank}, any rule choosing either the centre triangulation or one of \(N\) independent draws hits that mode with probability at most \(1-(1-p_*)^N\le Np_*<3N/(n-1)!\), so constant success probability requires \(N=\Omega((n-1)!)\).

\section{Sorting Gaussian means on the line gives a \texorpdfstring{\(1/2\)}{1/2}-approximation and is exact for four sites or symmetric means}\label{sec:line}
One-dimensional Gaussian sites with a common variance admit an elementary deterministic \(1/2\)-approximation. On the line, a Delaunay triangulation joins consecutive sites into a labelled path, which corresponds to two opposite left-to-right orders. We first recall why sorting the centres maximizes the probability of an oriented order, yielding factor \(1/2\) for unoriented paths, and then prove exact optimality for symmetric multisets of means and for every configuration of four sites.

For independent Gaussian sites \(X_1,\ldots,X_n\) of positive variance, write
\(p(\pi)=\Prob\{X_{\pi(1)}<\cdots<X_{\pi(n)}\}\)
for each permutation \(\pi\) of their labels, let \([\pi]\) denote the corresponding unoriented path, and let \(\pi^{\mathrm{rev}}\) denote the reversed permutation. For \(n\ge2\), the path has probability \(P([\pi])=p(\pi)+p(\pi^{\mathrm{rev}})\). Set \(P_*=\max_\pi P([\pi])\).

\begin{proposition}\label{prop:line}
Let \(X_i\sim N(\mu_i,\sigma^2)\), \(1\le i\le n\), be independent, with \(n\ge2\) and a common \(\sigma>0\). If \(\pi_0\) orders the means nondecreasingly, then
\[
 P([\pi_0])\ge\frac12 P_*.
\]
Equality \(P([\pi_0])=P_*\) holds when \(n=4\), and for every \(n\) when the multiset of means is symmetric about a point. The path \([\pi_0]\) can be constructed deterministically with \(O(n\log n)\) comparisons, and in polynomial bit time for rational means.
\end{proposition}
\begin{proof}
Savage~\cite{savage} studied rank-order probabilities. Palmer, Hill and Scheding~\cite[Theorem~1 and Appendix~I]{palmer} prove that sorting the means gives an optimal oriented order. We recall the short density comparison to deduce the path guarantee. Whenever \(\mu_i\le\mu_j\), the normal densities satisfy
\[
 \frac{f_i(x)f_j(y)}{f_j(x)f_i(y)}
 =\exp\!\left(\frac{(\mu_j-\mu_i)(y-x)}{\sigma^2}\right)\ge1
 \qquad(x<y).
\]
Correcting an adjacent inversion and integrating over ordered configurations therefore cannot decrease the order probability. Repeating adjacent swaps yields \(p(\pi)\le p(\pi_0)\) for every permutation \(\pi\) (allowing tied means), and for distinct means every corrected inversion gives a strict increase, so an oriented mode determines their sorted order. Because an unoriented labelled Delaunay path corresponds to two opposite left-to-right orders, its probability is the sum \(P([\pi])=p(\pi)+p(\pi^{\mathrm{rev}})\). Summing the oriented bound over \(\pi\) and \(\pi^{\mathrm{rev}}\) gives \(P([\pi])\le2p(\pi_0)\le2P([\pi_0])\) for every \(\pi\), hence
\[
 P_*\le2p(\pi_0)\le2P([\pi_0]).
\]
Sorting rational means and linking consecutive labels takes \(O(n\log n)\) comparisons and polynomial bit time without evaluating any probabilities.

For the exact cases, translate and rescale to unit variance and centred, sorted means \(d_1\le\cdots\le d_n\) with \(\sum_i d_i=0\). Write \(c=(d_{\pi(1)},\ldots,d_{\pi(n)})\) and let \(Z\) be the increasing order statistics of \(n\) independent standard normal variables. On the ordered cone, the Gaussian likelihood ratio gives \(p(\pi)=e^{-\|d\|^2/2}\E e^{c\cdot Z}/n!\). Because \(Z\) and \(-Z^{\mathrm{rev}}\) share the same distribution, adding the reversed permutation yields
\[
 P([\pi])=\frac{2e^{-\|d\|^2/2}}{n!}\,\E\cosh(c\cdot Z).
\]
When the multiset of means is symmetric about zero, \(d^{\mathrm{rev}}=-d\), so the rearrangement inequality gives \(-d\cdot Z\le c\cdot Z\le d\cdot Z\). Evenness of \(\cosh\) and its monotonicity on \([0,\infty)\) immediately prove \(P([\pi])\le P([\mathrm{id}])\).

For \(n=4\), sorting remains optimal even for asymmetric means. For a single ordered vector \(z\), \(\cosh(c\cdot z)\) can exceed \(\cosh(d\cdot z)\), so we pair each term with the reversed permutation. Since \(Z\overset{d}{=}-Z^{\mathrm{rev}}\) and \(\cosh\) is even, averaging \(\cosh(c\cdot Z)\) and \(\cosh(c^{\mathrm{rev}}\cdot Z)\) preserves the expectation in \(P([\pi])\). Put
\[
 x=c\cdot z,\qquad y=c^{\mathrm{rev}}\cdot z,\qquad
 a=d\cdot z,\qquad b=-d^{\mathrm{rev}}\cdot z.
\]
By the rearrangement inequality, every permutation \(c'\) of \(d\) satisfies \(-b\le c'\cdot z\le a\), and in particular \(x,y\in[-b,a]\). Because \(\sum_i d_i=0\), the average of \(c'\cdot z\) over all permutations vanishes, so \(a\ge0\) and \(b\ge0\). Using \(\sum_i c_i=0\), we have \(c_1+c_4=\tfrac12[(c_1-c_2)+(c_4-c_3)]\), and every pairwise difference of entries of \(d\) has magnitude at most \(d_4-d_1\), so \(|c_1+c_4|\le d_4-d_1\). Together with \(c_2+c_3=-(c_1+c_4)\) and \(2z_1\le z_2+z_3\le2z_4\), this gives
\[
 |x+y|=|c_1+c_4|\,|z_1+z_4-z_2-z_3|
 \le(d_4-d_1)(z_4-z_1)\le a+b,
\]
where the last inequality uses \(a+b=(d_4-d_1)(z_4-z_1)+(d_3-d_2)(z_3-z_2)\). Since \(x,y\in[-b,a]\) also implies \(|x-y|\le a+b\), we obtain \(\max(|x|,|y|)\le\max(a,b)\) and \(|x|+|y|=\max(|x+y|,|x-y|)\le a+b\). Relabelling the pairs so that \(|x|\ge|y|\) and \(a\ge b\), if \(|y|\le b\) then both entries are already dominated. Otherwise, shifting the excess \(|y|-b\) from \(|y|\) to \(|x|\) moves the pair to \((|x|+|y|-b,b)\) with \(|x|+|y|-b\le a\), which can only increase the sum of the convex even function \(\cosh\). In every case \(\cosh x+\cosh y\le\cosh a+\cosh b\). Taking expectations and using \(Z\overset{d}{=}-Z^{\mathrm{rev}}\) concludes \(P([\pi])\le P([\mathrm{id}])\).
\end{proof}

For common variance, sorting therefore gives a factor of at least \(1/2\) for every \(n\) and is exact in the cases above. The polynomial-time complexity of finding a most probable unoriented path for arbitrary means and \(n\) remains open.

\section{Encoding an independent set in Delaunay diagonals}\label{sec:socg-gaussian}

\begin{figure}[!htb]
\centering\includegraphics[width=.96\linewidth]{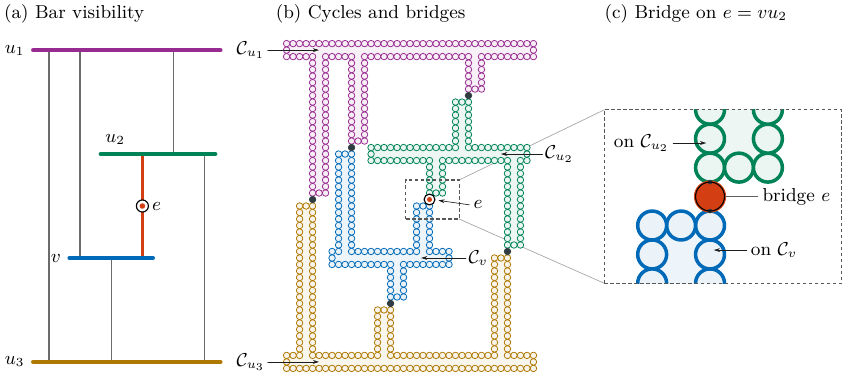}
\caption{From bars to cycles. In (a) and (b), the four bars of \(K_4\) become cycles and its six edges become bridges. In (c), the bridge for \(e=vu_2\) touches both cycles at opposite points.}\label{fig:vertex-chains}
\end{figure}

Four sites on an empty circle form a quadrilateral whose two diagonals encode a binary choice. Given a planar cubic graph \(G=(V,E)\), we represent each vertex by a cycle of tangent circles and each edge by a bridge (Fig.~\ref{fig:vertex-chains}), tuning four sites per circle so cycle neighbours couple at order \(t\) and bridges at order \(t^2\). Expanding Gaussian orthant probabilities shows that every modal sign pattern solves \textsc{Maximum Independent Set}, and triangulating the exterior at small scale preserves the gap between correct and failing outputs.

\begin{theorem}\label{thm:socg-gaussian}
For independent planar Gaussians with rational generic centres and common covariance \(I_2\), finding a most probable complete labelled Delaunay triangulation is NP-hard. For every fixed \(0<\epsilon<1\), the same holds for returning any triangulation of probability at least \(2^{-n^{1-\epsilon}}\) times the maximum, where \(n\) counts sites. Each reduction takes polynomial bit time and one solver call.
\end{theorem}
The proof starts with the NP-hard restriction of \textsc{Maximum Independent Set} to 2-connected planar cubic graphs~\cite[Theorem~4.1(a)]{mohar_face}. We build the decision cells for any planar cubic graph and then compute their sign probabilities under Gaussian displacements.

\subsection{Decision cells and the complete triangulation}\label{sec:geometry}\label{app:geometry}
We first construct a subdivision with one quadrilateral, called a \emph{decision cell}, on each empty circle. The four sites of a cell determine its two possible diagonals. We then identify which choices share a displacement, and complete the exterior so that the cell diagonals determine the entire triangulation.

Earlier stability tests handle points in disks~\cite{ely}, in rectangles~\cite{khanban}, and with dependent displacements~\cite{gitik}. Related algorithms identify Voronoi regions guaranteed under displacement~\cite{evans} and certify moves preserving a Delaunay triangulation~\cite{filtering}. Here we need to control the faces surrounding the cells while each cell retains two possible diagonals.

\subsubsection{From graph vertices to cycles of circles}\label{app:contacts}
To represent a vertex \(v\) of \(G\) without relying on long-range interactions, we replace \(v\) by a cycle \(\mathcal C_v\) of tangent unit circles, each carrying one decision cell. Two consecutive cells share the site at their point of tangency, so a displacement of that shared site affects both diagonal tests at once and will favour agreement around \(\mathcal C_v\). When all cells of \(\mathcal C_v\) agree, their common diagonal sign will record whether \(v\) enters the independent set. To couple two adjacent vertices \(u\) and \(v\), we insert one more unit circle, called the \emph{bridge} of the edge \(uv\), touching one circle of \(\mathcal C_u\) and one of \(\mathcal C_v\). Fig.~\ref{fig:vertex-chains} shows all six bridges for the four vertex cycles of \(K_4\).

\begin{lemma}\label{lem:geometry}
Given a planar cubic graph \(G=(V,E)\) with \(N=|V|\), one can construct in polynomial time \(m=O(N^3)\) unit circles with distinct centres in \(2\mathbb Z^2\). The circles consist of one induced tangency cycle \(\mathcal C_v\) for each vertex \(v\), together with one bridge for each edge \(uv\) joining vertices \(u\) and \(v\). The bridge of \(uv\) touches exactly one circle of \(\mathcal C_u\) and one of \(\mathcal C_v\). Each cycle turns through a right angle at its incident circle. The two points of tangency are opposite ends of a diameter of the bridge. All other tangencies are between consecutive circles of a single cycle. Each cycle circle touches at most one bridge. In each cycle we designate a circle that touches no bridge. Its two neighbours touch it at opposite ends of a diameter. All disk interiors are disjoint, and no point of the plane lies on three of the circles.
\end{lemma}
\begin{proof}[Proof of Lemma~\ref{lem:geometry}]
We turn a visibility drawing into cycles by thickening the bars into rectangles and running arms along the edge corridors. The facing corners of two arms leave room for a single bridge. Since \(G\) is cubic, \(N\ge4\). Add edges to obtain a maximal planar graph, which is \(3\)-connected. The \(\varepsilon\)-visibility theorem of Tamassia and Tollis~\cite[Section~4, Theorem~4]{visibility}, which requires \(2\)-connectivity, gives horizontal bars at distinct heights and an open vertical band of lines of sight for every edge. Keep the bands for the edges of \(G\), and replace endpoint abscissas and heights by their integer ranks. This coordinatewise monotone change preserves the bands and makes their endpoints integers.

Number the edges \(e=1,\ldots,E\). Choose an integer \(j_e\) with \((j_e,j_e+1)\) in the band of \(e\), multiply bar coordinates by \(100(E+1)\), and put \(x_e=100(E+1)j_e+100e\). Distinct bar heights, endpoints and chosen abscissas are at least \(100\) apart. Every bar strictly between the endpoints of \(e\) misses its open band and is at horizontal distance at least \(100\) from \(x_e\).

For \(e=uv\) with heights \(h_u<h_v\), set \(y_e=(h_u+h_v)/2\), an integer. Thicken each bar by ten in the \(\ell_\infty\) metric. Add to the polygon for \(u\) the arm
\([x_e-20,x_e]\times[h_u+10,y_e]\), and to that for \(v\) the arm
\([x_e,x_e+20]\times[y_e+2,h_v-10]\). The two facing corners are \(q_e=(x_e,y_e)\) and \(r_e=(x_e,y_e+2)\). Rectangles have height twenty, arms width twenty and length at least thirty-eight. Arms of distinct edges are at least \(60\) apart. The separations of \(100\) keep all other pieces of different vertices at least ten apart and keep nonincident sides of each polygon at least twenty apart.

Each rectangle with its attached arms is an orthogonal polygon homeomorphic to a disk. Two boundary lattice points are at distance one exactly when they are consecutive along the boundary: nonincident sides are separated, and the nearest points on two incident sides other than their common corner are at distance \(\sqrt2\). The boundary points therefore form an induced cycle. The only boundary points at distance one from \(\beta_e=(x_e,y_e+1)\) are \(q_e,r_e\). Its other two lattice neighbours have ordinate \(y_e+1\), absent from the two arms, and the other pieces are separated by at least ten. Different \(\beta_e\)'s are separated by their abscissas. When every boundary is traversed counterclockwise, the predecessor of \(q_e\) is the lattice point below it and the predecessor of \(r_e\) is the one above it, while \(\beta_e\) lies above \(q_e\) and below \(r_e\). At both corners the bridge contact is therefore opposite to the predecessor contact. The successor of \(q_e\) is the lattice point on its left and the successor of \(r_e\) is the one on its right. Hence the cells at the corners \(q_e\), for the different edges, are translates of each other, and the cell at \(r_e\) is the image of the cell at \(q_e\) under a point reflection. The two lattice neighbours of \(\beta_e\) with ordinate \(y_e+1\) carry no circle, so the west point of every bridge is unused and all bridges are translates of each other.

Double all coordinates, place a unit circle at each boundary point and add the unit circle centred at \(2\beta_e\) for each edge. The preceding distances give exactly the tangencies in the statement, with the two bridge contacts opposite. Each cycle has at least eighty cells and only three bridge incidences. If the thickened bar of a vertex is \([\ell-10,r+10]\times[h-10,h+10]\), its designated cell is the one at the lattice point \((\ell,h+10)\) of the upper side. Every arm is at horizontal distance at least eighty from this point, so its two cycle neighbours lie on that side and its contacts with them are opposite. The counterclockwise orientation runs westwards along the upper side, so every designated cell has its successor contact at its west point, its predecessor contact at its east point and \(\mathbf C\) at its north point. The designated cells of all cycles, with their four sites, are therefore translates of each other. Every pair of distinct centres has distance at least two. Their open disks are disjoint. Three such centres cannot have a common unit-distance point: each pair would then have distance exactly two, and three points of a unit circle cannot be pairwise at distance two. Thus no site belongs to three circles. The \(O(N)\) pieces lie in a box of side \(O(N^2)\), and each polygon has a constant number of sides. Its boundary therefore contains \(O(N^2)\) lattice points, giving \(O(N^3)\) over all vertices. Including the \(E=3N/2\) bridges, \(m=O(N^3)\), and the construction takes polynomial time.
\end{proof}

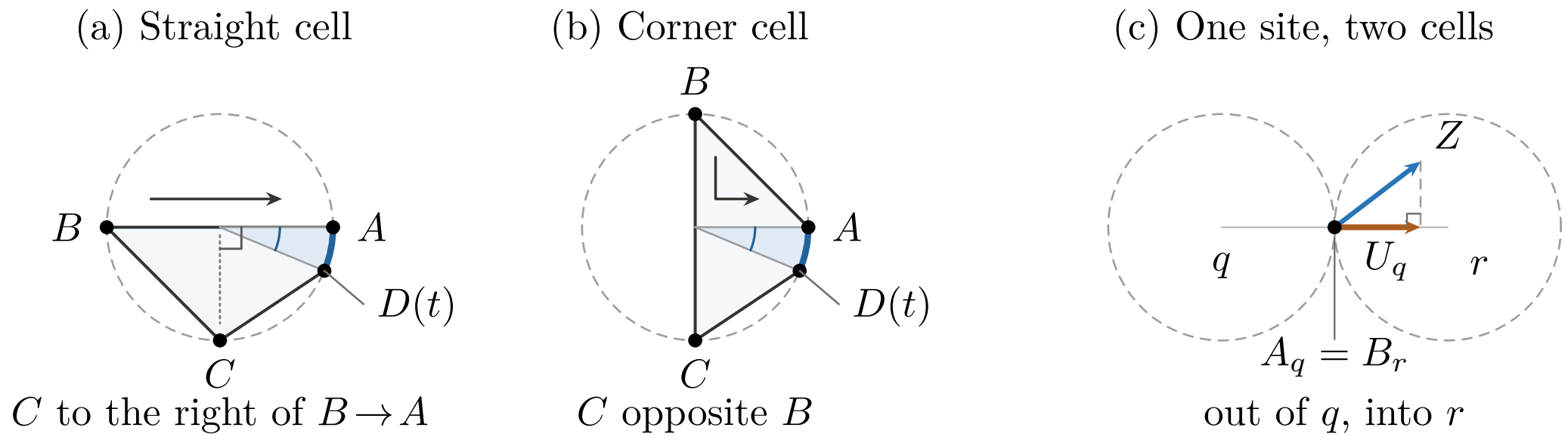
\begin{figure}[!htbp]
\centering
\begingroup

\definecolor{ndtBlue}{HTML}{2875B7}
\definecolor{ndtOrange}{HTML}{DB7627}
\tikzset{
 ndt picture/.style={font=\small,text=black,line cap=round,line join=round},
 ndt construction/.style={draw=black!38,line width=.5pt,dash pattern=on 2.4pt off 2pt},
 ndt edge/.style={draw=black!80,line width=.75pt},
 ndt selected/.style={draw=ndtBlue,line width=1.35pt},
 ndt arrow/.style={draw=black!80,line width=.65pt,-{Stealth[length=4pt,width=3.4pt]}},
 ndt dot/.style={circle,fill=black,inner sep=0pt,minimum size=3.5pt},
 ndt part/.style={anchor=west,font=\small}
}
\begin{tikzpicture}[ndt picture,x=.99cm,y=.99cm]
\foreach \x/\corner/\title in {0/0/{(a) Straight cell},4.2/1/{(b) Corner cell}}{
 \begin{scope}[shift={(\x,0)}]
 \node[anchor=west] at (-1.42,1.75) {\title};
 \coordinate (A) at (1,0); \coordinate (C) at (0,-1);
 \coordinate (D) at ({12/13},{-5/13});
 \ifnum\corner=1 \coordinate (B) at (0,1); \else \coordinate (B) at (-1,0); \fi
 \draw[ndt construction] (0,0) circle (1);
 \draw[ndt edge,fill=ndtBlue!5] (A)--(B)--(C)--(D)--cycle;
 \fill[ndtBlue!15] (0,0)--(1,0) arc[start angle=0,end angle=-22.619865,radius=1]--cycle;
 \draw[black!39,line width=.48pt] (0,0)--(A) (0,0)--(D);
 \draw[ndtBlue!80!black,line width=.58pt] (.53,0) arc[start angle=0,end angle=-22.619865,radius=.53];
 \draw[ndtBlue!85!black,line width=1.6pt] (1,0) arc[start angle=0,end angle=-22.619865,radius=1];
 \foreach \p in {A,B,C,D} \node[ndt dot] at (\p) {};
 \node[right=2pt] at (A) {$A$}; \node[below=1pt] at (C) {$C$};
 \ifnum\corner=1
  \node[above=1pt] at (B) {$B$};
  \draw[ndt arrow] (.18,.62)--(.18,.25)--(.57,.25);
  \node at (0,-1.66) {$C$ opposite $B$};
 \else
  \node[left=2pt] at (B) {$B$};
  \draw[ndt arrow] (-.61,.25)--(.55,.25);
  \draw[black!50,line width=.55pt,densely dotted] (0,-.07)--(0,-.85);
  \draw[black!65,line width=.55pt] (0,-.19)--(.19,-.19)--(.19,0);
  \node at (0,-1.66) {$C$ to the right of $B\!\to\!A$};
 \fi
 \draw[black!55,line width=.5pt] (D)--(1.27,-.68);
 \node[anchor=west] at (1.25,-.70) {$D(t)$};
 \end{scope}}
\begin{scope}[shift={(9.85,0)}]
 \node[anchor=west] at (-2.1,1.75) {(c) One site, two cells};
 \foreach \x in {-1,1}{\draw[ndt construction] (\x,0) circle (1);}
 \draw[black!28] (-1,0)--(1,0);
 \node at (-1,-.34) {$q$}; \node at (1.28,-.34) {$r$};
 \draw[ndt arrow,draw=ndtBlue,line width=1.2pt] (0,0)--(.76,.58) node[above right=-1pt,text=black] {$Z$};
 \draw[ndt arrow,draw=ndtOrange!78!black,line width=1.45pt] (0,0)--(.76,0);
 \draw[ndt construction] (.76,.58)--(.76,0);
 \draw[black!58,line width=.45pt] (.64,0)--(.64,.12)--(.76,.12);
 \node[anchor=north,inner sep=1.4pt] at (.46,-.07) {$U_q$};
 \draw[black!55,line width=.45pt] (0,-.08)--(0,-.99);
 \node at (0,-1.15) {$A_q=B_r$};
 \node[ndt dot] at (0,0) {};
 \node at (0,-1.66) {out of $q$, into $r$};
\end{scope}
\end{tikzpicture}

\endgroup
\caption{The sites of a decision cell at \(t=1/5\). In (a), travel from \(B\) to \(A\) puts the unused cardinal site \(C\) on the right. In (b), the turn puts \(C\) opposite \(B\). The marked arc places \(D(t)\) at angle \(2\arctan t\) from \(A\) towards \(C\). In (c), the orange component \(U_q\) of the displacement \(Z\) at \(A_q=B_r\) points out of cell \(q\) and into cell \(r\).}\label{fig:cell-geometry}
\end{figure}

We now place the four sites of each cycle cell. Traverse \(\mathcal C_v\) counterclockwise and consider a circle \(q\) with centre \(o_q\). Enter \(q\) at its contact \(\mathbf B_q\) with its predecessor \(q^-\) and leave at its contact \(\mathbf A_q\) with its successor. At a corner where the cycle turns through a right angle, place \(\mathbf C_q\) opposite \(\mathbf B_q\) across a diameter, so that whenever a bridge touches this corner, \(\mathbf C_q\) is its contact with \(q\). Along a straight segment, place \(\mathbf C_q\) at the cardinal point facing outward to the right of travel. Fig.~\ref{fig:cell-geometry}(a) and (b) show these two shapes.

The fourth site, \(\mathbf D_q(t)\), belongs to \(q\) alone. Place it near \(\mathbf A_q\) on the arc toward \(\mathbf C_q\) that avoids \(\mathbf B_q\): in local coordinates with origin \(o_q\) and \(\mathbf A_q=(1,0)\),
\[
 \mathbf D_q(t)=\left(\frac{1-t^2}{1+t^2},
              \varsigma_q\frac{2t}{1+t^2}\right),
\]
where \(\varsigma_q\in\{-1,1\}\) selects the arc and \(0<t<1/4\) is a rational scale shared by all cycle cells. The four sites lie in cyclic order \(\mathbf A_q,\mathbf B_q,\mathbf C_q,\mathbf D_q\), so the quadrilateral has two competing diagonals, \(\mathbf A_q\mathbf C_q\) and \(\mathbf B_q\mathbf D_q\). Each shared contact carries one label and one random displacement, as in Fig.~\ref{fig:cell-geometry}(c), while distinct sites receive independent displacements from the fixed law \(\nu\), including the completion sites added in Section~\ref{sec:geom-stability}.

To ensure that every decision cell is an actual face of the unperturbed Delaunay subdivision, its closed unit disk must contain no site other than its own four vertices. Take any other circle in the construction, with centre \(o_q+2\mathbf u\) for some \(\mathbf u\in\mathbb Z^2\setminus\{0\}\), and any site \(o_q+2\mathbf u+\mathbf v\) on it, where \(\|\mathbf v\|=1\). Its squared distance from \(o_q\) satisfies
\[
 \|2\mathbf u+\mathbf v\|^2
 =1+4\bigl(\|\mathbf u\|^2+\langle\mathbf v,\mathbf u\rangle\bigr)\ge1,
\]
because \(\langle\mathbf v,\mathbf u\rangle\ge-\|\mathbf u\|\ge-\|\mathbf u\|^2\). Equality holds only when \(\|\mathbf u\|=1\) and \(\mathbf v=-\mathbf u\), which is the tangency point already shared by the two cells. Apart from a shared tangency site, every site of another cell therefore lies strictly outside the closed circumdisk of \(q\).

\begin{figure}[!htb]
\centering
\includegraphics[width=.96\linewidth]{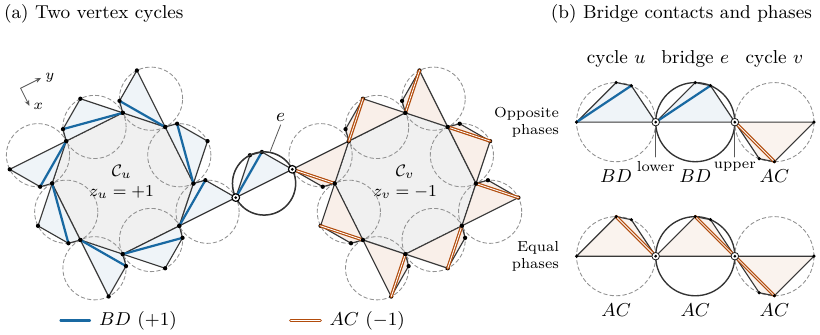}
\caption{A bridge between cycle phases. In (a), the contacts are ringed and all 17 cells use $t=1/5$. In (b), the opposite phases shown favour $BD$ by agreement below and opposition above. Equal phases cancel the order-$t^2$ preference for a bridge sign ($AC$ shown).}\label{fig:socg-block}
\end{figure}

\subsubsection{Linearizing the diagonal test and coupling cycles}\label{app:cofactors}
At the base configuration the four vertices of a cell are cocircular. To see which diagonal is selected under a small random displacement, we compute the first-order change of the cell's incircle determinant. Because \(\mathbf D_q(t)\) sits close to \(\mathbf A_q\), the four vertices respond with very different sensitivities: \(\mathbf A_q\) and \(\mathbf D_q\) carry weight near one, whereas \(\mathbf B_q\) and \(\mathbf C_q\) carry weight of order \(t\). To see how one edge couples two cycles, consider the bridge \(x\) joining the cycles for \(u\) and \(v\) in Fig.~\ref{fig:socg-block}. The weights favour a common diagonal sign on each cycle and opposite signs across the bridge.

Fix a decision cell \(q\) with base sites \(\mathbf A_q,\mathbf B_q,\mathbf C_q,\mathbf D_q\). Its two diagonals intersect in their interiors. Equating the convex combinations for that intersection and normalizing the coefficient of \(\mathbf D_q\) to \(-1\) produces the unique affine dependence
\begin{equation}\label{eq:affine}
 a_q\mathbf A_q-b_q\mathbf B_q+c_q\mathbf C_q-\mathbf D_q=0,
 \qquad a_q-b_q+c_q-1=0,
\end{equation}
with positive weights \(a_q,b_q,c_q\). Geometrically, each weight is the ratio of the area of the triangle spanned by the other three vertices to the area of \(\mathbf A_q\mathbf B_q\mathbf C_q\)~\cite[Section~3.2]{khanimov}. Substituting the explicit coordinates yields Table~\ref{tab:coeff}, and reflecting the cell preserves all three weights while negating \(\varsigma_q\).

\begin{table}[ht]
\caption{Exact coefficients of the affine dependence \eqref{eq:affine} for a decision cell rotated so that \(\mathbf A\) is at east. \(N,W,S\) denote the other cardinal points. Entries in parentheses give the reflected case, which has the same coefficients. The first row gives the corner cycle cells, in which \(\mathbf B\) is a quarter turn from \(\mathbf A\) and \(\mathbf C\) is opposite to \(\mathbf B\). The second row gives the straight cells.}\label{tab:coeff}
\centering
\begin{tabular}{ccrccc}
\toprule
\(\mathbf B\)&\(\mathbf C\)&\(\varsigma\)&\((1+t^2)a\)&\((1+t^2)b\)&\((1+t^2)c\)\\
\midrule
\(N\) (\(S\))&\(S\) (\(N\))&\(-1\) (\(1\))&\(1-t^2\)&\(t(1-t)\)&\(t(1+t)\)\\
\midrule
\(W\)&\(S\) (\(N\))&\(-1\) (\(1\))&\(1-t\)&\(t(1-t)\)&\(2t\)\\
\bottomrule
\end{tabular}
\end{table}

Table~\ref{tab:coeff} shows that \(a_q=1+O(t)\), \(b_q=t(1+O(t))\) and \(c_q=O(t)\). At every corner cell, the weight of \(\mathbf C_q\) is \(c(t)=t(1+t)/(1+t^2)\), so the two cycle cells touching any bridge share this exact weight.

\begin{figure}[H]
\centering
\includegraphics[width=\linewidth]{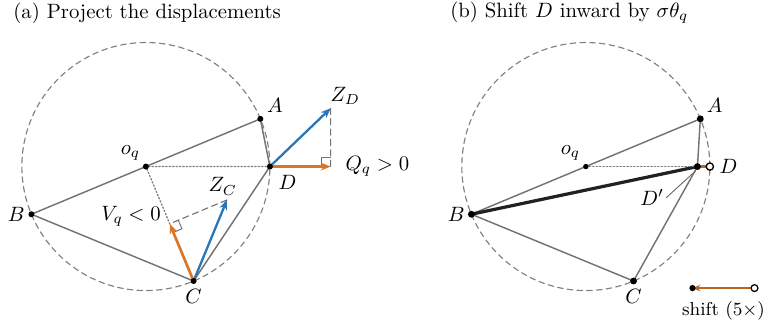}
\caption{Radial displacement and an inward shift of the fourth centre. In (a), projecting the blue displacements onto the dotted radii gives the orange components \(V_q<0\) at \(C\) and \(Q_q>0\) at \(D\), in units of \(\sigma\). In (b), the shifted centre \(D'\) lies inside the circle through \(A,B,C\) and outside triangle \(ABC\), so the quadrilateral takes diagonal \(BD'\). The detail enlarges the radial shift \(\sigma\theta_q\) fivefold.}\label{fig:radial-field}
\end{figure}

The exact diagonal test of cell \(q\) is its normalized incircle determinant \(\Gamma_q\). Take coordinates relative to \(o_q\) and order the four rows as \((\mathbf P_1,\ldots,\mathbf P_4)=(\mathbf A_q,\mathbf B_q,\mathbf C_q,\mathbf D_q)\), giving site \((x_i,y_i)\) the row \((x_i,y_i,x_i^2+y_i^2,1)\). Multiplying this determinant by \(\iota_q/4\), where \(\iota_q=\sgn\operatorname{orient}(\mathbf A_q,\mathbf B_q,\mathbf C_q)\) at the base, defines \(\Gamma_q\). Under a small generic perturbation, \(\Gamma_q>0\) places \(\mathbf D_q\) inside the circumcircle of the other three sites and selects diagonal \(\mathbf B_q\mathbf D_q\) (Fig.~\ref{fig:radial-field}(b)), whereas \(\Gamma_q<0\) selects \(\mathbf A_q\mathbf C_q\).

Let \(\mathbf n_i=\overrightarrow{o_q\mathbf P_i}\) be the outward unit normal at base vertex \(i\). Displacing each vertex by \(h\mathbf z_i\) changes \(\Gamma_q\) to first order by \(h\) times
\begin{equation}\label{eq:variation}
 a_q\langle\mathbf n_1,\mathbf z_1\rangle-b_q\langle\mathbf n_2,\mathbf z_2\rangle
 +c_q\langle\mathbf n_3,\mathbf z_3\rangle-\langle\mathbf n_4,\mathbf z_4\rangle .
\end{equation}
Each inner product records the outward radial motion of a vertex, while tangential motion enters only at quadratic order. Consequently, the weights of \eqref{eq:affine} govern the first-order response with alternating signs around the circle.

To check the normalization, put the origin at \(o_q\). Displace each site from its base position \(\mathbf n_i\) to \(\mathbf n_i+h\mathbf z_i\). Subtracting the last column of the incircle determinant from its third leaves the entries \(\|\mathbf n_i+h\mathbf z_i\|^2-1\), all zero at \(h=0\). The cofactors of the other columns therefore vanish there, and the derivative of the third entry at \(h=0\) is the scalar product \(2\langle\mathbf n_i,\mathbf z_i\rangle\). Thus, after multiplication by \(\iota_q\), the determinant has first variation \(2\sum_iK_i\langle\mathbf n_i,\mathbf z_i\rangle\), where \(K_i\) is the cofactor in the third column multiplied by \(\iota_q\).

To identify these cofactors, replace the third column by any other column. The resulting determinant is zero, so \(\sum_iK_i\mathbf n_i=0\) and \(\sum_iK_i=0\). The affine dependence of four distinct concyclic points is unique up to scale. Equation~\eqref{eq:affine} therefore makes \(K\) proportional to \((a_q,-b_q,c_q,-1)\). The fourth cofactor is
\(K_4=-\iota_q\operatorname{orient}(\mathbf A_q,\mathbf B_q,\mathbf C_q)=-2\), since three distinct cardinal points of the unit circle have absolute orientation determinant two. Hence \(K=2(a_q,-b_q,c_q,-1)\). Dividing the first variation by four gives \eqref{eq:variation}, with the same signs and scale.

In base units the circles have radius one, and each site receives a random displacement \(\sigma Z_i\) with independent \(Z_i\sim\nu\). After choosing a small common scale \(\sigma>0\) in Section~\ref{sec:geom-stability}, dividing all coordinates by \(\sigma\) preserves the triangulation and restores unit noise scale \(Z_i\). In units of \(\sigma\), write \(V_q\) and \(Q_q\) for the outward radial components of the displacements at \(\mathbf C_q\) and \(\mathbf D_q\), respectively (Fig.~\ref{fig:radial-field}(a)). By \eqref{eq:variation}, these two vertices contribute \(c_qV_q-Q_q\) to the linearized test.

To bias a cell toward diagonal \(\mathbf B_q\mathbf D_q\), we shift the nominal position of \(\mathbf D_q\) inward by \(\sigma\theta_q\) along its radius, that is, by \(-\sigma\theta_q(\mathbf D_q-o_q)\) for a real parameter \(\theta_q\) called the cell's \emph{field} (Fig.~\ref{fig:radial-field}(b)). For the same random draw, the outward radial shift of \(\mathbf D_q\) from its base position becomes \(Q_q-\theta_q\), which its weight \(-1\) converts into an additive term \(+\theta_q\) in the linear test.

We can now trace agreement along the cycle \(\mathcal C_u\) of our two-vertex example. At the contact \(\mathbf A_q=\mathbf B_r\) between consecutive cells \(q\) and \(r\) of \(\mathcal C_u\) (Fig.~\ref{fig:cell-geometry}(c)), let \(U_q\) be the radial component of the shared displacement measured outward from \(o_q\). Viewed from the centre \(o_r\) of the next cell, the same displacement points inward and has outward radial component \(-U_q\). Equation~\eqref{eq:variation} therefore contributes \(+a_qU_q\) to cell \(q\) and \((-b_r)(-U_q)=+b_rU_q\) to cell \(r\): moving the contact away from \(o_q\) increases both linear tests together. At the entry contact shared with \(q^-\), the same reversal contributes \(+b_qU_{q^-}\) to cell \(q\). Collecting all four vertices and the field gives the linearized cycle test
\begin{equation}\label{eq:test}
 \Lambda_q=a_qU_q+b_qU_{q^-}+c_qV_q-Q_q+\theta_q.
\end{equation}

\begin{figure}[!htbp]
\centering
\includegraphics[width=\linewidth]{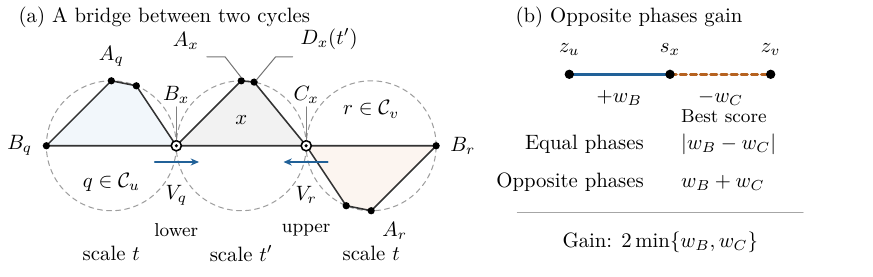}
\caption{Why a bridge favours opposite cycle phases. In (a), a clockwise quarter-turn places the lower contact \(B_x\) on the left and the upper contact \(C_x\) on the right. Each is shared with a cycle cell, \(q=q_B(x)\) or \(r=q_C(x)\). The arrows point outward from these cells and fix the positive directions of \(V_q,V_r\) in \eqref{eq:bridgetest}. Only the bridge uses \(A_x,D_x\). In (b), with zero fields and weights from \eqref{eq:weightsdef}, optimizing \(s_x\) gives \(|w_Bz_u-w_Cz_v|\), larger by \(2\min\{w_B,w_C\}\) when the phases differ.}\label{fig:bridge}
\end{figure}

Next, follow the bridge \(x\) connecting a corner cell \(q=q_B(x)\in\mathcal C_u\) to a corner cell \(r=q_C(x)\in\mathcal C_v\) (Fig.~\ref{fig:bridge}(a)). Our obstacle is to favour opposite phases on \(\mathcal C_u\) and \(\mathcal C_v\) while keeping bridge effects weaker than cycle agreement so that nonconstant cycles remain less probable than constant ones. We build the bridge as one more corner cell, using its own rational parameter \(0<t'\le t\). For standard Gaussian noise we simply set \(t'=t\) in Section~\ref{sec:socg-gaussian}, because bridge correlations already scale as \(t^2\) while cycle agreement scales as \(t\). Keeping \(t'\le t\) available lets Section~\ref{sec:heavy} guarantee separation of bridge and cycle weights across the entire class under \eqref{eq:L}.

\begin{figure}[!htb]
\centering
\includegraphics[width=\linewidth]{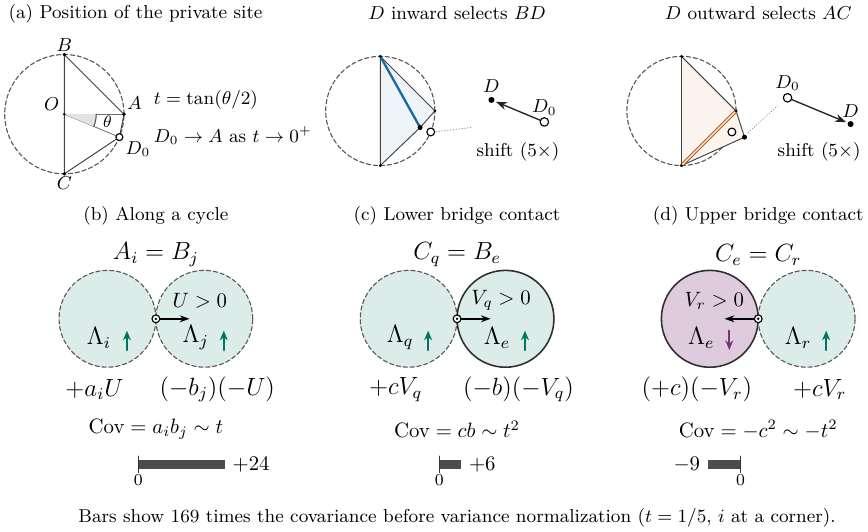}
\caption{Radial shifts and covariances. In (a), the insets magnify the shifts fivefold. In (b)--(d), black arrows move the shared sites. Up and down arrows give the sign of each test's change.}\label{fig:socg-diagonals}
\end{figure}

In the lattice layout, the bridge's two contacts are its south point \(\mathbf B_x=\mathbf C_{q_B(x)}\) and its north point \(\mathbf C_x=\mathbf C_{q_C(x)}\). Place \(\mathbf A_x\) at the unused west point of \(x\) and \(\mathbf D_x(t')\) on the arc from \(\mathbf A_x\) toward \(\mathbf C_x\) avoiding \(\mathbf B_x\). Rotating by a half-turn matches Table~\ref{tab:coeff} (corner row), giving bridge weights
\[
 (a',b',c')=\frac{1}{1+t'^2}
       \bigl(1-t'^2,\ t'(1-t'),\ t'(1+t')\bigr).
\]
Write \(U_x,Q_x\) for the outward radial projections at the unshared sites \(\mathbf A_x,\mathbf D_x\). At both contacts, the outward normal of the bridge opposes the outward normal of the incident cycle cell. In \eqref{eq:variation}, the lower contact carries weight \(-b'\), so reversing the normal gives \(+b'V_{q_B(x)}\), whereas the upper contact carries weight \(+c'\), giving \(-c'V_{q_C(x)}\). The linearized bridge test is therefore
\begin{equation}\label{eq:bridgetest}
 \Lambda_x=a'U_x+b'V_{q_B(x)}-c'V_{q_C(x)}-Q_x.
\end{equation}
Both shared projections appear with positive weight \(+c(t)\) in their respective cycle tests \eqref{eq:test}. Consequently, \(V_q\) shifts \(\Lambda_x\) in the same direction as \(\Lambda_q\), whereas \(V_r\) shifts \(\Lambda_x\) in the direction opposite to \(\Lambda_r\) (as in Fig.~\ref{fig:socg-diagonals}(c) and (d), with bridge weights \(b',c'\) replacing \(b,c\) and cycle weight \(c(t)\) replacing \(c\)). When \(\mathcal C_u\) and \(\mathcal C_v\) take opposite phases, say \(+1\) on \(\mathcal C_u\) and \(-1\) on \(\mathcal C_v\) as in Fig.~\ref{fig:socg-block}, setting \(s_x=+1\) matches the preferred sign at both contacts simultaneously (Fig.~\ref{fig:bridge}(b)). When \(\mathcal C_u\) and \(\mathcal C_v\) share the same phase, every choice of \(s_x\) conflicts with one endpoint. Across the entire construction, radial projections measured once per distinct site are independent, and by rotational invariance each has the law of \(U\), the projection of \(Z\sim\nu\) onto a fixed unit vector.

When all cells of a cycle \(\mathcal C_v\) share the same sign, we call that sign the \emph{phase} of \(v\): phase \(+1\) selects \(v\) and phase \(-1\) excludes it. We place a nonzero field only on the designated cell of each cycle, leaving all other cycle cells and all bridges with zero field. A \emph{block} consists of all \(m\) decision cells for \(G\) with one scale pair \((t,t')\) and one field vector \(\theta\). Write \(p_{t,t'}^\theta(s)\) for the probability that the \(m\) linear tests \eqref{eq:test}--\eqref{eq:bridgetest} attain the sign vector \(s\in\{-1,1\}^m\). Fig.~\ref{fig:socg-diagonals}(a) depicts the two local diagonals of one cell, and Fig.~\ref{fig:socg-block} illustrates our two-vertex example with one bridge under opposite phases.

\subsubsection{Completing the subdivision and stabilizing the geometry}\label{sec:geom-stability}\label{app:completion-grid}
Two geometric obstacles separate the linear sign vector \(s\) from a complete labelled Delaunay triangulation returned by the modal oracle. First, the unperturbed layout contains nontriangular faces outside the decision cells (such as the shaded octagons in Fig.~\ref{fig:socg-block}) as well as an unconstrained outer hull, and multiple triangulations of those exterior regions would split the probability of one sign vector across several complete types. Second, real realizations obey the exact nonlinear determinants \(\Gamma_q\) on generic rational centres, while the combinatorial analysis uses the linearized tests \(\Lambda_q\) on cocircular base points. We resolve the first obstacle by inserting guard and completion sites, and the second by choosing a small dyadic scale \(\sigma\) and tiny generic shifts that preserve all noncell faces and match exact signs to linear signs away from near-ties.

\begin{samepage}
\begin{lemma}\label{lem:completion}
Consider any finite collection of these decision cells, including separated copies built with different parameters. In polynomial time one can add rational sites outside their closed circumdisks so that the convex hull is triangular, each closed circumdisk still contains exactly the four vertices of its cell, and the decision cells are the only nontriangular Delaunay faces. With \(M\) cells the final number of sites satisfies \(n\le12M+4\). All coordinates have polynomial bit length in the input coordinates and \(M\).
\end{lemma}
\end{samepage}

\begin{proof}
We first enclose all cells in a triangle of three new sites, called guards, and then insert one site inside each unwanted nontriangular face, outside every closed cell circumdisk. Each added site is chosen from a rational grid so that it avoids all earlier sites, all lines through pairs of earlier sites, and all circles through noncollinear triples of earlier sites.

Translate the layout into \([-D,D]^2\) for an integer \(D\) exceeding its diameter plus ten. Choose one guard from each of the unit boxes around \((-4D,-3D)\), \((4D,-3D)\) and \((0,4D)\). Their triangle strictly encloses every closed cell circumdisk (Fig.~\ref{fig:completion}(c) and (d)). The resulting \(n_0\le4M+3\) sites have a triangular hull. Testing all noncollinear triples for empty circumcircles and grouping cocircular triples identifies all faces of this intermediate Delaunay subdivision in polynomial time.

\begin{figure}[!tbp]
\centering
\includegraphics[width=\linewidth]{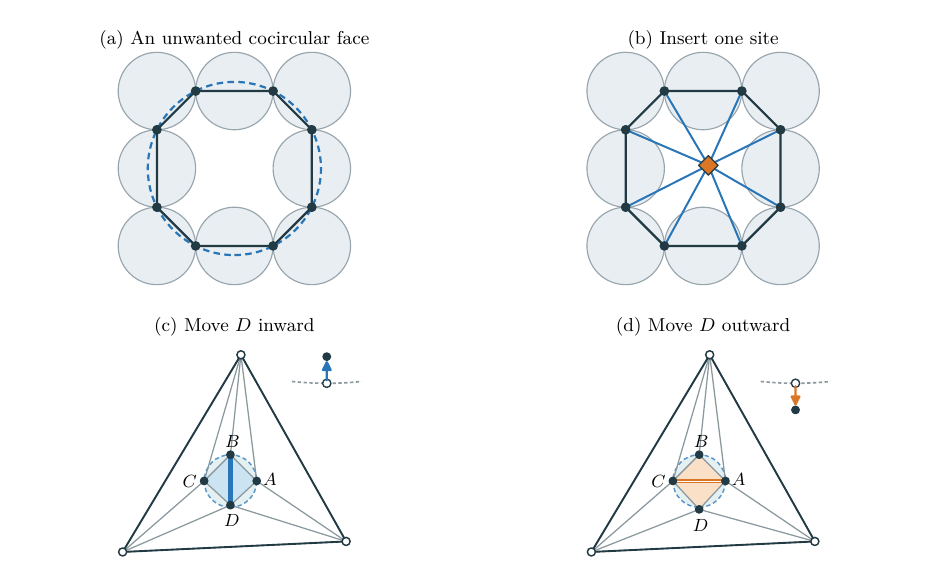}
\caption{Triangulating an unwanted face preserves the decision cells. In (a), the eight contact sites surround an empty octagonal face of Fig.~\ref{fig:socg-block}(a). The dashed circumcircle has no site in its interior. In (b), a new site (diamond) lies inside that face and outside every shaded unit disk. Its eight incident triangles replace the octagon. In (c) and (d), a separate seven-site example shows both diagonal choices inside a triangular hull of three guards. With the other six sites fixed, shifting \(D\) inward or outward by \(8\%\) selects \(BD\) or \(AC\). The triangulations share seven faces. Each inset marks the original \(D\) by an open point and its new position by a filled point.}\label{fig:completion}
\end{figure}

Take any unwanted nontriangular face \(F\). The closed cell circumdisks cannot cover the connected open set \(\operatorname{int}F\): their interiors are disjoint and their tangency points lie on the boundary of their union, so covering \(\operatorname{int}F\) would place \(\operatorname{int}F\) inside a single closed circumdisk, whose only sites are the four vertices of one decision cell, contradicting that \(F\) is unwanted. Hence \(\operatorname{int}F\) contains an open disk disjoint from all closed cell circumdisks. We select one such rational disk per unwanted face before any insertions, keeping all disk bit lengths dependent only on the initial subdivision, and insert one grid point in each disk (Fig.~\ref{fig:completion}(a) and (b)).

A horizontal sweep gives an open disk with rational centre and radius in \(\operatorname{int}F\) outside the closed disks for each unwanted face \(F\) of Lemma~\ref{lem:completion}. The critical ordinates of this open set, at vertices, circle extrema and pairwise intersections of sides and circles, are algebraic of bounded degree and polynomial bit length. Root separation lets us choose a horizontal line at a rational height between two consecutive critical ordinates, meeting the open set. On this line we choose a rational point in the set. Dividing the slack in the defining inequalities at this point by a bound on their gradients gives an interior square, hence an open disk, with dyadic centre and radius of polynomial bit length.

\emph{Grid count. If \(S\) has at most \(\bar n\) points and \(k=\bar n^3+1\), every \(k\)-by-\(k\) square grid has a point outside \(S\), off the lines through two and the circles through three noncollinear points of \(S\).} Indeed, a line contains at most \(k\) grid points and a circle at most \(2k\), so at most \(\bar n+k\binom{\bar n}2+2k\binom{\bar n}3\le k\bar n^3<k^2\) grid points are excluded. With \(\bar n=12M+4\), bounding the number of points throughout Lemma~\ref{lem:completion}, grids fixed in the guard boxes and disks supply the guards and inserted points. Lemma~\ref{lem:stability} uses the same count with \(\bar n=n\).

After all insertions, no guard or inserted site can belong to a face with four or more vertices, because the last chosen vertex of such a face would lie on the circle through three earlier vertices, which the grid selection rules out. Every remaining nontriangular face therefore uses only original sites, so its open circumdisk was already empty before the insertions and it was already a face then. Since every unwanted face received an inserted site strictly inside its circumdisk, only the decision cells remain nontriangular, the guards still form the hull, and each cell circumdisk still contains only its four vertices.

To bound \(n\), triangulate the subdivision with the guards in place. Having \(n_0\) sites and three hull vertices, it has \(2n_0-5\) bounded triangles. Each unwanted face contains at least one such triangle, so at most \(2n_0-5\) sites are inserted, yielding \(n\le 3n_0-5\le12M+4\).
\end{proof}

Fig.~\ref{fig:completion}(c) and (d) show how one cell's diagonal then fixes the entire triangulation. Let \(f_3\) be the number of triangular faces in the completed base configuration \(P\). Adding one diagonal per decision cell yields \(f_3+2M\) triangles, and Euler's formula for \(n\) sites with three hull vertices gives
\begin{equation}\label{eq:euler}
 f_3+2M=2n-5.
\end{equation}
We call the following strict orientation and incircle determinants at \(P\) the \emph{certificates} of the completed configuration. First, every nonguard site lies strictly inside the guard triangle, and each of the \(f_3\) fixed triangles keeps its orientation and empty circumcircle. Second, within each decision cell, all four vertex triples keep their orientations (preserving convexity and cyclic order), and every site outside the cell stays strictly outside all four triple circumcircles. Because the four triple circumcircles coincide at \(P\) and can split apart under perturbation, exteriority is recorded for all four.

Whenever a generic perturbed configuration satisfies all certificates, each \(\Gamma_q\) selects one local diagonal and two empty triangles inside cell \(q\), because the cell test excludes the fourth vertex and the exteriority certificates exclude all outside sites. Together with the \(f_3\) fixed triangles, these \(2M\) triangles already supply all \(2n-5\) bounded faces allowed by \eqref{eq:euler} inside the preserved triangular hull. Thus each sign vector \(s\in\{-1,1\}^M\) determines a unique complete labelled triangulation \(T(s)\), and checking whether the labelled edge \(\mathbf B_q\mathbf D_q\) is present recovers \(s_q\) uniquely.

We now make this correspondence hold quantitatively for perturbed sites. Given a radius threshold \(R\ge1\), call site \(i\) \emph{large} if \(\|Z_i\|>R\) and \emph{nonlarge} otherwise. Separating large displacements from linear near-ties lets us use the exact same construction both for a global Gaussian tail bound (Section~\ref{sec:socg-gaussian}) and for a local per-site bound under heavy tails (Section~\ref{sec:transfer}). The constant \(r_*/400\) in \eqref{eq:sigma} is chosen small enough to preserve every nonlarge certificate and to confine perturbed cell circumdisks inside radius \(11/10\) when we later track individual large sites in Lemma~\ref{lem:localdisks}.

\begin{lemma}[Local geometric stability]\label{lem:stability}
Consider \(B\ge1\) separated blocks of \(m\) decision cells each, translated by even integer vectors into boxes at mutual distance at least four, where each block has its own rational parameters \(0<t'\le t<1/4\) and fields satisfying \(|\theta_q|<1/2\) on designated cells and zero elsewhere. Complete their union by Lemma~\ref{lem:completion} into a base configuration \(P\) of \(n\le12M+4\) rational sites (where \(M=Bm\)), and let \(B_P\) bound the binary encoding length of every coordinate of \(P\) and of every rational field \(\theta_q\). Fix a noise law \(\nu\) on \(\R^2\) whose every one-dimensional projection \(\langle\mathbf n,Z\rangle\) (\(\|\mathbf n\|=1\)) has density bounded by a fixed rational \(L_g\ge1\).

For any rational \(R\ge1\) and \(0<\tau<1\) with binary encoding lengths at most \(B_R\) and \(B_\tau\), one can compute in time polynomial in \(B_P,n,B_R,B_\tau\) rational bounds \(0<r_*\le1\) and \(C_P\ge1\), a dyadic scale
\begin{equation}\label{eq:sigma}
 \sigma\le\frac{\min(r_*/400,\ \tau)}{2C_P(R+2)^2},
\end{equation}
and rational shifts \(\zeta_i\) so that the rescaled centres
\begin{equation}\label{eq:finalcentres}
 \mu_i=\frac{P_i+\sigma d_i+\zeta_i}{\sigma}
\end{equation}
(where \(d_{\mathbf D_q}=-\theta_q(\mathbf D_q-o_q)\) and \(d_i=0\) elsewhere) lie in general position and have bit length \(O(B_P+\log n+B_R+B_\tau)\). For independent \(Z_i\sim\nu\) and \(X_i=\mu_i+Z_i\), write \(\sigma X_i=P_i+\Delta_i\). Outside a null event of nongeneric realizations:
\begin{enumerate}
\item Every nonlarge site satisfies \(\|\Delta_i\|<\sigma(R+2)\le r_*/(800C_P)\le1/800\), and every certificate involving only nonlarge sites keeps its base sign.
\item Every decision cell \(q\) whose four sites are nonlarge satisfies \(|\Gamma_q(\sigma X)/\sigma-\Lambda_q|<\tau\).
\item The near-tie event \(\mathcal N_\tau=\{\exists q:\,|\Lambda_q|\le\tau\}\) satisfies \(\Prob(\mathcal N_\tau)\le2L_gM\tau\). Outside \(\mathcal N_\tau\), every cell with four nonlarge sites has \(\sgn\Gamma_q(\sigma X)=\sgn\Lambda_q\).
\end{enumerate}
\end{lemma}

\begin{proof}
We compute exact Taylor bounds of polynomial bit length at \(P\), pick the generic shifts \(\zeta_i\) from a fine rational grid, and verify the three perturbation bounds. Each certificate and each normalized test \(\Gamma_q\) is a determinant of degree at most four in the coordinates of at most four sites of \(P\). Evaluating all certificates exactly at \(P\) gives a rational lower bound \(0<r_*\le1\) on their nonzero magnitudes, and expanding every certificate and every \(\Gamma_q\) in displacements about \(P\) gives a rational upper bound \(C_P\ge1\) on the sums of absolute values of their nonconstant Taylor coefficients. Because each determinant involves at most four sites of bit length \(B_P\), every nonzero value has denominator of \(O(B_P)\) bits and every Taylor coefficient has \(O(B_P)\) bits. Thus \(\log(1/r_*)+\log C_P=O(B_P+\log n)\), and both bounds are computed in polynomial time. Whenever the relevant sites move by vectors of norm at most \(u\le1\), every certificate changes by at most \(C_Pu\) and the nonlinear remainder of \(\Gamma_q\) has magnitude at most \(C_Pu^2\).

Take \(\sigma\) to be the largest dyadic power satisfying \eqref{eq:sigma}. Then
\[
 \log_2(1/\sigma)=O(B_P+\log n+B_R+B_\tau),
\]
which is polynomial in the input parameters. Since \(|\theta_q|<1/2\) on designated cells and zero elsewhere, \(\|d_i\|<1/2\) at every site. Next choose rational shifts \(\zeta_i\), one site at a time, from a square grid of \((n^3+1)\) points per side inside the open box \(\|\zeta\|_\infty<\sigma\tau/32\), so that \(P_i+\sigma d_i+\zeta_i\) avoids all earlier shifted points, their pairwise lines and their triple circumcircles. The grid count in the proof of Lemma~\ref{lem:completion} guarantees that such a grid point always exists and can be found in polynomial time. Dividing by \(\sigma\) in \eqref{eq:finalcentres} preserves general position, restores unit noise scale in \(X_i=\mu_i+Z_i\), and keeps the coordinate bit lengths within the stated bound.

In the coordinates of \(P\), the total displacement of site \(i\) is \(\Delta_i=\sigma(d_i+Z_i)+\zeta_i\). Since \(\|d_i\|<1/2\) and \(\|\zeta_i\|<\sqrt2\,\sigma\tau/32<\sigma/16\), every nonlarge site (\(\|Z_i\|\le R\)) satisfies \(\|\Delta_i\|<\sigma(R+2)\). By \eqref{eq:sigma}, using \(R+2\ge3>1\),
\[
 \sigma(R+2)\le\frac{r_*}{800C_P}\le\frac{1}{800},\qquad
 C_P\sigma(R+2)\le\frac{r_*}{2},\qquad
 C_P\sigma(R+2)^2\le\frac{\tau}{2}.
\]
The second inequality preserves the sign of every certificate involving only nonlarge sites, proving item~1. For item~2, consider any cell \(q\) with four nonlarge sites. Its nonlinear part in \(\Gamma_q(P+\Delta)/\sigma\) has magnitude at most \(C_P\sigma(R+2)^2\le\tau/2\). By \eqref{eq:affine}, \eqref{eq:variation} and Table~\ref{tab:coeff}, the four gradient norms of \(\Gamma_q\) at \(P\) sum to \(a_q+b_q+c_q+1=2+2b_q\le3\), so the shifts \(\zeta_i/\sigma\) alter the linear part by less than \(3\sqrt2\,\tau/32<\tau/2\). Adding the two errors gives \(|\Gamma_q(\sigma X)/\sigma-\Lambda_q|<\tau\).

Finally, for item~3, the outward radial projection at the private site \(\mathbf D_q\) appears only in \(\Lambda_q\), with coefficient \(-1\), and has density bounded by \(L_g\). Conditioning on all other displacements bounds \(\Prob(|\Lambda_q|\le\tau)\) by \(2L_g\tau\) for each cell, and a union bound over all \(M\) cells gives \(\Prob(\mathcal N_\tau)\le2L_gM\tau\). Whenever \(|\Lambda_q|>\tau\) and item~2 holds, \(\Gamma_q(\sigma X)/\sigma\) has the same sign as \(\Lambda_q\).
\end{proof}

\subsection{Gaussian diagonal choices at two probability scales}
For the direct Gaussian proof, use the same rational parameter \(0<t<1/4\) on all circles, including bridges, so \(t^{\prime}=t\) in \eqref{eq:bridgetest}. Table~\ref{tab:coeff} gives the coefficients of every linear test. In particular the coefficients at \(\mathbf A\) and \(\mathbf D\) have magnitude close to one, while those at \(\mathbf B\) and \(\mathbf C\) have order \(t\).

Give distinct sites independent \(N_2(0,I_2)\) displacements. Tangent circles have opposite outward normals. When two cells share a site, its displacement projected onto their common radial line has unit variance, so the covariance of their tests is the product of its two signed coefficients. Along a cycle, outward motion \(U\) from \(i\) contributes \(a_iU\) in \(i\) and \((-b_j)(-U)=+b_jU\) in \(j\), favouring equal consecutive signs (Fig.~\ref{fig:socg-diagonals}(b)). At the lower and upper arm corners \(q,r\), outward projections \(V_q,V_r\) have cycle weight \(c\) and contribute \(bV_q-cV_r\) to the bridge test (Fig.~\ref{fig:socg-diagonals}(c) and (d)). Bridge covariances \(cb=t^2+O(t^3)\) below and \(-c^2=-t^2+O(t^3)\) above are thus an order smaller than \(a_i b_j=t+O(t^2)\) on the cycle, while other pairs are uncorrelated. Every test has variance \(a^2+b^2+c^2+1=2+O(t)\). Shifting \(D\) outward by \(t^2\) noise units on one straight cell per cycle away from the arms adds bias \(-t^2\).

\subsection{Why optimal Gaussian sign patterns encode independent sets}\label{sec:gaussian-score}
Normalizing each test by its standard deviation yields a Gaussian vector with covariance \(I+tL+t^2H\) and mean \(t^2h\). Both \(L\) and \(H\) are symmetric with zero diagonal and bounded row sums (each circle touches at most three others): \(L\) carries \(1/2+O(t)\) on consecutive cycle pairs, \(H\) carries \(1/2+O(t)\) below and \(-1/2+O(t)\) above on pairs joining a bridge to its endpoints, and \(h\) equals \(-1/\sqrt2+O(t)\) on designated straight cells and \(0\) elsewhere.

For any sign pattern \(s\in\{-1,1\}^m\), let \(F_s(x,y)\) be \(2^m\) times its orthant probability under \(N(yh,\,I+xL+yH)\), keeping \(L,H,h\) fixed at \(t\). At \((0,0)\) coordinates are independent standard normals, and differentiating Gaussian orthant probabilities~\cite{plackett} at \((0,0)\) gives
\[
 F_s(0,0)=1,\qquad
 \partial_x F_s=\frac2\pi\sum_{i<j}L_{ij}s_i s_j,\qquad
 \partial_y F_s=\frac2\pi\sum_{i<j}H_{ij}s_i s_j+\sqrt{\frac2\pi}\sum_i h_i s_i .
\]
Because \(L,H\) have zero diagonal, differentiating the log density at \((0,0)\) gives \(z_i z_j\) per off-diagonal pair and \(z_i\) for the mean. Under the independent standard Gaussian law conditioned on orthant \(s\), the coordinates remain independent and have conditional means \(s_i\sqrt{2/\pi}\).

Naive expansion around \((0,0)\) faces a two-scale dilemma: repairing a cycle shifts \(F_s\) by \(\Omega(t)\), whereas bridges and biases act at order \(t^2\), which an \(O(m^2t^2)\) remainder would swamp. One uniform bound handles both scales: for \(|x|+|y|\le c_0/m\) (with a small absolute \(c_0>0\)), every second partial derivative of \(F_s\) is \(O(m^2)\). Indeed, put \(J=(I+xL+yH)^{-1}\). Bounded row sums give \(\|J-I\|=O(c_0/m)\), and \(h\) has \(O(m)\) bounded entries so \(\|yh\|=O(c_0/\sqrt m)\). In the density \(\phi\), the determinant factor is \(e^{O(c_0)}\), and completing the square shows that \(\phi(z)/\phi_0(z)\le\exp(O(c_0)(1+\|z\|^2/m))\) relative to the standard Gaussian \(\phi_0\). Differentiating \(\log\det J^{-1}\) via traces and using \(\partial J=-J(\partial J^{-1})J\) bounds all first and second partial derivatives of \(\log\phi\) by \(O(m+\|z\|^2)\), so \(\partial_u\partial_v\phi=(\partial_u\partial_v\log\phi+(\partial_u\log\phi)(\partial_v\log\phi))\phi\) adds a factor \(O((m+\|z\|^2)^2)\). Under \(\phi_0\) conditioned on any orthant, \(\|z\|^2\sim\chi^2_m\) has moment generating function \((1-2u)^{-m/2}\). Evaluating its first two derivatives at \(u=O(c_0/m)\) uniformly bounds the integral by \(O(m^2)\) and justifies differentiating under the integral sign.

Set \(t=2^{-K\lceil\log_2(m+1)\rceil}\) for a fixed large integer \(K\). Expanding to \((t,0)\) then \((t,t^2)\) yields
\[
 F_s(t,0)=1+t\partial_x F_s(0,0)+O(m^2t^2),\qquad
 F_s(t,t^2)=F_s(t,0)+t^2\partial_y F_s(0,0)+O(m^2t^3).
\]
Crucially, the second formula retains the exact cycle probability \(F_s(t,0)\): its error bounds the variation of \(\partial_y F_s\) over a segment of length \(O(t)\), multiplied by \(t^2\).

First suppose a cycle is nonconstant. Setting all its signs equal repairs at least two disagreeing adjacent pairs, raising \(\partial_x F_s(0,0)\) by an absolute constant while changing the \(y\)-term by \(O(mt^2)\), so the repaired pattern raises \(F_s(t,t^2)\) by \(\Omega(t)-O(m^2t^2)>0\). Second, suppose every cycle has constant phase \(z_v\in\{-1,1\}\), and set \(S=\{v:z_v=+1\}\). At \(y=0\) the centred Gaussian cycles and fair bridge signs are independent. Negating any cycle preserves its probability, so \(F_s(t,0)\) is the same for all constant-cycle patterns. In \(\partial_y F_s(0,0)\), bridge \(e\) from lower \(u\) to upper \(v\) contributes \(s_e(z_u-z_v)/\pi+O(t)\), so the optimal bridge sign earns \(2/\pi\) on each cut edge and \(0\) otherwise (up to \(O(t)\)). Vertex biases contribute \(-\sum_v z_v/\sqrt\pi+O(mt)=(|V|-2|S|)/\sqrt\pi+O(mt)\). Since \(G\) is cubic, the cut has \(3|S|-2|E(S)|\) edges. Adding a vertex with \(k\) selected neighbours changes the cut size by \(3-2k\) and costs one bias penalty. With optimal bridge signs the scaled probability is a constant plus
\[
 \frac{4t^2}{\pi}\biggl(\alpha\,|S|-|E(S)|\biggr)+O(m^2t^3),\qquad
 \alpha=\frac{3-\sqrt\pi}{2}\in(0,1).
\]
The diagonal probabilities thus realize the classical independent-set objective~\cite[Section~4.2]{lucas}. On \(K_4\), one vertex scores \(\alpha\) and two score \(2\alpha-1<\alpha\). Removing either endpoint of an internal edge raises the score by at least \(1-\alpha\). Among independent sets, a maximum \(S^*\) beats each smaller \(S\) by \(\alpha(|S^*|-|S|)\ge\alpha\). Every pattern failing to encode a maximum independent set thus loses \(\Omega(t^2)\) in \(F_s(t,t^2)\). Large \(K\) makes \(m^2t\) arbitrarily small and keeps \(F_s(t,t^2)\le2\), so for fixed dyadic \(b>0\) every failing pattern has linear probability at most \((1-\beta)p_*\), with \(\beta=bt^2\) and block maximum \(p_*\ge2^{-m}\).

\subsection{Transferring and amplifying the Gaussian probability gap}\label{sec:gaussian-transfer}
To transfer linear signs to complete triangulations and amplify the gap, place \(r\) copies of the layout in boxes at distance at least four via even-integer translations, giving \(M=rm\) cells (\(r=1\) for exact search).

For these \(M\) cells, Lemma~\ref{lem:completion} constructs a triangulated exterior and a triangular hull using \(n\le12M+4\) sites. Equation~\eqref{eq:euler} shows that the fixed faces and the two faces inside each cell account for every bounded face. Thus the signs of the \(M\) tests identify the complete labelled triangulation whenever the stability certificates hold.

\emph{Global perturbation scale and gap amplification.} Work in base coordinates at scale \(\sigma\), writing \(P_i\) for base sites and \(d_i\) for deterministic shifts in noise units (nonzero only at designated cells). Some draw exceeds norm \(R\) with probability at most \(ne^{-R^2/2}\). Conditional on the other site displacements, each unstandardized linear test \(\Lambda\) is a translate of the independent standard normal term from its private site \(D\), so some test lies in \([-\tau,\tau]\) with probability at most \(2M\tau\). Given dyadic \(\delta>0\), pick integer \(R=O(1+\log(n/\delta))\) and dyadic \(0<\tau<1\) with \(ne^{-R^2/2}+2M\tau\le\delta\). Let \(0<d\le1\) bound below all nonzero base determinant magnitudes and \(C\ge1\) bound the sums of absolute Taylor coefficients of all predicates and unstandardized cell determinants \(\Gamma\) at \(P_i\). Both have polynomial bit length since each determinant has fixed degree in at most eight coordinates. Choose \(\sigma\) as the largest dyadic power below \(\min(d,\tau)/(4C(R+2)^2)\), so \(\sigma(R+2)<1\), then pick generic nominal sites \(P_i+\sigma d_i+\zeta_i\) with \(\|\zeta_i\|\le\sigma\tau/16\) by the grid rule. On the truncation event, every site stays within \(\sigma(R+2)\) of \(P_i\): predicate changes stay below \(C\sigma(R+2)<d\), and after dividing each \(\Gamma\) by \(\sigma\), the degree-\(\ge2\) terms stay below \(C\sigma(R+2)^2<\tau/4\) while the shift from \(\zeta_i/\sigma\) stays below \(3\tau/16\) since \(a+b+c+1=2+2b<3\). Dividing nominal sites by \(\sigma\) gives generic rational centres with unit Gaussian noise whose triangulation matches the linear signs outside an event of probability at most \(\delta\). Each type probability therefore agrees with the linear model within \(\delta\), and any type outside the image of the sign map has probability at most \(\delta\).

For exact search (\(r=1\)), \(\delta<\beta 2^{-m}/4\) makes the linear gap of one block exceed \(2\delta\), so every geometric mode decodes to a maximum independent set. For relative approximation, we apply the product principle of Suri, Verbeek and Y{\i}ld{\i}z~\cite[Lemma~7 and Theorem~4]{suri} to independent linear copies and transfer via \(\delta\). Pick \(r=2^{k\lceil\log_2(m+1)\rceil}\) with fixed integer \(k\) so \(r\beta\ge 8(12rm+4)^{1-\epsilon}+16\) (possible since \(\beta^{-1}=O(m^{2K})\)), set \(c=2^{-\lfloor r\beta/2\rfloor}\le 2^{-n^{1-\epsilon}}\), and take \(\delta=c2^{-M}/8\). Any pattern failing on all \(r\) copies has linear probability at most \((1-\beta)^r p_*^r<cp_*^r/4\), where \(p_*\) is the single-block linear maximum. A modal linear pattern thus keeps geometric probability at least \((1-c/8)p_*^r\), while any type outside the image or failing on all copies stays at most \(3cp_*^r/8\). Every qualifying output has probability at least \(2^{-n^{1-\epsilon}}(1-c/8)p_*^r\ge c(1-c/8)p_*^r>3cp_*^r/8\), so it encodes a maximum independent set in some copy. Discarding nonconstant or nonindependent readings and returning a largest remaining set completes both one-call reductions in polynomial bit time.

\section{The most probable convex hull}\label{sec:gaussian-hull}

Because a hull records only boundary sites, we encode each item along a core polygon's edges. Moving a candidate outward opens a triangular cap that can capture a random \emph{probe}, and balancing opening cost against capture makes the ideal score peak at the target.

\begin{theorem}\label{thm:gaussian-hull}
Given the rational centres and positive rational scales \(s_i\) of independent planar Gaussian sites with covariances \(s_i^2 I_2\), finding a most probable set of extreme labels is NP-hard. The same holds for their cyclic order up to rotation and reversal. Each reduction uses one call and works for every mode.
\end{theorem}

\begin{proof}
Take a \textsc{Subset Sum} instance with positive binary weights \(a_1,\ldots,a_m\) and target \(1\le T\le A=\sum_i a_i\)~\cite[problem 18, pp.~95, 100]{karp}. We choose opening probabilities and cap sizes so the ideal score peaks at \(T\), then give the fixed vertices positive variance.

\emph{Core polygon and candidates.} Let \(K\) be the hull of fixed vertices \(20(j,j(j-m))\), \(0\le j\le m\), and upper corners of \([0,20m]\times[0,20m^2]\). It contains the disk of radius \(10m\) about the rectangle centre \(o\), giving \(1-B\le e^{-50m^2}<2^{-m-5}\) for the probe \(P\sim N_2(o,I_2)\) and \(B=\Pr(P\in K)\). On each lower edge \(i\), with midpoint \(M_i\) and outward unit normal \(u_i\), put an independent candidate \(X_i=M_i+s_i(d_i u_i+Z_i)\), \(Z_i\sim N_2(0,I_2)\). Here \(d_i\) sets opening probability \(\Phi(d_i)\) and \(s_i\) scales the cap for a fixed outward draw (Fig.~\ref{fig:hull-caps}(a)). A hull is \emph{canonical for \(S\)} if its extreme sites are \(K\)'s vertices and candidates in \(S\). For cyclic output, take \(K\)'s boundary order and insert each \(i\in S\) between the consecutive endpoints of edge \(i\).

\begin{figure}[!htbp]
\centering\includegraphics[width=\linewidth]{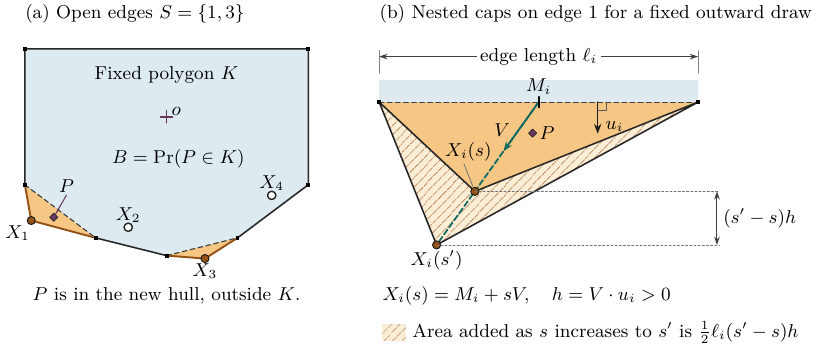}
\caption{Tuning the probability of a cap. In (a), the probe \(P\) lands in the cap of \(X_1\). In (b), edge 1 is rotated. For each outward \(V\), integrate the probe density over the hatched region. The conditional average gives \(w_i(s')-w_i(s)\). Bisection chooses a rational \(s_i\) with \(w_i(s_i)\) close to \(\lambda a_i\).}\label{fig:hull-caps}
\end{figure}

\emph{Target score.} Choose \(0<\lambda A\le1/8\) and set \(c=(B+\lambda T)^{-1}\). Give edge \(i\) opening odds \(p_i/(1-p_i)=e^{-c\lambda a_i}\) and target expected capture \(\lambda a_i\) conditional on opening. By independence, opening precisely \(S\) has probability \(Qe^{-cx}\), with \(x=\lambda\sum_{i\in S}a_i\) and \(Q=\prod_i(1-p_i)\ge2^{-m}\) (\(p_i<1/2\)). The target cap contributions sum to \(x\), giving ideal capture \(B+x\) and ideal hull score \(QF(x)\), where
\[
 F(x)=(B+x)e^{-cx},\qquad F'(x)=ce^{-cx}(\lambda T-x).
\]
Relative capture gain \(1/(B+x)\) equals cost \(c\) at \(x=\lambda T\) (Fig.~\ref{fig:hull-score}(c)). Integrating \(F'\) with \(ce^{-cx}\ge1/4\) gives \(F(\lambda T)-F(x)\ge(x-\lambda T)^2/8\), so for integer weights each target subset beats any non-target in ideal score by at least \(G=2^{-m}\lambda^2/8\), with \(QF(\lambda T)>QB\) (Fig.~\ref{fig:hull-score}(b)).

\begin{figure}[!hbp]
\centering\includegraphics[width=\linewidth]{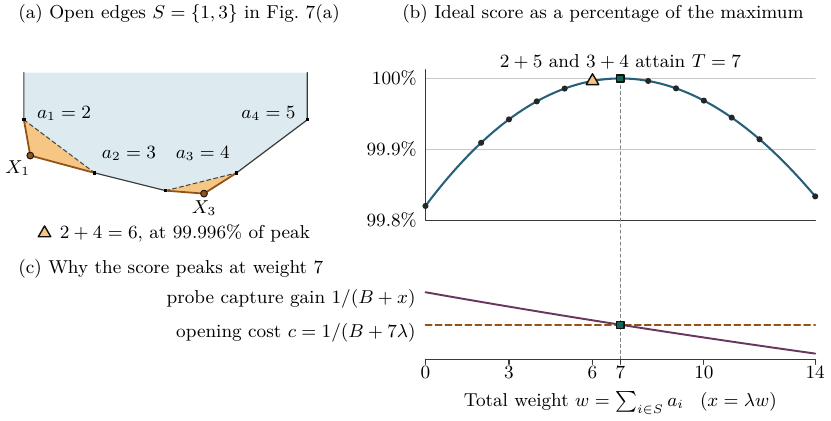}
\caption{The peak at weight 7. In (a), the open edges of Fig.~\ref{fig:hull-caps}(a) have total weight 6. In (b), symbols mark attainable weights. The triangle marks 6 and the square marks 7 (\(2+5=3+4\)). In (c), the falling relative gain in probe capture meets the fixed relative opening cost at 7, maximizing the ideal score. Here \(B=7/8\) and \(\lambda=1/128\) are illustrative.}\label{fig:hull-score}
\end{figure}

\emph{Calibrating shifts and scales in polynomial time.} Set \(U=100(m+1)^2\), \(k=16(U^2+m+\lceil\log_2(A+1)\rceil+32)\), \(\lambda=2^{-k}\), \(\beta=\lambda^4\), \(\rho=1/(100U)\), and \(s_{\max}=\rho/(16k)\). For \(|d_i|\le1/4\) and \(s_i\le s_{\max}\), the probability that at least one candidate leaves its radius-\(\rho\) disk around \(M_i\) is at most \(me^{-(16k-1)^2/2}<\beta\). Because vertices are multiples of \(20\), all vertices and midpoints are integer points on a convex boundary: collinear triples share an edge, and every noncollinear triple has determinant at least \(1\) in magnitude, shifting by at most \(4U\rho+4\rho^2<1\). Whenever every candidate stays in its midpoint disk, outward candidates preserve convexity in boundary order with disjoint cap interiors outside \(K\), and inward ones stay in \(K\).

Since \(c\lambda A\le1/7\) and \(\Phi(-1/4)<(1+e^{1/7})^{-1}\), bisection finds rational \(d_i\in[-1/4,0]\) with \(|\Phi(d_i)-p_i|\le\beta\). Let \(w_i(s)\) be the expected probe probability of cap \(i\) conditional on opening. For each outward draw \(V=d_i u_i+Z_i\) of height \(h=V\cdot u_i>0\), the apex moves along a ray from \(M_i\) (Fig.~\ref{fig:hull-caps}(b)). As \(s\) grows, the caps nest, and increasing \(s\) by \(\Delta s\) adds area \(\ell_i h\Delta s/2\) for an edge of length \(\ell_i\). The probe density is at most \(1/(2\pi)\) and the conditional mean height is below \(2\), so \(w_i\) increases from zero and is \(U\)-Lipschitz. With probability above \(1/100\), the standard draw has normal component in \([1,2]\) and tangential component in \([-1,1]\). Then \(h\ge1/2\) and every cap point lies within distance \(U/4\) of \(o\), so the probe density exceeds \(2^{-U^2-4}\) and \(w_i(s_{\max})\ge s_{\max}2^{-U^2-14}>\lambda A\). At each midpoint, enclose \(w_i\) within radius \(\beta/8\). If the interval contains \(\lambda a_i\), return the midpoint. Otherwise bisect by monotonicity down to width \(\beta/(2U)\). This yields rational \(s_i\ge\lambda/(2U)\) with \(|w_i(s_i)-\lambda a_i|\le\beta\).

To compute \(w_i(s)\), write the outward draw as \(V=h u_i+v e_i\) in the unit normal and tangent directions. At fraction \(t\) from base to apex, the position and area Jacobian are
\[
 M_i+(1-t)\ell_i y e_i+tsV,\qquad \ell_i s h(1-t),\qquad
 0\le t\le1,\quad |y|\le1/2 .
\]
Integrate the probe density times this Jacobian against \(\varphi(h-d_i)\varphi(v)\) (\(\varphi=\Phi'\)) over \(h>0,v\in\R\) and divide by \(\Phi(d_i)>1/3\). For \(q\) bits of accuracy, truncating \(h\) and \(|v|\) at \(10(q+10)\) leaves a Gaussian tail error below \(2^{-q-5}\). On the rescaled unit box, the integrand is a polynomial times \(e^{-H}\), where \(H=O((U+q+10)^2)\) is a polynomial of fixed degree. Choosing Taylor degree \(D=O((U+q+10)^2)\) keeps the integrated remainder below \(2^{-q-5}\) and reduces the integral to polynomially many monomial integrals over rational bounds in four variables. Because \(H\) has fixed degree, submultiplicativity bounds the sum of absolute coefficients by \((U+q+10)^{O(D)}\), so dyadic precision \(O(q+D\log(U+q+10))\) suffices. The same method evaluates \(B\) on a triangulation of \(K\) (two variables) and both \(\Phi\) and the opening odds (one variable), with \(O(\log(U/\beta))\) bisection steps and \(q=O(\log(U/\beta))\) bits per step.

\emph{Transfer to positive-variance sites.} Multiply the pattern indicator for \(S\) by the sum of the probe indicators of \(K\) and its open caps. Independence preserves each open candidate's conditional law, so the expectation is the pattern probability times \(B+\sum_{i\in S} w_i(s_i)\). Inside all midpoint disks, this product equals the canonical hull indicator almost surely under either convention. Outside that event it is at most \(m+1\), so leaving those disks changes the probability by at most \((m+2)\beta\). Since \(B+\sum_i w_i(s_i)<2\), telescoping the \(m\) opening probabilities and adding the \(m\) cap errors costs at most \(3m\beta\). A noncanonical hull requires a missed disk or \(P\notin K\). By independence and the translation bound of Section~\ref{sec:centres}, rounding each \(d_i u_i\) within \(\beta/(2m+4)\) to a rational vector costs under \(\beta\) in total variation.

Finally, give each fixed vertex independent Gaussian noise of positive rational scale. Let \(N=2m+4\), \(s_*=\min(1,s_1,\ldots,s_m)\ge\lambda/(2U)\), and \(b=\beta/(100N^3)\). For any triple with at least one random site, sample a random site last. The first two sites lie at distance at least \(s_*b\) except with probability at most \(b^2/2\) (by Gaussian disk density, or unit separation when both are fixed). Conditional on a pair at distance at least \(s_*b\), the signed orientation determinant \(\Delta\) is Gaussian with standard deviation at least \(s_*^2b\), so \(\delta=s_*^2b^2\) gives \(\Pr(|\Delta|\le\delta)\le 2\delta/(\sqrt{2\pi}\,s_*^2b)=\sqrt{2/\pi}\,b\). Including closer pairs gives at most \(b^2/2+\sqrt{2/\pi}\,b\) per triple. All-fixed triples have nonzero integer determinants, so a union bound keeps near-collinear failure below \(\beta/10\). With probability at least \(1-\beta/10\), all positions before perturbing the fixed vertices and all new standard draws have norm at most \(J=10(U+\lceil\log_2(N/b)\rceil+10)\). Taking the largest power of two \(\tau\le\delta/(32J^2)\) keeps every determinant shift below \((8\tau+4\tau^2)J^2<\delta\), preserving all orientations and both hull conventions in polynomial bit length.

Combining the errors from leaving the midpoint disks, calibration, rounding, and vertex perturbation places each canonical hull probability within \((5m+4)\beta<G/4\) of its ideal score, while all noncanonical hulls together have probability below \(1-B+3\beta<2^{-m-4}\). In a yes instance, every target hull exceeds \(2^{-m}B-G/4>2^{-m-4}\) and beats every non-target canonical hull by more than \(G/2\). In a no instance, no subset sums to \(T\). Reading the candidate labels from any returned mode therefore decides \textsc{Subset Sum} with one call.
\end{proof}

\section{Delaunay hardness for radial laws, the confinement criterion and the Gaussian specialization}\label{sec:model}\label{sec:extensions}
To extend planar Delaunay hardness beyond Gaussian displacements, the reduction still needs, for every \(n\), an origin-centred disk whose radius has polynomial bit length and that contains all \(n\) independent displacements simultaneously with probability at least \(1/2\). Let \(Z\) denote one displacement. This section introduces the logarithmic tail condition that governs such radii, states the general radial-law hardness theorem, proves that the condition is equivalent both to a finite logarithmic moment and to polynomial-bit confinement, and records how the exact-mode Gaussian result of Section~\ref{sec:socg-gaussian} specializes from Theorem~\ref{thm:main} while the decaying factor \(2^{-n^{1-\epsilon}}\) follows from the replication argument of Section~\ref{sec:amplification}.
\begin{definition}\label{def:L}
Let \(Z\) have law \(\nu\) on \(\R^2\), and put \(T_\nu(R)=\Prob\{\|Z\|>R\}\) for every real \(R\ge0\). We say that \(\nu\) satisfies condition \emph{(L)} if there are a fixed integer \(C_L\ge1\) and a fixed rational \(\gamma>0\) such that
\begin{equation}\tag{L}\label{eq:L}
 T_\nu(2^s)\le C_Ls^{-\gamma}
 \qquad\text{for every integer }s\ge1.
\end{equation}
\end{definition}

\Needspace{9\baselineskip}
\begin{theorem}\label{thm:main}
Fix a probability law \(\nu\) on \(\R^2\) with Lebesgue density
\(\rho(z)=\varphi(\|z\|)\), where \(0\le\rho\le M_\rho<\infty\), satisfying \eqref{eq:L}. Given rational centres \(\mu_1,\ldots,\mu_n\) in general position, finding any most probable complete labelled Delaunay triangulation of the sites \(X_i=\mu_i+Z_i\), with independent displacements \(Z_i\) of law \(\nu\), is NP-hard. Construction and decoding take polynomial time, and the reduction makes one modal oracle call.
\end{theorem}

The reduction is completed in Section~\ref{sec:transfer-candidates}, using the geometry of Section~\ref{sec:geometry} and the block probabilities from Section~\ref{sec:heavy}. Once the law \(\nu\) is fixed, finitely many rational constants depending on it suffice for a single reduction that handles every planar cubic input graph. The algorithm uses those constants directly. It makes one query, and every answer, including any tied mode, decodes to a maximum independent set.

It remains open whether one algorithm could take a description of the law as an additional input. We also leave open whether comparing the probabilities of two given types exactly is in NP, meaning that every yes answer has a polynomially verifiable certificate.

Proposition~\ref{prop:equiv} characterizes (L) through a moment of the logarithm of the displacement length and through a geometric property. With probability at least \(1/2\), one disk whose radius has polynomial bit length contains all \(n\) displacements. In the reduction, Section~\ref{sec:transfer} uses (L) directly to choose radii at the higher confidence levels needed to compare triangulation probabilities.

\Needspace{7\baselineskip}
\begin{proposition}\label{prop:equiv}
Condition \eqref{eq:L} holds if and only if
\(\E\bigl[(\log(1+\|Z\|))^a\bigr]<\infty\) for some \(a>0\).
Equivalently, there is a polynomial \(p\) such that for every \(n\ge1\) the independent displacements \(Z_1,\ldots,Z_n\) with law \(\nu\) satisfy
\[
 \Prob\!\left\{\max_{1\le i\le n}\|Z_i\|\le 2^{p(n)}\right\}\ge\tfrac12 .
\]
\end{proposition}
\begin{proof}
We express both conditions through the tail of the displacement length. Put \(W=\log_2(1+\|Z\|)\). Under (L), for \(u\ge4\), the inequalities \(2^u-1\ge2^{u-1}\) and monotonicity give
\[
 \Prob(W>u)\le T_\nu(2^{\lfloor u-1\rfloor})
 \le C_L\lfloor u-1\rfloor^{-\gamma}\le C_L2^\gamma u^{-\gamma}.
\]
Choose \(0<a<\gamma\). The tail identity
\(\E W^a=a\int_0^\infty u^{a-1}\Prob(W>u)\,du\)
proves finiteness. Conversely, if \(\E W^a<\infty\), choose a rational \(0<\gamma<a\). Markov's inequality gives
\(T_\nu(2^s)\le\Prob(W>s)\le\E W^a\,s^{-a}\le C_Ls^{-\gamma}\)
for every integer \(s\ge1\), after increasing the integer \(C_L\). Changing the logarithm base multiplies the moment by a constant.

For confinement, (L) permits \(R_n=2^{s_n}\) with
\(s_n=\lceil(2C_Ln)^{1/\gamma}\rceil\). An integer polynomial bounds this exponent, and a union bound gives the required probability. Conversely, independence and confinement give \((1-T_\nu(2^{p(n)}))^n\ge1/2\), hence \(T_\nu(2^{p(n)})\le(\ln 2)/n\).
Choose constants \(C\ge1\) and an integer \(k\ge1\) with \(p(n)\le Cn^k\). For large \(s\), take \(n=\lfloor(s/C)^{1/k}\rfloor\). Then \(p(n)\le s\), so monotonicity gives \(T_\nu(2^s)=O(s^{-1/k})\). Enlarging the constant for small \(s\) proves (L).
\end{proof}

\subsection*{Gaussian Delaunay hardness from the radial law}
For \(Z\sim N_2(0,I_2)\), the density is radial and bounded, and \(\Pr(\|Z\|>r)=e^{-r^2/2}\) satisfies~\eqref{eq:L}. Theorem~\ref{thm:main} therefore gives the exact-mode claim of Theorem~\ref{thm:socg-gaussian}. For every fixed \(0<\epsilon<1\), the replication argument in Section~\ref{sec:amp-replication} recovers a maximum independent set from any answer of probability at least \(2^{-n^{1-\epsilon}}p_*\), where \(n\) counts sites after replication. Both reductions use rational generic centres, take polynomial bit time and make one query.

\section{From local responses to a score on the graph}\label{sec:responses}
The Gaussian argument in Section~\ref{sec:socg-gaussian} obtained the sign probabilities by differentiating an orthant integral. For two centred Gaussian tests of correlation \(r\), the sign product has expectation \(2\arcsin(r)/\pi\)~\cite[Section~27]{sheppard}. For a general displacement law the coordinates shared by two tests still determine their dependence. We first record the conditional response of one test to a shared coordinate. The next section bounds the size of these responses at the scales used in the reduction.

For any symmetric random variable \(W\) with no atoms (in particular, any symmetric variable having a density), define its centred response function \(L_W(x)=\Prob(W<x)-\tfrac12\). Because \(W\) has no atoms and has the same law as \(-W\), we have \(\Prob(W<-x)=\Prob(W>x)=1-\Prob(W<x)\), so \(L_W\) is odd and nondecreasing with \(L_W(0)=0\). Let \(\Lambda_q^0\) be the zero-field linear test of cell \(q\) from \eqref{eq:test}. Holding the predecessor contribution fixed at \(x=b_qU_{q^-}\), we write \(\Lambda_q^0=x-W_q\), where \(W_q=Q-a_qU-c_qV\) for independent copies \(U,V,Q\) of \(U\) (symmetric and atomless because \(Q\) is). Then \(\Prob(\Lambda_q^0>0\mid x)-\tfrac12=\Prob(W_q<x)-\tfrac12=L_{W_q}(x)\), and the conditional mean of the sign is \(\E[\sgn\Lambda_q^0\mid x]=2L_{W_q}(x)\). Setting all fields to zero in the first three quantities, define the cycle weight \(\kappa_q(t)\), the two bridge weights \(w_B(t,t')\) and \(w_C(t,t')\), and the bias \(\mathcal M(\theta)\) of a designated cell \(d\) under a field \(\theta\) by
\begin{equation}\label{eq:weightsdef}
 \begin{aligned}
 \kappa_q(t)&=4\E\bigl[\bigl(F(a_{q^-}U)-\tfrac12\bigr)\,L_{W_q}(b_qU)\bigr],\\
 w_B(t,t')&=\E[\sgn\Lambda_x\,\sgn\Lambda_{q_B(x)}],\qquad
 w_C(t,t')=-\E[\sgn\Lambda_x\,\sgn\Lambda_{q_C(x)}],\\
 \mathcal M(\theta)&=\E[\sgn\Lambda_d]=\sgn(\theta)\,\Prob\bigl(|\Lambda_d^0|\le|\theta|\bigr).
 \end{aligned}
\end{equation}
Keeping only \(a_{q^-}U_{q^-}-Q_{q^-}\) in the predecessor test and conditioning on \(U_{q^-}\) gives conditional mean signs \(2(F(a_{q^-}U_{q^-})-\tfrac12)\) for that main part and \(2L_{W_q}(b_qU_{q^-})\) for \(\Lambda_q^0\). Because these two signs are conditionally independent given \(U_{q^-}\), multiplying their conditional means and averaging over \(U_{q^-}\overset{d}{=}U\) yields the factor of four in \(\kappa_q(t)=\E[\sgn(a_{q^-}U-Q')\,\sgn(a_qU'+b_qU+c_qV-Q)]\), where \(U,Q',U',V,Q\) are independent copies of \(U\). Similarly, \(w_B\) and \(-w_C\) are the exact sign correlations of bridge \(x\) with its lower endpoint \(q_B(x)\) and upper endpoint \(q_C(x)\), while \(\mathcal M(\theta)\) is the mean sign of a designated cell under field \(\theta\). By Lemma~\ref{lem:geometry} and Section~\ref{sec:geometry}, all bridges have identical weights \((a',b',c')\), all bridge endpoints are corners with identical weights, and all designated cells are straight cells with identical weights. Thus \(w_B\), \(w_C\) and \(\mathcal M\) do not depend on the choice of bridge or cycle.

\subsection{The contribution of the bridges and the fields}\label{sec:gauss-block}
\begin{figure}[!htbp]
\centering
\resizebox{\linewidth}{!}{
\begin{tikzpicture}[x=1cm,y=1cm,font=\small,text=black,>=Stealth,line cap=round,line join=round]
\definecolor{repairblue}{HTML}{2875B7}
\definecolor{repairorange}{HTML}{DB7627}
\useasboundingbox (0,-.35) rectangle (15.5,5.0);
\node[anchor=west] at (.2,4.75) {(a) A nonconstant cycle};
\node[anchor=west] at (8.30,4.75) {(b) Replace it by one phase};
\foreach \xx/\after in {3.15/0,11.35/1}{
  \begin{scope}[shift={(\xx,2.5)}]
  \draw[black!48,line width=.65pt] (0,0) circle (1.17);
  \foreach \k in {0,...,11}{
    \pgfmathtruncatemacro{\sg}{(\k<6 || \k>10 || \after==1) ? 1 : 0}
    \ifnum\sg=1
      \node[draw=repairblue,fill=repairblue!11,circle,inner sep=1.5pt,minimum size=15pt,line width=.65pt] at ({30*\k}:1.17) {$+$};
    \else
      \node[draw=repairorange,fill=repairorange!13,circle,inner sep=1.5pt,minimum size=15pt,line width=.65pt] at ({30*\k}:1.17) {$-$};
    \fi
  }
  \ifnum\after=0
    \foreach \cut in {165,315}{
      \draw[black,line width=1.35pt] (\cut:.86)--(\cut:1.34);
      \draw[repairorange!75!black,line width=2pt] (\cut:1.34)--(\cut:1.53);
    }
  \else
    \draw[repairblue,line width=.8pt,dash pattern=on 1.7pt off 1.4pt]
      (177:.75) arc (177:303:.75);
  \fi
  \foreach \ang/\idx in {0/1,150/2,210/3}{
    \ifnum\idx=3
      \draw[repairorange!85!black,line width=1.75pt] (\ang:1.34)--(\ang:1.74);
    \else
      \draw[black!58,line width=.7pt] (\ang:1.34)--(\ang:1.74);
    \fi
    \node[draw=black!65,fill=white,rectangle,rounded corners=1pt,inner sep=2.5pt] at (\ang:1.98) {$s_{x_{\idx}}$};
    \draw[black!40,line width=.55pt] (\ang:2.26)--(\ang:2.61);
    \fill[black!35] (\ang:2.65) circle (1.5pt);
  }
  \node at (0,0) {$\mathcal C_v$};
  \draw[black!60,->,line width=.5pt] (.7,1.64)--(.14,1.27);
  \node[anchor=west,font=\footnotesize] at (.7,1.64) {field $-\lambda$};
  \ifnum\after=0
    \node[font=\footnotesize] at (0,-1.68) {two boundaries};
  \else
    \node[font=\footnotesize] at (0,-1.68) {five signs changed};
  \fi
  \end{scope}
}
\draw[->,black!65,line width=.75pt] (6.1,2.45)--(8.35,2.45);
\node[align=center,font=\footnotesize] at (7.23,3.02) {repair $\mathcal C_v$};
\node[align=center,font=\footnotesize] at (7.23,1.80) {fixed bridge signs};
\node[anchor=west,font=\footnotesize] at (.2,-.18) {Cycle gain: at least two repaired boundaries};
\node[anchor=east,font=\footnotesize] at (15.3,-.18) {Only the link to $x_3$ changes here};
\end{tikzpicture}}
\caption{Repairing a nonconstant cycle. In (a), five consecutive minus signs form two phase boundaries. In (b), changing all five to plus removes both boundaries with every bridge sign fixed. Only the contribution from the highlighted link to \(x_3\) changes in this example. In general, at most three bridge contributions and one field contribution can change.}\label{fig:cycle-repair}
\end{figure}
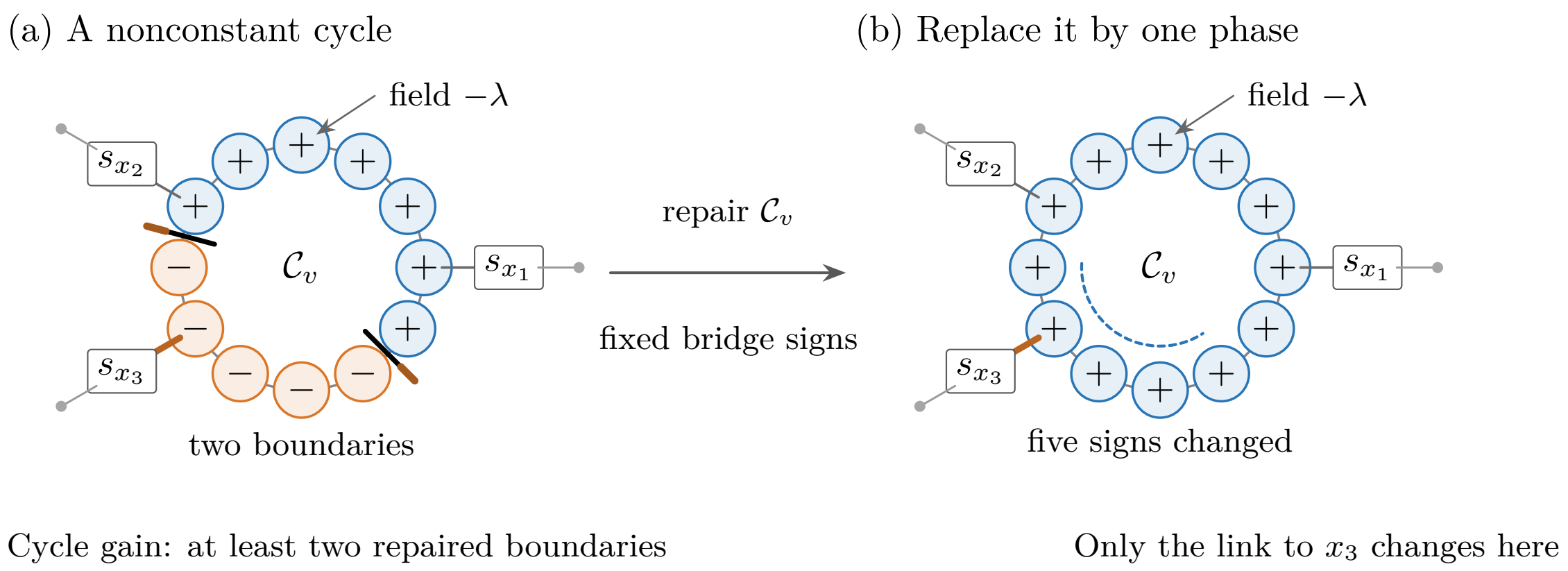

To characterize the most probable sign patterns from the local weights in \eqref{eq:weightsdef}, the proof proceeds in two stages. First, for any pattern with a nonconstant cycle, assigning the same sign to all cells of that cycle strictly increases probability under the hypotheses of Lemma~\ref{lem:block} (Fig.~\ref{fig:cycle-repair}), as Section~\ref{sec:heavy-block} proves. Second, once every cycle is constant, we express how bridges and designated fields combine into an independent-set objective on \(G\).

\begin{figure}[!htbp]
\centering
\includegraphics[width=\linewidth]{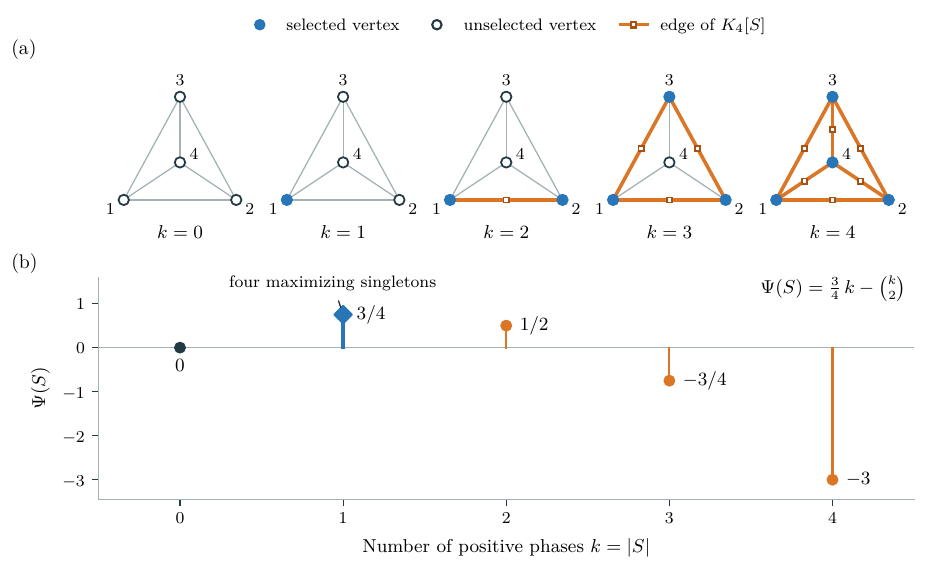}
\caption{Maximizing the independent-set score on \(K_4\). In (a), one representative is shown for each \(k=|S|\). Filled vertices form \(S\), and the highlighted edges belong to the induced graph \(K_4[S]\). In (b), taking \(\mathcal M(\lambda)=3W/2\) within \eqref{eq:field} gives \(\alpha=3/4\). After optimizing the bridge signs, \(\Psi(S)=(3/4)k-\binom{k}{2}\) reaches its maximum \(3/4\) at all four singletons.}\label{fig:k4-score}
\end{figure}

For constant cycles let \(z_v\) be the common sign on the cycle for vertex \(v\), and put \(S=\{v:z_v=+1\}\). The first-order contribution of a bridge from \(u\) to \(v\) has the form \(s_x(w_Bz_u-w_Cz_v)\). Maximizing over its sign gives an advantage \(2W\) for opposite phases, where \(W=\min(w_B,w_C)\). The cubic graph has \(3|S|-2|E(S)|\) edges leaving \(S\). Each positive cycle also pays a field cost of \(2\mathcal M(\lambda)\) relative to the negative phase. After optimizing every bridge, the part of the leading score that depends on \(S\) is \(4W\Psi(S)\), where
\begin{equation}\label{eq:IS}
 \Psi(S)=\alpha|S|-|E(S)|,\qquad
 \alpha=\frac32-\frac{\mathcal M(\lambda)}{2W}.
\end{equation}
Without a field (\(\mathcal M(\lambda)=0\)), \(\alpha=3/2>1\) would add more for every selected vertex than an internal edge subtracts. Choosing \(\lambda\) so that its bias falls in the target window
\begin{equation}\label{eq:field}
 \frac75W\le\mathcal M(\lambda)\le\frac{13}{5}W
\end{equation}
places the vertex reward strictly inside \((0,1)\), namely \(\frac15\le\alpha\le\frac45\). Every selected vertex then contributes \(\alpha\) to \(\Psi(S)\), while every edge between two selected vertices subtracts \(1>\alpha\). Consequently, if \(S\) contains an edge, removing either endpoint increases \(\Psi(S)\) by at least \(1-\alpha\ge1/5\). If \(S\) is independent but not maximum, replacing \(S\) by a maximum independent set increases \(\Psi(S)\) by at least \(\alpha\ge1/5\). In both cases, choosing optimal bridge signs for the improved phases raises the leading term of \(2^mp_{t,t'}^\theta\) by at least \(4W/5\). Fig.~\ref{fig:k4-score} illustrates this independent-set objective~\cite[Section~4.2]{lucas} on \(K_4\), and the same algebraic score \eqref{eq:IS}--\eqref{eq:field} will also govern Section~\ref{sec:heavy}.

\section{Decoding a maximum independent set from optimal cycle phases}\label{sec:heavy}\label{app:analytic}
Return to our concrete pair of adjacent vertices \(u\) and \(v\) joined by a bridge \(x\) (Fig.~\ref{fig:socg-block}), and consider what changes when we pass from standard Gaussian noise to a general rotation-invariant law satisfying Theorem~\ref{thm:main}. Two difficulties arise across this wider class. First, we no longer have closed-form formulas like Sheppard's to evaluate \(\kappa_q,w_B,w_C,\mathcal M\) or to calibrate a single block's field in advance. Second, a single scale \(t'=t\) does not guarantee that bridge interactions stay weaker than cycle agreement for every law under \eqref{eq:L}. While some laws with infinite variance, including the planar Cauchy law of Section~\ref{sec:examples}, still achieve single-scale separation between bridge and cycle weights, very slow tails can concentrate enough mass above \(1/t\) (where both responses in \eqref{eq:test} and \eqref{eq:bridgetest} saturate near \(\pm1\)) to keep the bridge weights comparable to \(\kappa_q(t)\). Our single-scale estimates then no longer guarantee that repairing a nonconstant cycle outweighs its three incident bridges. Section~\ref{sec:heavy-weights} resolves both difficulties for the entire class of Theorem~\ref{thm:main} by allowing a smaller bridge scale \(t'\le t\) and bounding all four weights from the common symmetric scalar law of the projections with bounded density. Section~\ref{sec:heavy-block} then extends the sign expansion and the decoding lemma for one block to any linear-test model of the form \eqref{eq:test}--\eqref{eq:bridgetest} driven by independent copies of such a symmetric scalar law.

\subsection{Bounded responses and weight bounds}\label{sec:heavy-weights}
To bound the weights in \eqref{eq:weightsdef} without moment assumptions, we measure each test's sensitivity through truncated averages that stay finite for every probability law. We first verify that a bounded radial planar density yields a common symmetric projection law \(U\) with a bounded one-dimensional density \(g\), then use uniform continuity of \(g*g\) to match upper and lower bounds on every local weight. Everything in this section depends on the planar noise only through the representation of the linear tests \eqref{eq:test}--\eqref{eq:bridgetest} in independent copies of this common symmetric scalar variable \(U\) with bounded density \(g\), together with the coefficient ranges \eqref{eq:ranges}. Merely having bounded directional projection densities in every direction does not suffice on its own: without a common scalar law, projections in different directions can have different shapes, and without symmetry, even zero-field marginal probabilities need not equal \(1/2\). Rotational invariance supplies the required scalar representation directly in Theorem~\ref{thm:main}, and directional rescaling recovers it for fixed invertible linear images in Section~\ref{sec:elliptical}.

Let \(g\) and \(F\) be the density and distribution function of a unit projection \(U\) of \(Z\sim\nu\). By rotational invariance, this law is symmetric and independent of direction. Integrating the planar density \(\rho(z)=\varphi(\|z\|)\) along a line and substituting \(u=x^2+y^2\) yields
\begin{equation}\label{eq:abel}
 g(x)=\int_{x^2}^\infty\frac{\varphi(\sqrt u)}{\sqrt{u-x^2}}\,du,
 \qquad \int_0^\infty\varphi(\sqrt u)\,du=\frac{1}{\pi}.
\end{equation}
Because \(0\le\varphi\le M_\rho\), the integral in \eqref{eq:abel} is maximized when the mass \(1/\pi\) sits at height \(M_\rho\) on an interval of length \(1/(\pi M_\rho)\) adjacent to the singularity at \(u=x^2\). Hence \(g\) is bounded, with
\begin{equation}\label{eq:greg}
 \|g\|_\infty\le2\sqrt{M_\rho/\pi},\qquad
 |F(x)-F(y)|\le\|g\|_\infty|x-y|,\qquad F(-x)=1-F(x).
\end{equation}
The difference of two independent projections has density \(h=g*g\). Since \(g\in L^1\cap L^\infty\), the convolution \(h\) is uniformly continuous, and Cauchy--Schwarz gives \(0\le h\le h(0)\). Put \(J=h(0)=\int_{-\infty}^\infty g(x)^2\,dx>0\), and note that \(J\le\|g\|_\infty\int_{-\infty}^\infty g(x)\,dx=\|g\|_\infty\).

\begin{figure}[!tbp]
\centering
\includegraphics[width=\linewidth]{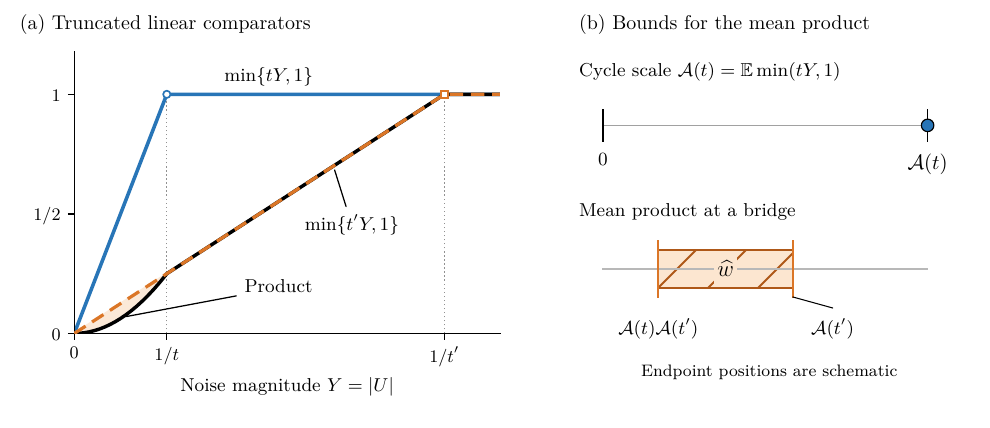}
\caption{Two thresholds for the same projected displacement. In (a), the truncated linear comparators are \(\min(tY,1)\) and \(\min(t'Y,1)\), with \(t=1/5\) and \(t'=1/20\). Their product (black) coincides with the second curve once \(Y\ge1/t\). In (b), averaging bounds the bridge scale \(\widehat w\) between \(\A(t)\A(t')\) and \(\A(t')\), both at most the cycle scale \(\A(t)\). The interval positions are schematic.}\label{fig:responses}
\end{figure}

Set \(Y=|U|\). When a test variable multiplies a projection of magnitude \(y\) by a coefficient of order \(t\), its centred response \(L_W\) grows linearly for small \(ty\) and saturates at \(1/2\) for large \(ty\), behaving like \(\min(ty,1)\). We therefore define the two truncated scale averages
\[
 \A(t)=\E\min(tY,1),\qquad
 \widehat w(t,t')=\E\bigl[\min(tY,1)\min(t'Y,1)\bigr].
\]
Both averages vanish as their arguments tend to zero, even when \(Y\) has no finite positive moment. Putting \(a_0=\E\min(Y,1)>0\), monotonicity and subadditivity of \(s\mapsto\min(s,1)\) give
\begin{equation}\label{eq:AB}
 a_0t\le\A(t)\le1\qquad(0<t\le1),\qquad
 \A(2t)\le2\A(t).
\end{equation}
Under condition \eqref{eq:L}, splitting the expectation of \(\min(2^{-j}Y,1)\) at \(Y=2^{j/2}\) and using \(Y\le\|Z\|\) yields
\begin{equation}\label{eq:Aupper}
 \A(2^{-j})\le2^{-j/2}+C_L\lfloor j/2\rfloor^{-\gamma}
             \le C_Aj^{-\gamma}\qquad(j\ge3)
\end{equation}
for a fixed integer \(C_A\). Section~\ref{sec:transfer-candidates} uses \eqref{eq:Aupper} to bound the length of the candidate scale list without evaluating \(\A\). Fig.~\ref{fig:responses} shows how choosing \(t'<t\) keeps the bridge scale \(\widehat w(t,t')\le\A(t')\) well below the cycle scale \(\A(t)\). We abbreviate \(\A=\A(t)\), \(\A'=\A(t')\) and \(\widehat w=\widehat w(t,t')\) when the scales are fixed.

To prepare for the elliptical extension in Section~\ref{sec:elliptical}, we rely on Table~\ref{tab:coeff} only through fixed coefficient ranges. For \(0<t'\le t<1/4\) and \(C_0=2\), every cycle cell and bridge satisfies
\begin{equation}\label{eq:ranges}
 \begin{gathered}
 |a_q-1|\le C_0t,\qquad \frac{t}{C_0}\le b_q\le C_0t,\qquad 0\le c_q\le C_0t,\\
 |a'-1|\le C_0t',\qquad \frac{t'}{C_0}\le b',c'\le C_0t',
 \end{gathered}
\end{equation}
with \(c_q=c(t)\ge t/C_0\) at the two corner endpoints of every bridge. Recall from Lemma~\ref{lem:geometry} that all bridges share the same coefficients, all bridge endpoints share the same corner coefficients, and all designated cells share the same straight-cell coefficients. Every constant derived in this section depends only on \(\nu\) and \(C_0\ge2\).

Conditioning each bridge correlation in \eqref{eq:weightsdef} on the shared projection \(V\) expresses both bridge weights through the unified response function \(L_W(y)=\Prob(W<y)-\tfrac12\):
\begin{equation}\label{eq:bridgeweights}
 w_B=4\E\bigl[L_{W_1}(c(t)Y)\,L_{W_2}(b'Y)\bigr],\qquad
 w_C=4\E\bigl[L_{W_1}(c(t)Y)\,L_{W_3}(c'Y)\bigr],
\end{equation}
where \(W_1=Q-a_qU-b_qU'\) (for an endpoint \(q\)), \(W_2=Q-a'U+c'V'\) and \(W_3=Q-a'U-b'V'\), all variables being independent copies of \(U\). Because each \(L_{W_i}\) is odd and nondecreasing, replacing the signed shared projection by \(Y=|U|\) shows immediately that \(w_B,w_C>0\), while the minus sign in \eqref{eq:bridgetest} produces the negative correlation \(-w_C\) at the upper endpoint.

To compare every response \(L_W(cy)\) against \(\min(ty,1)\), we need the density of \(W\) to stay close to \(J=h(0)\) on a fixed neighbourhood of zero as the coefficients approach their \(t=0\) limits.

\begin{lemma}\label{lem:near}
For \(0<\varepsilon\le1/2\), let \(\mathcal W_\varepsilon\) be the set of random variables \(W=Q-aU-\gamma_1V_1-\gamma_2V_2\), where \(Q,U,V_1,V_2\) are independent copies of \(U\), \(|a-1|\le\varepsilon\) and \(|\gamma_1|,|\gamma_2|\le\varepsilon\). Each \(W\) is symmetric with a continuous density \(k_W\le\|g\|_\infty\), and \(\sup_{W\in\mathcal W_\varepsilon}\|k_W-h\|_\infty\to0\) as \(\varepsilon\to0^+\). Consequently there are \(\varepsilon_1\in(0,1/2]\) and \(y_1\in(0,1]\), depending only on \(\nu\), such that
\begin{equation}\label{eq:near}
 \tfrac{11}{12}J\le k_W(y)\le\tfrac{13}{12}J
 \qquad(|y|\le y_1,\ W\in\mathcal W_{\varepsilon_1}).
\end{equation}
\end{lemma}

\phantomsection\label{app:near}
To compare the density \(k_W\) of a test variable \(W=Q-aU-\gamma_1V_1-\gamma_2V_2\in\mathcal W_\varepsilon\) with the reference density \(h=g*g\) of \(Q-U\), we split the perturbation into two pieces. First we bound the effect of dilating \(U\) to \(aU\) through continuity of dilations in \(L^1(\R)\) and Young's inequality. Then we bound the translation by \(S=\gamma_1V_1+\gamma_2V_2\) by splitting at a small cutoff \(r>0\) and using continuity of translations in \(L^1(\R)\).

\begin{proof}[Proof of Lemma~\ref{lem:near}]
Because \(W\) is the sum of \(Q\) and an independent symmetric random variable, \(W\) is symmetric and admits the density \(k_W(y)=\E g(y+aU+\gamma_1V_1+\gamma_2V_2)\), which satisfies \(0\le k_W\le\|g\|_\infty\). Let \(g_a(z)=a^{-1}g(z/a)\) be the density of \(aU\), and put \(k_a=g*g_a\) for the density of \(Q-aU\). As the convolution of an integrable function and a bounded function, \(k_a\) is uniformly continuous, and therefore so is \(k_W(y)=\E k_a(y+S)\), where \(S=\gamma_1V_1+\gamma_2V_2\). We work with these continuous versions throughout.

Since translations and dilations act continuously on \(L^1(\R)\), the \(L^1\)-modulus of continuity \(\omega_1(r)=\sup_{|u|\le r}\|g(\cdot+u)-g\|_1\) tends to zero as \(r\to0^+\), and \(\|g_a-g\|_1\to0\) as \(a\to1\). Young's convolution inequality yields
\[
 \|k_a-h\|_\infty\le\|g\|_\infty\|g_a-g\|_1.
\]
For \(a\ge1/2\) (which holds whenever \(\varepsilon\le1/2\)), we have \(\|g_a\|_\infty=a^{-1}\|g\|_\infty\le2\|g\|_\infty\), so Young's inequality also gives the uniform modulus bound
\[
 |k_a(y)-k_a(y')|\le\|g_a\|_\infty\,\omega_1(|y-y'|)\le2\|g\|_\infty\,\omega_1(|y-y'|).
\]
Averaging \(k_a(y+S)-k_a(y)\) over \(S\) and splitting according to whether \(|S|\le r\) or \(|S|>r\), we obtain for every \(r>0\) and every \(y\in\R\),
\[
 |k_W(y)-k_a(y)|
 \le2\|g\|_\infty\,\omega_1(r)+2\|g\|_\infty\Prob(|S|>r)
 \le2\|g\|_\infty\,\omega_1(r)+4\|g\|_\infty\Prob(\varepsilon Y>r/2),
\]
where the last step uses \(|S|\le\varepsilon(|V_1|+|V_2|)\) and a union bound (writing \(Y=|U|\)). Given any \(\eta>0\), choose \(r>0\) so that \(2\|g\|_\infty\,\omega_1(r)\le\eta/3\), and then choose \(\varepsilon\in(0,1/2]\) small enough that both \(4\|g\|_\infty\Prob(\varepsilon Y>r/2)\le\eta/3\) and \(\|g\|_\infty\|g_a-g\|_1\le\eta/3\) whenever \(|a-1|\le\varepsilon\). By the triangle inequality, \(\sup_{W\in\mathcal W_\varepsilon}\|k_W-h\|_\infty\le\eta\), proving uniform convergence. Setting \(\eta=J/24\) and choosing \(y_1\in(0,1]\) by continuity of \(h\) at zero so that \(|h(y)-J|\le J/24\) for \(|y|\le y_1\) yields \eqref{eq:near}.
\end{proof}

No rate of convergence is needed in Lemma~\ref{lem:near}. For any \(W\in\mathcal W_{\varepsilon_1}\), \(y\ge0\) and coefficient \(c\in[t/C_0,C_0t]\), integrating \eqref{eq:near} on \([0,\min(cy,y_1)]\) and combining \(k_W\le\|g\|_\infty\) with \(L_W\le1/2\) yields
\begin{equation}\label{eq:response}
 c_\nu\min(ty,1)\le L_W(cy)\le C_\nu\min(ty,1),
\end{equation}
with the two constants
\[
 c_\nu=\tfrac{11}{12}J\min(y_1,1/C_0),\qquad
 C_\nu=\max(1,C_0\|g\|_\infty).
\]
The same bounds hold with \(t'\) for coefficients in \([t'/C_0,C_0t']\). Notice two elementary relations that simplify our constants below: taking \(y=1/t\) in \eqref{eq:response} gives \(c_\nu\le L_W(c/t)\le1/2\), and \(C_0\ge2\) gives \(c_\nu\le\tfrac{11}{24}J<\tfrac12 J\).

\begin{lemma}\label{lem:weights}
There is \(t_1\in(0,1/4)\), depending only on \(\nu\) and \(C_0\), such that for \(0<t'\le t\le t_1\) and every cycle cell \(q\),
\begin{equation}\label{eq:weightbounds}
 \A\A'\le\widehat w\le\A',\qquad
 4c_\nu^2\widehat w\le w_B,w_C\le 4C_\nu^2\widehat w,\qquad
 \frac{c_\nu}{2}\A\le\kappa_q\le2C_\nu\A.
\end{equation}
Moreover \(\mathcal M\) is odd and nondecreasing, \(|\mathcal M(\theta)|\le2\|g\|_\infty|\theta|\) for every \(\theta\), and
\begin{equation}\label{eq:biasbounds}
 \tfrac{11}{6}J\theta\le\mathcal M(\theta)\le\tfrac{13}{6}J\theta\quad(0\le\theta\le y_1),\qquad
 \mathcal M(\tfrac54\theta)\le\tfrac32\mathcal M(\theta)\quad(0\le\theta\le\tfrac45y_1).
\end{equation}
\end{lemma}

\begin{proof}
We establish the correlation bounds using positive covariance of nondecreasing functions and then bound \(\mathcal M\) directly from \eqref{eq:near}. For any two nondecreasing functions \(f_1,f_2\) and an independent copy \(Y'\) of \(Y\), the product \((f_1(Y)-f_1(Y'))(f_2(Y)-f_2(Y'))\) is nonnegative, so taking expectations gives \(\E[f_1(Y)f_2(Y)]\ge\E f_1(Y)\,\E f_2(Y)\). Applying this to \(f_1(y)=\min(ty,1)\) and \(f_2(y)=\min(t'y,1)\) gives \(\widehat w\ge\A\A'\), while \(\min(tY,1)\le1\) gives \(\widehat w\le\A'\).

Choose \(t_1\in(0,1/4)\) with \(t_1\le\varepsilon_1/C_0\), reduced if necessary so that \(\Prob(|Q|\le a|U|)\ge1/4\) whenever \(|a-1|\le C_0t_1\). This holds for small \(t_1\) because \(\Prob(|Q|\le|U|)=1/2\) for independent copies \(Q,U\) and the probability varies continuously with \(a\). By \eqref{eq:ranges}, the variables \(W_1,W_2,W_3\) of \eqref{eq:bridgeweights}, the variable \(W_q=Q-a_qU-c_qV\) of \eqref{eq:weightsdef}, and \(-\Lambda_d^0=Q-a_dU-b_dU'-c_dV\) all belong to \(\mathcal W_{\varepsilon_1}\).

Substituting \eqref{eq:response} into \eqref{eq:bridgeweights} gives \(4c_\nu^2\widehat w\le w_B,w_C\le 4C_\nu^2\widehat w\). For \(\kappa_q\), both factors inside \eqref{eq:weightsdef} are odd in \(U\), so we may replace \(U\) by \(Y\), turning both into nondecreasing functions of \(Y\). Using \(|F-\frac12|\le\frac12\) and \eqref{eq:response} yields \(\kappa_q\le2C_\nu\A\). Applying the covariance inequality in the opposite direction gives
\[
 \kappa_q\ge4\bigl(\E F(a_{q^-}Y)-\tfrac12\bigr)\,\E L_{W_q}(b_qY).
\]
By symmetry of \(Q\), \(\E F(aY)-\frac12=\Prob(0\le Q\le aY)=\frac12\Prob(|Q|\le aY)\ge\frac18\), and the second factor is at least \(c_\nu\A\) by \eqref{eq:response}, proving \(\kappa_q\ge\frac12 c_\nu\A\). Finally, for \(\theta\ge0\), \(\mathcal M(\theta)\) integrates the density of \(-\Lambda_d^0\in\mathcal W_{\varepsilon_1}\) over \([-\theta,\theta]\), which is nondecreasing and at most \(2\|g\|_\infty\theta\). Integrating \eqref{eq:near} yields the first pair of bounds in \eqref{eq:biasbounds}, and the second follows because \(\frac54\cdot\frac{13}{11}<\frac32\).
\end{proof}

\subsection{Cycle agreement and decoding a maximum independent set}\label{sec:heavy-block}
We now state the general sign-correlation expansion for one block (proved in Section~\ref{app:expansion}) and show that the two-step decoding argument of Section~\ref{sec:gauss-block} carries over once the bridge scale satisfies \(\widehat w(t,t')\le c_1\A(t)\). Keeping the expansion separate from this two-scale inequality allows the same result to supply both the Gaussian specialization in Section~\ref{sec:socg-gaussian} and the heavy-tailed block bound below. Both lemmas assume that the independent projections entering \eqref{eq:test}--\eqref{eq:bridgetest} share a common symmetric scalar law \(U\) with bounded density \(g\) and that the coefficients satisfy \eqref{eq:ranges}. The constants in \(O_\nu\) depend only on the law of \(U\) and on \(C_0\ge2\).

\begin{lemma}\label{lem:expansion}
Suppose the independent projections in \eqref{eq:test}--\eqref{eq:bridgetest} follow a common symmetric scalar law \(U\) with bounded density \(g\), the coefficients satisfy \eqref{eq:ranges}, \(0<t'\le t<1/4\), the fields vanish outside the designated cells of the cycles, and \(|\theta_q|\le\A\). If \(m\A\) is below a sufficiently small constant depending only on the law of \(U\) and \(C_0\), then, uniformly over all sign vectors and these fields,
\begin{equation}\label{eq:all}
 2^mp_{t,t'}^\theta(s)
 =\prod_{v\in V(G)}\pi_v(s)
  +w_B\sum_x s_xs_{q_B(x)}-w_C\sum_x s_xs_{q_C(x)}
  +O_\nu(m^2\A\,\widehat w).
\end{equation}
For each cycle, at zero field,
\begin{equation}\label{eq:cycle}
 \pi_v^0(s)=1+\sum_{q\in\mathcal C_v}\kappa_q s_qs_{q^-}
           +O_\nu(|\mathcal C_v|^2\A^2),
\end{equation}
and negating all signs of \(\mathcal C_v\) leaves \(\pi_v^0(s)\) unchanged. A field \(\theta\) at the designated cell changes \(\pi_v(s)\) by at most \(6\|g\|_\infty|\theta|\), and the two constant sign vectors of the cycle satisfy
\begin{equation}\label{eq:fieldgap}
 \pi_v(+)-\pi_v(-)=2\mathcal M(\theta)+O_\nu(|\mathcal C_v|\,\A\,|\theta|).
\end{equation}
\end{lemma}

\phantomsection\label{app:expansion}
We now prove Lemma~\ref{lem:expansion} by expanding \(2^mp_{t,t'}^\theta(s)\) in the sign correlations of subsets of cells. Bridge-free subsets sum identically to \(\prod_v\pi_v(s)\), and every pair consisting of a bridge and one endpoint produces the exact weight \(w_B\) or \(-w_C\). To bound all remaining subsets, we condition on the shared projections \(U_q,V_q\) of all cycle cells and use the common symmetric scalar law \(U\) (with distribution function \(F\) and bounded density \(g\)) to express each cycle cell's conditional mean sign through \(F\), splitting it into a zero-mean odd main part plus a small correction bounded by truncated responses. Three sign-reversal symmetries of this common symmetric law then eliminate every unmatched term, and the remaining expectations factor cleanly across disjoint projections. We carry out this argument in four steps covering sign correlations, symmetries and majorants, a single zero-field cycle, and bridge subsets together with designated fields.

\begin{proof}[Proof of Lemma~\ref{lem:expansion}]
\emph{Step 1: Sign correlations.}
For any subset \(L\) of the \(m\) cells, write \(s_L=\prod_{q\in L}s_q\) and \(\varrho_L=\E\prod_{q\in L}\sgn\Lambda_q\), taking the empty product equal to one. Because every projection has a bounded density, each linear test is nonzero almost surely, so the indicator of \(\{\sgn\Lambda_q=s_q\}\) equals \((1+s_q\sgn\Lambda_q)/2\) almost surely. Multiplying over all \(m\) cells and taking expectation gives
\begin{equation}\label{eq:signexpansion}
 2^mp_{t,t'}^\theta(s)=\sum_L s_L\varrho_L,
\end{equation}
where \(L\) ranges over all subsets of cells. The same identity restricted to one cycle yields \(\pi_v(s)=\sum_{L\subseteq\mathcal C_v}s_L\varrho_L\). Distinct cycles share no sites, so \(\varrho_L=\prod_{v\in V(G)}\varrho_{L\cap\mathcal C_v}\) whenever \(L\) contains no bridge. Summing over all bridge-free subsets therefore contributes \(\prod_v\pi_v(s)\) exactly, fields included. Because bridges and their endpoint cells carry zero field, \eqref{eq:weightsdef} identifies \(\varrho_{\{x,q_B(x)\}}=w_B\) and \(\varrho_{\{x,q_C(x)\}}=-w_C\), producing the two bridge sums in \eqref{eq:all}. It remains to bound every other subset containing a bridge and to expand each individual cycle factor.

\emph{Step 2: Main parts, corrections and symmetries.}
Condition on the shared projections \(\{U_q,V_q\}\) of all cycle cells. Every remaining projection (\(Q_q\) on each cycle cell and \(U_x,Q_x\) on each bridge) enters only a single test, so all \(m\) signs are conditionally independent. Writing \(\chi_q=\E[\sgn\Lambda_q\mid\{U_r,V_r\}]\) for the conditional mean sign of cell \(q\), we have \(\varrho_L=\E\prod_{q\in L}\chi_q\).

For a cycle cell \(q\), \eqref{eq:test} gives \(\chi_q=2F(a_qU_q+b_qU_{q^-}+c_qV_q+\theta_q)-1\). Decompose \(\chi_q=\xi_q+\Delta_q\), where the \emph{main part} \(\xi_q=2F(a_qU_q)-1\) is an odd function of \(U_q\) alone with \(|\xi_q|\le1\), and \(\Delta_q\) is the \emph{correction}. For a bridge \(x\) with endpoints \(q=q_B(x)\) and \(r=q_C(x)\), integrating out \(U_x,Q_x\) in \eqref{eq:bridgetest} gives \(\chi_x=2\Prob(Q-a'U<b'V_q-c'V_r\mid V_q,V_r)-1\), an odd function of \((V_q,V_r)\). Because \(Q-a'U\) is symmetric with density bounded by \(\|g\|_\infty\), its distribution function satisfies the same Lipschitz bound \eqref{eq:greg} as \(F\). Put \(u(y)=\min(ty,1)\) and \(u'(y)=\min(t'y,1)\), and recall \(C_\nu=\max(1,C_0\|g\|_\infty)\). Combining \(|F(z+\delta)-F(z)|\le\min(\|g\|_\infty|\delta|,1)\) with the coefficient ranges \eqref{eq:ranges} yields the majorants
\[
 |\Delta_q|\le M_q=2C_\nu\bigl[u(|U_{q^-}|)+u(|V_q|)\bigr]+2\|g\|_\infty|\theta_q|,\qquad
 |\chi_x|\le M_x=2C_\nu\bigl[u'(|V_q|)+u'(|V_r|)\bigr].
\]

Expanding \(\chi_q=\xi_q+\Delta_q\) for every cycle cell in \eqref{eq:signexpansion} writes \(2^mp_{t,t'}^\theta(s)\) as a sum of terms \(s_{K\cup H\cup X}\,T(K,H,X)\), where \(K\) and \(H\) are disjoint sets of cycle cells (represented by corrections \(\Delta_q\) and main parts \(\xi_q\), respectively), \(X\) is a set of bridges, and
\[
 T(K,H,X)=\E\Bigl[\prod_{q\in K}\Delta_q\prod_{q\in H}\xi_q\prod_{x\in X}\chi_x\Bigr].
\]
Three symmetries cancel almost all terms:
\begin{enumerate}
\item[(a)] If some cell \(q\in H\) is not the predecessor of any cell in \(K\), then \(U_q\) appears only in the odd factor \(\xi_q\), whose expectation is zero, so \(T(K,H,X)=0\).
\item[(b)] If some bridge \(x\in X\) has neither endpoint in \(K\), then its shared projections \((V_{q_B(x)},V_{q_C(x)})\) appear only in the odd factor \(\chi_x\), so negating them together gives \(T(K,H,X)=0\).
\item[(c)] Negating all projections simultaneously reverses every main part \(\xi_q\), every bridge factor \(\chi_x\), and every correction \(\Delta_q\) of a cell with \(\theta_q=0\). Thus \(T(K,H,X)=0\) whenever \(|K|+|H|+|X|\) is odd and no cell of \(K\) carries a nonzero field.
\end{enumerate}

For any term not cancelled by rule~(a), every element of \(H\) is the predecessor of an element of \(K\), so each choice of \(K\) allows at most \(2^{|K|}\) choices of \(H\). Bounding \(|\xi_q|\le1\) leaves the expectation of the product of majorants \(M_q\) (\(q\in K\)) and \(M_x\) (\(x\in X\)). Distinct cycle majorants \(M_q\) depend on disjoint projections, because \(M_q\) uses only \(U_{q^-}\) and \(V_q\) and \(q\mapsto q^-\) is bijective on each cycle. Each bridge majorant \(M_x\) shares projections only with the majorants of its two endpoints, and by Lemma~\ref{lem:geometry} each cycle cell touches at most one bridge. The expectation therefore factors over each bridge \(x\in X\) (grouped with whichever of its endpoints lie in \(K\)) and the remaining cells of \(K\). For a cycle cell \(q\), and a bridge \(x\) with zero-field endpoints \(q\) and \(r\),
\begin{equation}\label{eq:majorants}
 \E M_q\le6C_\nu\A,\qquad
 \E[M_xM_q]\le16C_\nu^2\widehat w,\qquad
 \E[M_xM_qM_r]\le64C_\nu^3\widehat w\A.
\end{equation}
The first inequality uses \(|\theta_q|\le\A\) and \(\|g\|_\infty\le C_0\|g\|_\infty\le C_\nu\). Expanding the independent factors gives \(\E[M_xM_q]=4C_\nu^2(\widehat w+3\A\A')\) and \(\E[M_xM_qM_r]=32C_\nu^3\A(\widehat w+\A\A')\), and applying \(\A\A'\le\widehat w\) (which holds because \(u(|V|)\) and \(u'(|V|)\) are both nondecreasing functions of \(|V|\), hence positively correlated) completes \eqref{eq:majorants}. Because \(m\A\) is below a small constant depending on \(\nu\) and \(C_0\), we have \(4C_\nu\A\le1\) and \(e^{12C_\nu m\A}\le3/2\).

\emph{Step 3: One cycle at zero field.}
Restrict to \(K,H\subseteq\mathcal C_v\) and \(X=\emptyset\) with zero field. By rule~(a), the only term not cancelled with \(K=\emptyset\) is the empty product \(1\). For \(K=\{q\}\), the term with \(H=\emptyset\) vanishes by rule~(c), leaving \(H=\{q^-\}\), whose expectation is \(\E[\Delta_q\xi_{q^-}]\). Averaging \(\Delta_q\) over \(U_q\) and \(V_q\) leaves \(2L_{W_q}(b_qU_{q^-})\) while \(\xi_{q^-}=2F(a_{q^-}U_{q^-})-1\), so this term equals \(\kappa_q\) by \eqref{eq:weightsdef}. For \(l=|K|\ge2\), there are at most \(|\mathcal C_v|^l/l!\) choices of \(K\) and at most \(2^l\) choices of \(H\), each bounded by \((6C_\nu\A)^l\). Summing over \(l\ge2\) gives at most \(\sum_{l\ge2}(12C_\nu|\mathcal C_v|\A)^l/l!=O_\nu(|\mathcal C_v|^2\A^2)\), proving \eqref{eq:cycle}. Negating all projections of \(\mathcal C_v\) reverses all zero-field tests, so \(\pi_v^0(-s)=\pi_v^0(s)\).

\emph{Step 4: Bridge subsets and designated fields.}
Consider terms with \(X\neq\emptyset\). By rule~(b), every term not cancelled has \(K\) meeting the endpoints of every bridge in \(X\). Fix \(X\) and sum \(|T(K,H,X)|\) over all \(K\) and allowed \(H\) (where each cell of \(K\) may include or omit its predecessor in \(H\)). By the factorization above, this sum is bounded by a product over all bridges \(x\in X\) and all remaining cycle cells. Each bridge \(x\in X\) with endpoints \(q,r\) contributes the three ways \(K\) can meet \(\{q,r\}\):
\[
 2\E[M_xM_q]+2\E[M_xM_r]+4\E[M_xM_qM_r]\le128C_\nu^2\widehat w,
\]
using \eqref{eq:majorants} and \(4C_\nu\A\le1\). Every other cycle cell \(q'\) contributes \(1+2\E M_{q'}\le1+12C_\nu\A\). Hence the terms for a fixed \(X\) total at most \((128C_\nu^2\widehat w)^{|X|}e^{12C_\nu m\A}\). Summing over the at most \(m^j/j!\) sets \(X\) of size \(j\ge2\) gives \(O_\nu(m^2\widehat w^2)=O_\nu(m^2\A\widehat w)\), since \(\widehat w\le\A'\le\A\).

For a single bridge \(X=\{x\}\), terms where \(K\) includes at least one cycle cell outside \(\{q_B(x),q_C(x)\}\) sum to at most \(128C_\nu^2\widehat w\,(e^{12C_\nu m\A}-1)=O_\nu(m\A\widehat w)\) by the same factorization. Terms with \(K=\{q_B(x),q_C(x)\}\) give at most four terms of size at most \(64C_\nu^3\widehat w\A\). Only \(K=\{q\}\) with \(q\in\{q_B(x),q_C(x)\}\) remains: \(H=\emptyset\) is the exact correlation \(w_B\) or \(-w_C\) already extracted into \eqref{eq:all}, and \(H=\{q^-\}\) has three zero-field factors (main parts never involve a field, even if \(q^-\) is designated) and thus vanishes by rule~(c). Summing over at most \(m\) bridges yields \(O_\nu(m^2\A\widehat w)\), proving \eqref{eq:all}.

Finally, consider a cycle \(\mathcal C_v\) whose designated cell \(d\) carries field \(\theta\), and define the event \(\mathcal E=\{|\Lambda_d^0|\le|\theta|\}\). Using the same projections with and without \(\theta\), the sign vectors differ at most at \(d\) and only on \(\mathcal E\). Conditioned on all projections other than \(Q_d\), the other signs are fixed and \(\Prob(\mathcal E\mid\text{others})\le2\|g\|_\infty|\theta|\). The event that all cells outside \(d\) match \(s\) is the union of the two zero-field sign vectors agreeing with \(s\) outside \(d\). By the count in Step~3, each has probability at most \(2^{-|\mathcal C_v|}e^{12C_\nu|\mathcal C_v|\A}\le\frac32\,2^{-|\mathcal C_v|}\), so their union has probability at most \(3\cdot2^{-|\mathcal C_v|}\). Multiplying by \(2^{|\mathcal C_v|}\cdot2\|g\|_\infty|\theta|\) shows that \(\theta\) changes \(\pi_v(s)\) by at most \(6\|g\|_\infty|\theta|\).

For the two constant sign vectors of \(\mathcal C_v\), negating all projections shows that \(\pi_v(-)\) under field \(\theta\) equals \(\pi_v(+)\) under field \(-\theta\). Up to a null set, the events that all cycle signs are positive under \(\theta\) and under \(-\theta\) differ precisely when \(\mathcal E\) occurs and every cell in \(\mathcal C_v\setminus\{d\}\) has positive sign. Expanding the indicator of those \(|\mathcal C_v|-1\) signs as in \eqref{eq:signexpansion} gives
\[
 \pi_v(+)-\pi_v(-)
 =2\sgn(\theta)\sum_{L\subseteq\mathcal C_v\setminus\{d\}}
  \E\Bigl[\mathbf 1_{\mathcal E}\prod_{q\in L}\sgn\Lambda_q\Bigr].
\]
The empty set \(L=\emptyset\) contributes \(2\sgn(\theta)\Prob(\mathcal E)=2\mathcal M(\theta)\) by \eqref{eq:weightsdef}. For nonempty \(L\), condition again on \(\{U_q,V_q\}\). The event \(\mathcal E\) and the signs in \(L\) are conditionally independent, and the conditional probability \(\eta=\Prob(\mathcal E\mid\{U_q,V_q\})\) satisfies \(0\le\eta\le2\|g\|_\infty|\theta|\), depends only on \((U_d,U_{d^-},V_d)\), and is invariant under negating all projections. Expanding each \(\chi_q=\xi_q+\Delta_q\) for \(q\in L\) produces the same expectations \(T(K,H,\emptyset)\) multiplied inside by \(\eta\). Since \(d\notin L\), we have \(d\notin H\), so rule~(a) still cancels every cell of \(H\) except possibly \(d^-\) (because \(U_{d^-}\) enters \(\eta\)). The lone term with \(K=\emptyset\) and \(H=\{d^-\}\) equals \(\E[\eta\xi_{d^-}]=0\), because negating all projections preserves \(\eta\) and reverses \(\xi_{d^-}\). Every remaining nonzero term therefore has \(|K|\ge1\), at most \(2^{|K|+1}\) choices of \(H\), and magnitude at most \(2\|g\|_\infty|\theta|\,(6C_\nu\A)^{|K|}\). Summing over \(K\neq\emptyset\) as in Step~3 bounds the error by \(O_\nu(|\mathcal C_v|\,\A\,|\theta|)\), proving \eqref{eq:fieldgap}.
\end{proof}

Because \(\kappa_q\le2C_\nu\A\) and \(|\theta_q|\le\A\), each cycle factor satisfies \(\pi_v(s)=1+O_\nu(|\mathcal C_v|\A)\), so the product over any subset of cycles equals \(1+O_\nu(m\A)\). To keep the single-cycle repair of Section~\ref{sec:gauss-block} (Fig.~\ref{fig:cycle-repair}) profitable across the full class of Theorem~\ref{thm:main}, we enforce \(\widehat w(t,t')\le c_1\A(t)\) for a fixed rational constant \(c_1\) depending only on the law of \(U\) and \(C_0\), for which \(\A(t')\le c_1\A(t)\) suffices by \eqref{eq:weightbounds}. Once every cycle is constant, putting the common negative field \(-\lambda\) in \eqref{eq:field} recovers the same independent-set objective \(\Psi(S)\) of \eqref{eq:IS} (Fig.~\ref{fig:k4-score}).

\begin{lemma}\label{lem:block}
Fix a rational constant \(c_1\) with
\[
 0<c_1\le\frac{c_\nu J}{250\|g\|_\infty C_\nu^2},
\]
and suppose \(0<t'\le t\le t_1\), \(\widehat w(t,t')\le c_1\A(t)\), the common negative field \(-\lambda\) at each designated cell satisfies \eqref{eq:field}, and \(m^2\A(t)\) is below a sufficiently small constant depending only on \(\nu\) and \(C_0\). Call a sign vector \emph{correct} if it is constant on every cycle and its positive cycles form a maximum independent set of \(G\) (bridge signs are unrestricted). Then every mode of \(p_{t,t'}^\theta\) is correct. Moreover, for any fixed rational \(0<d_\nu\le\tfrac12 c_\nu^2\) (which satisfies \(d_\nu\le1/2\)), every incorrect sign vector \(s\) satisfies
\begin{equation}\label{eq:blockgap}
 p_{t,t'}^\theta(s)\le
 \bigl(1-d_\nu\A(t)\A(t')\bigr)\max_r p_{t,t'}^\theta(r),
\end{equation}
and its probability is at least \(2d_\nu\A(t)\A(t')2^{-m}\) below a mode.
\end{lemma}

\begin{proof}
We first verify that the single upper bound on \(c_1\) guarantees \(c_1<1\), keeps \(\lambda<\A\) so that Lemma~\ref{lem:expansion} applies, and bounds the combined bridge and field cost during a cycle repair. Recall that \(c_\nu\le1/2\), \(c_\nu<J/2\), \(J\le\|g\|_\infty\) and \(C_\nu\ge1\). Immediately \(c_1\le\frac{1/2}{250}<1\). By \eqref{eq:field} and \eqref{eq:weightbounds}, \(\mathcal M(\lambda)\le\frac{13}{5}W\le\frac{52}{5}C_\nu^2\widehat w\). Because \(\widehat w\le\A\) is small, this lies below \(\mathcal M(y_1)\), so monotonicity gives \(\lambda<y_1\). By \eqref{eq:biasbounds}, \(\frac{11}{6}J\lambda\le\mathcal M(\lambda)\le\frac{52}{5}C_\nu^2\widehat w\), and since \(\frac{52/5}{11/6}=\frac{312}{55}<6\), we obtain
\[
 \lambda\le\frac{6C_\nu^2}{J}\widehat w
 \le\frac{6C_\nu^2}{J}c_1\A
 \le\frac{6c_\nu}{250\|g\|_\infty}\A<\A.
\]
Lemma~\ref{lem:expansion} therefore applies. Since \(m\A\) is small, every product of cycle factors over a subset of vertices lies in \([7/8,9/8]\), and \(2^mp_{t,t'}^\theta(r)\le2\) for every sign vector \(r\).

Consider first a sign vector \(s\) that is nonconstant on some cycle \(\mathcal C_v\), and form \(s'\) by the single-cycle repair of Section~\ref{sec:gauss-block} (Fig.~\ref{fig:cycle-repair}). Using \(\kappa_q\ge\frac12 c_\nu\A\) from \eqref{eq:weightbounds} in \eqref{eq:cycle} gives \(\pi_v^0(s')-\pi_v^0(s)\ge2c_\nu\A-O_\nu(|\mathcal C_v|^2\A^2)\). Including the field shift of at most \(6\|g\|_\infty\lambda\) on each of \(\pi_v(s')\) and \(\pi_v(s)\) and multiplying by the product of the remaining cycle factors (between \(7/8\) and \(9/8\)) gives a gain in the product of cycle factors of at least
\[
 \tfrac78(2c_\nu\A)-\tfrac98(12\|g\|_\infty\lambda)-O_\nu(m^2\A^2)
 =\tfrac74 c_\nu\A-\tfrac{27}{2}\|g\|_\infty\lambda-O_\nu(m^2\A^2).
\]
Only three incident bridge links change, costing at most \(6\max(w_B,w_C)\le24C_\nu^2\widehat w\). Combining this with the field term and using \(1\le\|g\|_\infty/J\) and the choice of \(c_1\) yields
\[
 6\max(w_B,w_C)+\tfrac{27}{2}\|g\|_\infty\lambda
 \le C_\nu^2\Bigl(24+\frac{81\|g\|_\infty}{J}\Bigr)\widehat w
 \le\frac{105\|g\|_\infty C_\nu^2}{J}c_1\A
 <\frac{c_\nu}{2}\A.
\]
Subtracting this loss and absorbing the \(O_\nu(m^2\A\widehat w)\) remainder of \eqref{eq:all} into \(O_\nu(m^2\A^2)\) leaves a net gain of at least \(\frac54 c_\nu\A-O_\nu(m^2\A^2)\ge c_\nu\A\) in \(2^mp_{t,t'}^\theta\), for sufficiently small \(m^2\A\).

Suppose next that every cycle is constant under \(s\), with phases \(z_v\in\{-1,1\}\) whose positive set \(S\) is not a maximum independent set. Writing \(\bar\pi_v=(\pi_v(+)+\pi_v(-))/2\), \eqref{eq:fieldgap} gives \(\pi_v(z_v)=\bar\pi_v-z_v\mathcal M(\lambda)+O_\nu(|\mathcal C_v|\A\lambda)\). Multiplying over all \(N\le m\) vertices yields
\[
 \prod_{v\in V(G)}\pi_v(z_v)
 =\prod_{v\in V(G)}\bar\pi_v-\mathcal M(\lambda)\sum_{v\in V(G)}z_v+O_\nu(m^2\A\,\widehat w).
\]
Here the cross errors \(m\A\cdot N\lambda\), \(m\A\lambda\) and \((N\lambda)^2\) each have order \(O_\nu(m^2\A\widehat w)\) because \(\lambda\le\min(\frac{6C_\nu^2}{J}\widehat w,\A)\). Substituting into \eqref{eq:all} recovers the exact score \(4W\Psi(S)\) of \eqref{eq:IS} up to an additive constant and \(O_\nu(m^2\A\widehat w)\). By the score improvement established after \eqref{eq:field} in Section~\ref{sec:gauss-block}, switching to an improved phase vector with optimal bridge signs raises the main terms by at least \(4W/5\) (regardless of the original bridge signs). Because \(W\ge4c_\nu^2\widehat w\), taking \(m^2\A\) sufficiently small keeps all errors below \(2W/5\), leaving a gain in \(2^mp_{t,t'}^\theta\) of at least
\[
 \frac25W\ge\frac85 c_\nu^2\widehat w\ge c_\nu^2\A\A'.
\]

Since \(c_\nu\le1/2\), any fixed rational \(0<d_\nu\le\frac12 c_\nu^2\) satisfies \(d_\nu\le1/8<1/2\) and \(2d_\nu\A\A'\le c_\nu^2\A\A'<c_\nu\A\). Both cases therefore improve every incorrect sign vector by at least \(2d_\nu\A\A'2^{-m}\). Every mode is correct, and dividing this absolute gap by \(\max_r p_{t,t'}^\theta(r)\le2\cdot2^{-m}\) proves \eqref{eq:blockgap}.
\end{proof}

\section{Controlling escaping sites in the reduction for heavy-tailed laws}\label{sec:transfer}
We now convert the heavy-tailed block analysis of Section~\ref{sec:heavy} into the polynomial-time reduction of Theorem~\ref{thm:main}, using one modal oracle call. Two obstacles prevent us from reusing the short Gaussian transfer of Section~\ref{sec:socg-gaussian} directly. First, under a logarithmic tail such as \((\ln R)^{-\eta}\), requiring all \(n\) sites to stay inside radius \(R\) except on an event of probability \(\beta2^{-M}/4\) would need exponentially many bits for \(R\) (Section~\ref{sec:examples}). We instead truncate at an inverse-polynomial tail probability per site and show that each large displacement disturbs at most four cells across at most two blocks (Section~\ref{sec:transfer-local}). Second, without closed-form formulas for \(\A(t)\), \(w_B\), \(w_C\) or \(\mathcal M\), the reduction cannot compute one suitable triple \((t,t',\lambda)\) directly. We therefore enumerate a polynomial list of candidate blocks guaranteed to contain a \emph{suitable} block satisfying Lemma~\ref{lem:block}, and decode every block without identifying which one is suitable (Section~\ref{sec:transfer-candidates}).

Algorithm~\ref{alg:reduction} summarizes the complete procedure in conceptual form. Throughout this section, \(m\) counts the cells of one block (including its bridges), \(B\) counts the candidate blocks in one list, \(M=Bm\) counts all cells in one list, and \(n\le12M+4\) counts all sites after joint completion. Condition \eqref{eq:L} keeps both the candidate list size and every coordinate bit length polynomial in the graph size.

\begin{algorithm}[tb]
\caption{One-call reduction for a fixed law satisfying \eqref{eq:L}}\label{alg:reduction}
\begin{algorithmic}[1]
\Require Planar cubic graph \(G\) \Comment{The reduction uses rational constants fixed for \(\nu\).}
\Ensure A maximum independent set of \(G\), using one modal oracle call
\State Construct the vertex cycles and bridges of \(G\) (Lemma~\ref{lem:geometry}) and record the cell count \(m\) of one block.
\State Enumerate a polynomial list of candidate triples \((t,t',\lambda)\) spanning all dyadic scale pairs up to a computable ceiling and a geometric grid of fields, ensuring that at least one listed block satisfies Lemma~\ref{lem:block} with a computable relative gap \(\beta\).
\State Place the candidate blocks in separated lattice boxes and complete their union into a single base configuration \(P\) with triangular hull (Lemma~\ref{lem:completion}).
\State Using \eqref{eq:L}, choose a confinement radius \(R\) giving per-site tail probability at most \(\beta/(256n)\), and choose a sign-error tolerance \(\tau\) whose total near-tie probability is at most \(\beta\,2^{-M}/8\).
\State Apply Lemma~\ref{lem:stability} to compute the dyadic geometric scale \(\sigma\), generic rational shifts \(\zeta_i\) and rescaled rational centres \(\mu_i\).
\State Query the modal solver once on \(\mu_1,\ldots,\mu_n\) to obtain a complete labelled triangulation \(T\).
\State Decode every block from its local cell triangles in \(T\), keep only blocks with constant cycles whose positive cycles form an independent set, and return a largest retained set (or \(\emptyset\) if none qualifies).
\end{algorithmic}
\end{algorithm}

\subsection{Escaping sites and the probability bound on unaffected blocks}\label{sec:transfer-local}
\begin{figure}[!htbp]
\centering
\includegraphics[width=\linewidth]{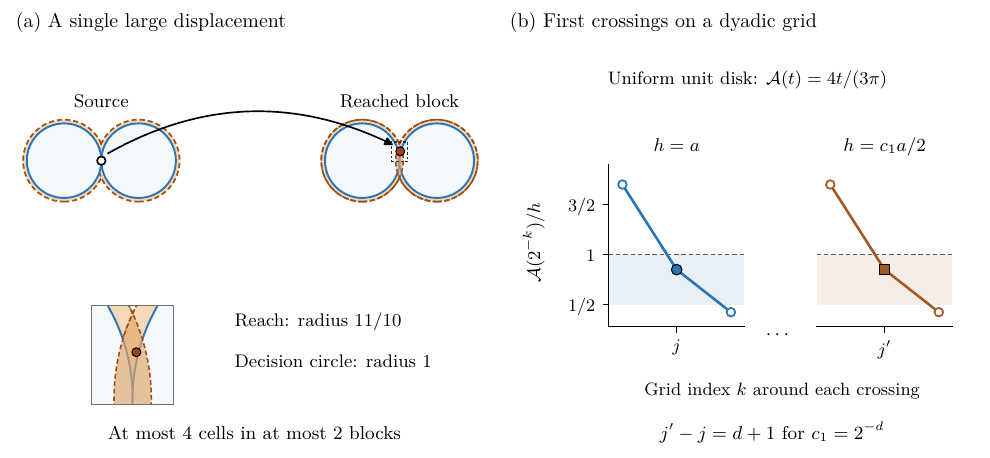}
\caption{Local effects of a large displacement and selection of the two scales. In (a), the hollow point moves into the overlap of the two enlarged disks, shown below at higher magnification. At most two incident cells and two reached cells in at most two blocks are affected. In (b), uniform noise on the unit disk gives \(\A(t)=4t/(3\pi)\) for \(t\le1\). Each group shows three consecutive indices relative to its own threshold \(h\). For \(a=2^{-L}\) with \(L\) large and an admissibly small dyadic \(c_1=2^{-d}\), the first crossings are \(j=L-1\) and \(j'=L+d\). The shaded bands give \(h/2<\A(2^{-k})\le h\) at each crossing.}\label{fig:relative-coupling}
\end{figure}

To avoid an exponential truncation radius under slow tails, we allow occasional large displacements anywhere in the configuration and bound how much of the triangulation they can obscure. The key geometric observation is that a nonlarge cell's triple circumdisks stay strictly inside a disk of radius \(11/10\) around its base centre (Lemma~\ref{lem:localdisks}). Deleting all large sites therefore restores the two selected triangles of every intact cell, and reinserting a single large site can erase faces in at most two reached cells (all in one block) in addition to the at most two cells containing that site in its source block (Fig.~\ref{fig:relative-coupling}(a)). Independence across the undamaged blocks then yields the product comparison of Proposition~\ref{prop:failure}, which serves both Theorem~\ref{thm:main} (with one copy) and Theorem~\ref{thm:factor} (with \(r\) copies).

Throughout this subsection, consider the configurations built by the reductions: one copy of the candidate list of Section~\ref{sec:transfer-candidates}, used for Theorem~\ref{thm:main}, or \(r\ge1\) copies of it in separated boxes, used for Theorem~\ref{thm:factor}. All their blocks are completed jointly into \(P\) and stabilized by Lemma~\ref{lem:stability} for a radius \(R=2^{s_R}\) with an integer \(s_R\ge1\) and a rational \(0<\tau<1\), and \(M\) and \(n\) count all cells and sites of the configuration. Recall that a site \(i\) is \emph{large} when \(\|Z_i\|>R\) and \emph{nonlarge} otherwise, and put \(q_R=T_\nu(R)=\Prob(\|Z_i\|>R)\). We discard the null event of nongeneric realizations.

\begin{lemma}\label{lem:localdisks}
Suppose the four sites of a decision cell are nonlarge. The circumdisk through any three of them is contained in the open disk of radius \(11/10\) about the base centre of the cell. After all large sites are deleted, the cell has its two selected triangular faces in the Delaunay triangulation. If neither face occurs in the complete triangulation before deletion, some large site lies within distance \(11/10\) of this centre.
\end{lemma}
\begin{proof}
We confine each perturbed circumcircle by bounding its incircle determinant outside radius \(11/10\). Translate the cell centre to the origin and write \(u=\sigma(R+2)\le r_*/(800C_P)\le r_*/800\), by item~1 of Lemma~\ref{lem:stability}. The three perturbed vertices have norms within \(u\) of one, and their orientation determinant keeps its base sign with magnitude at least \(r_*/2\). For any point \(x=(x_1,x_2)\), subtracting the fourth column from the squared-norm column of the incircle determinant of these three vertices and \(x\) writes it as
\[
 -(\|x\|^2-1)\operatorname{orient}(P_1+\Delta_1,P_2+\Delta_2,P_3+\Delta_3)+e(x),
 \qquad |e(x)|\le20u(\|x\|+1).
\]
Indeed, each of the first three entries in the modified third column has magnitude at most \(2u+u^2\), and its cofactor has magnitude at most \(2(1+u)(|x_1|+|x_2|)+2(1+u)^2\). Summing the three products with \(|x_1|+|x_2|\le\sqrt2\,\|x\|\) and \(u\le1/800\) gives \(|e(x)|\le20u(\|x\|+1)\). Whenever \(\|x\|\ge11/10\), we have \(\|x\|^2-1\ge(\|x\|+1)/10\), so the leading term has magnitude at least \(r_*(\|x\|+1)/20\), whereas \(|e(x)|\le r_*(\|x\|+1)/40\). The determinant therefore cannot vanish for \(\|x\|\ge11/10\), confining the circumcircle and its convex disk inside radius \(11/10\).

After deleting all large sites, every preserved certificate keeps the four vertices in convex cyclic order and every remaining external site outside all four triple circumcircles. The exact test \(\Gamma_q\) therefore selects two empty triangles, which appear as Delaunay faces. Reinserting sites can destroy an empty triangle only by placing a site inside its circumdisk, so if neither selected face remains in the full triangulation, some reinserted large site lies within distance \(11/10\) of the cell centre.
\end{proof}

Now fix a complete labelled triangulation \(T\) and a subset \(I\) of site labels, and consider the event that \(\DT(\sigma X)=T\) and that the set of large sites equals \(I\). Call a decision cell \emph{intact} if none of its four labels belongs to \(I\). For each intact cell, look for triangles of \(T\) on three of its four labels: if one exists, take the first in label order and read off the base diagonal (and hence sign) that it selects. If two triangles of \(T\) on the same intact cell select opposite diagonals, the event is empty, because deleting \(I\) preserves empty circles and would leave both conflicting faces in the unique triangulation of the intact quadrilateral.

Let \(k_b\) count the cells of block \(b\) whose signs cannot be read from \(T\) in this way. Suppose the event determined by \((T,I)\) is nonempty. Each label in \(I\) belongs to at most two decision cells, both in its own block. Every other unread cell is intact, so by Lemma~\ref{lem:localdisks} some large site lies within distance \(11/10\) of its base centre. No point of the plane lies in three disks of radius \(11/10\) centred in \(2\mathbb Z^2\) (Fig.~\ref{fig:relative-coupling}(a)): three such centres would lie pairwise within distance \(22/10\), so all three pairwise distances in \(2\mathbb Z^2\) would equal \(2\), which is impossible because \(2\mathbb Z^2\) contains no equilateral triangle. Moreover, separated blocks lie in boxes at mutual distance at least four, so two cells reached by the same large site belong to one block. Every nonempty event therefore satisfies
\[
 \sum_b k_b\le4|I|,\qquad
 \#\{b:k_b>0\}\le2|I|.
\]
Critically, neither bound depends on the total number of blocks: \(|I|\) large sites leave at most \(2|I|\) blocks undetermined regardless of how many blocks or repetitions are present.

\begin{lemma}\label{lem:block-product}
For a fixed complete type \(T\) and fixed set \(I\), call block \(b\) \emph{determined} if \(k_b=0\), and let \(s_b(T)\) be the sign vector read from \(T\). Writing \(p_b(s)\) for the linear sign distribution of block \(b\) and \(p_b^*=\max_s p_b(s)\),
\begin{equation}\label{eq:productcomparison}
 \begin{aligned}
 &\Prob\bigl\{\DT(\sigma X)=T,\ \{i:\|Z_i\|>R\}=I,\ \mathcal N_\tau^c\bigr\}\\
 &\hspace{1em}\le(16q_R)^{|I|}
 \prod_{k_b=0}p_b(s_b(T))\prod_{k_b>0}p_b^*.
 \end{aligned}
\end{equation}
If the event on the left is nonempty, at most \(2|I|\) blocks are undetermined.
\end{lemma}
\begin{proof}
If the event is empty, \eqref{eq:productcomparison} is immediate. Otherwise, every triangle of \(T\) whose three vertices lie outside \(I\) remains a Delaunay face after deleting \(I\), because removing sites cannot violate emptiness of its circumcircle. Hence every read intact cell has the same geometric decision sign as the face read from \(T\), and by item~3 of Lemma~\ref{lem:stability} this sign equals its linear sign outside \(\mathcal N_\tau\). Because read cells are intact, their linear tests depend only on displacements outside \(I\), so the event on the left of \eqref{eq:productcomparison} is contained in the intersection of two independent events: the read cells attain their signs from \(T\), and \(\|Z_i\|>R\) for all \(i\in I\). Distinct blocks share no sites, so the probability of the first event factors over blocks. In block \(b\), the partial vector read from \(T\) has at most \(2^{k_b}\) completions and thus probability at most \(2^{k_b}p_b^*\), reducing to \(p_b(s_b(T))\) when \(k_b=0\). Multiplying by \(q_R^{|I|}\) and applying \(\sum_b k_b\le4|I|\) yields \eqref{eq:productcomparison}.
\end{proof}

We now specify the polynomial-time \emph{decoder}, which inspects only the returned complete labelled triangulation \(T\) and the known base layout (never \(I\)). A block is \emph{read} if every cell of the block (including every bridge) has at least one triangle in \(T\) on three of its four labels and all such triangles in that cell select the same base diagonal. From each read block, extract the vertex set of cycles with constant positive sign, and discard the block if any cycle is nonconstant or if the positive vertex set is not independent in \(G\). Return a largest remaining independent set (using a fixed deterministic tie-breaking rule), or \(\emptyset\) if no block qualifies. Every determined block on a nonempty event of Lemma~\ref{lem:block-product} is read by this decoder, so any correct determined block produces a maximum independent set.

To compare complete triangulations both for one candidate list and for \(r\) repeated lists, write \(\bar p(T)=p_\mu(T)\) and \(P_*=\prod_b p_b^*\) for the modal probability of the full independent linear model. Independence across blocks and the \(2^M\) sign vectors give \(P_*\ge2^{-M}\), which lets Proposition~\ref{prop:failure} compare every failed geometric type directly against a linear maximizer.

\begin{proposition}\label{prop:failure}
Suppose the configuration consists of \(r\ge1\) copies of the candidate list, and each copy contains at least one block whose every incorrect sign vector satisfies \(p_b(s)\le(1-\beta)p_b^*\) with \(0<\beta\le1/2\). Then for every complete type \(T\) on which the decoder fails to return a maximum independent set,
\begin{equation}\label{eq:failure}
 \begin{aligned}
 \bar p(T)&\le e^{64nq_R}(1-\beta)^rP_*+2L_gM\tau,\\
 \max_{T'}\bar p(T')&\ge(1-4nq_R)P_*-2L_gM\tau,
 \end{aligned}
\end{equation}
where \(M\) and \(n\) count all cells and sites in the repeated configuration.
\end{proposition}
\begin{proof}
Pick one such block in each of the \(r\) copies, and fix a complete type \(T\) on which decoding fails together with a set \(I\) for which the event in \eqref{eq:productcomparison} is nonempty. By Lemma~\ref{lem:block-product}, at most \(2|I|\) blocks are undetermined, so at least \(r-2|I|\) of the \(r\) chosen blocks are determined. Every determined block is read by the decoder and must therefore be incorrect for decoding to fail. Bounding all other block factors in \eqref{eq:productcomparison} by \(p_b^*\) and using \((1-\beta)^{-2}\le4\) (since \(\beta\le1/2\)) yields
\[
 (16q_R)^{|I|}(1-\beta)^{r-2|I|}P_*
 \le(64q_R)^{|I|}(1-\beta)^rP_*,
\]
which remains valid even when \(r-2|I|<0\) because \((1-\beta)^{r-2|I|}>1\). Summing over all subsets \(I\subseteq\{1,\ldots,n\}\) gives \((1+64q_R)^n(1-\beta)^rP_*\le e^{64nq_R}(1-\beta)^rP_*\), and adding \(\Prob(\mathcal N_\tau)\le2L_gM\tau\) proves the first bound in \eqref{eq:failure} for every complete type \(T\) on which the decoder fails, regardless of its hull.

For the lower bound, fix a full linear sign vector \(s^*\) attaining \(P_*\). For any site \(i\), deleting the at most two cell tests that involve \(Z_i\) leaves an event independent of \(\|Z_i\|>R\) whose probability sums at most four full sign vectors, each bounded by \(P_*\). Thus the joint probability that \(s^*\) occurs and site \(i\) is large is at most \(4q_RP_*\). A union bound over all \(n\) sites and \(\mathcal N_\tau\) shows that \(s^*\) occurs with all sites nonlarge and outside \(\mathcal N_\tau\) with probability at least \((1-4nq_R)P_*-2L_gM\tau\). On this event, Lemma~\ref{lem:stability} and the completion count \eqref{eq:euler} ensure \(\DT(\sigma X)=T(s^*)\), proving the second bound in \eqref{eq:failure}.
\end{proof}

\subsection{Candidate blocks, precision and proof of Theorem~\ref{thm:main}}\label{sec:transfer-candidates}
It remains to build, in polynomial time and without evaluating \(\A\), \(w_B\), \(w_C\) or \(\mathcal M\), a list of candidate blocks guaranteed to contain at least one suitable block satisfying Lemma~\ref{lem:block} with a computable relative gap \(\beta\). To guarantee \(\widehat w(t,t')\le c_1\A(t)\), we seek dyadic scales \(t=2^{-j}\) and \(t'=2^{-j'}\) at the first indices where \(\A\) drops below two thresholds \(a\) and \(c_1a/2\) (Fig.~\ref{fig:relative-coupling}(b)). Listing every pair up to a computable ceiling \(j_{\max}\) supplied by \eqref{eq:Aupper}, paired with a geometric grid of fields whose step ratio \(5/4\) keeps the bias step below the width ratio \(13/7\) of \eqref{eq:field}, guarantees that one listed block is suitable.

Recall the fixed rational constant \(c_1\) from Lemma~\ref{lem:block} and the integer \(C_A\) from \eqref{eq:Aupper}. For a fixed integer \(K_\nu\ge3\) depending only on \(\nu\) and \(C_0\) (chosen sufficiently large below), set
\begin{equation}\label{eq:parameters}
 \ell=\lceil\log_2(m+1)\rceil,\qquad
 a=2^{-K_\nu\ell},\qquad
 j_{\max}=\min\{j\ge3:\,C_Aj^{-\gamma}\le c_1a/2\},
\end{equation}
where the minimum runs over integers. Choose \(K_\nu\) large enough that \(a<a_0/4\). Then \eqref{eq:AB} gives \(\A(1/4)\ge a_0/4>a\), so the first crossing of \(a\) occurs at an index \(j\ge3\), while \eqref{eq:Aupper} guarantees that both thresholds are crossed by \(j_{\max}\).

Let \(j\) and \(j'\) be the smallest integers with \(\A(2^{-j})\le a\) and \(\A(2^{-j'})\le c_1a/2\), respectively (Fig.~\ref{fig:relative-coupling}(b)). Because \(\A(2t)\le2\A(t)\) by \eqref{eq:AB}, these first crossings satisfy
\[
 a/2<\A(t)\le a,\qquad
 c_1a/4<\A(t')\le c_1a/2,\qquad
 t=2^{-j},\quad t'=2^{-j'}.
\]
Thus \(3\le j\le j'\le j_{\max}\) and \(\widehat w(t,t')\le\A(t')\le c_1\A(t)\). Moreover,
\[
 m^2\A(t)<2^{-\ell(K_\nu-2)}.
\]
Choosing one fixed large integer \(K_\nu\) makes \(m^2\A(t)\) smaller than every smallness threshold required by Lemma~\ref{lem:block}, uniformly in \(\ell\ge1\), and also puts \(t\le a/a_0\) below \(t_1\) of Lemma~\ref{lem:weights} and below the initial threshold \(2\|g\|_\infty t<4c_\nu^2a_0^2\) for the field grid below. Each requirement on \(K_\nu\) below has this exact form: a fixed multiple of \(2^{-K_\nu}\) must lie below a positive constant depending only on \(\nu\) and \(C_0\). Because \(\gamma>0\) is a fixed rational, \(j_{\max}=O_\nu(m^{K_\nu/\gamma})\) is polynomial in \(m\) and is computed by raising the defining inequality in \eqref{eq:parameters} to the denominator of \(\gamma\) and clearing denominators.

List all pairs \(3\le j\le j'\le j_{\max}\). For each pair \((t,t')=(2^{-j},2^{-j'})\), list the geometric field grid
\begin{equation}\label{eq:fieldgrid}
 \lambda_k=t^2t'(5/4)^k,\qquad k=0,1,\ldots,
\end{equation}
stopping at the first value at least \(1/4\), and assign field \(-\lambda_k\) to the designated cell of every cycle (zero elsewhere).

We verify that at the crossing pair \((j,j')\) located above, at least one grid value \(\lambda_k\) satisfies \eqref{eq:field}. At \(k=0\), Lemma~\ref{lem:weights} and \eqref{eq:AB} give \(W\ge4c_\nu^2\A\A'\ge4c_\nu^2a_0^2tt'\), so \(\mathcal M(\lambda_0)/W\le2\|g\|_\infty t/(4c_\nu^2a_0^2)<1<7/5\) by the choice of \(K_\nu\). At the final grid value (at least \(1/4\)), monotonicity gives \(\mathcal M(\lambda_{\text{last}})\ge\mathcal M(\min(y_1,1/4))\), which exceeds \(13W/5\) for large \(K_\nu\) because \(W\le4C_\nu^2\A'\le2C_\nu^2c_1a\). Let \(k\ge1\) be the first index with \(\mathcal M(\lambda_k)\ge7W/5\). Then \(\mathcal M(\lambda_{k-1})<7W/5<\mathcal M(4y_1/5)\) (again for large \(K_\nu\)), so \(\lambda_{k-1}\le4y_1/5\). By \eqref{eq:biasbounds},
\[
 \frac75W\le\mathcal M(\lambda_k)\le\frac32\mathcal M(\lambda_{k-1})<\frac{21}{10}W<\frac{13}{5}W,
\]
so the chosen field satisfies \eqref{eq:field}. Each grid has \(O(j+j'+1)\) values, all below \(5/16<1/2\), with bit length
\begin{equation}\label{eq:fieldbits}
 O(j+j'+1).
\end{equation}
The full list has \(B=O(j_{\max}^3)\) candidate blocks, polynomial in \(N\), and its construction evaluates neither \(\A\) nor any weight.

Call a listed block \emph{suitable} if \(a/2<\A(t)\le a\), \(c_1a/4<\A(t')\le c_1a/2\) and its field satisfies \eqref{eq:field}. By the argument above, the list contains at least one suitable block. On any suitable block, \(\A\A'>c_1a^2/8\), so Lemma~\ref{lem:block} guarantees that every incorrect sign vector \(s\) satisfies
\begin{equation}\label{eq:relativegap}
 p_b(s)\le(1-d_\nu\A\A')p_b^*\le(1-\beta)p_b^*,\qquad
 \beta=\frac{d_\nu c_1a^2}{8}.
\end{equation}
Because \(d_\nu\) and \(c_1\) are fixed rationals, \(\beta\in(0,1/2]\) is rational and computable, with \(\beta^{-1}=O_\nu(a^{-2})\) polynomial in \(m\).

Whenever \(\E U^2<\infty\), a single explicit block already satisfies Lemma~\ref{lem:block}, extending the single block of the Gaussian construction of Section~\ref{sec:socg-gaussian} to every finite-variance law covered by Theorem~\ref{thm:main}, although the reduction above still uses the candidate list. In this regime, \(\widehat w(t,t)\le t^2\E U^2\) is already quadratic while \(\A(t)\) is linear in \(t\), so a single scale \(t'=t\) suffices and the leading ratio of the field bias to the bridge weights is determined in advance by \(J\) and \(\E U^2\).

\begin{proposition}\label{prop:oneblock}
Suppose \(\E U^2<\infty\), and fix a rational \(\kappa_0\) with \(\frac{14}{5}J\,\E U^2<\kappa_0<\frac{26}{5}J\,\E U^2\). There is an integer \(K\), depending only on \(\nu\), \(C_0\) and \(\kappa_0\), such that for every \(m\) and \(\ell=\lceil\log_2(m+1)\rceil\) the single block with
\[
 t=t'=2^{-K\ell},\qquad \lambda=\kappa_0t^2
\]
satisfies the hypotheses of Lemma~\ref{lem:block}. For this block, \eqref{eq:relativegap} holds with \(\beta=d_\nu\underline a_0^2\,2^{-2K\ell}\) for any fixed rational lower bound \(0<\underline a_0\le a_0\). For the standard Gaussian, \(J\,\E U^2=1/(2\sqrt\pi)\), and one may take \(\kappa_0=1\), that is, \(\lambda=t^2\).
\end{proposition}
\begin{proof}
We check the three conditions of Lemma~\ref{lem:block} as \(t'=t\to0^+\). First, \(\widehat w(t,t)\le t^2\E U^2\le c_1a_0t\le c_1\A(t)\) once \(t\le c_1a_0/\E U^2\). Second, \(m^2\A(t)\le m^2t\,\E|U|<2^{(2-K)\ell}\E|U|\), which is arbitrarily small for large \(K\). Third, by Lemma~\ref{lem:near}, the densities of \(W_1,W_2,W_3\) in \eqref{eq:bridgeweights} converge uniformly to \(h\), so \(L_{W_i}(cy)/t\to Jy\) for every \(y\ge0\) whenever \(c/t\to1\), while \(L_{W_i}(cy)/t\le C_\nu y\) by \eqref{eq:response}. Since \(c(t)/t\), \(b'/t\) and \(c'/t\) all tend to \(1\) and \(\E Y^2=\E U^2<\infty\), dominated convergence gives \(w_B/t^2\to4J^2\E U^2\) and \(w_C/t^2\to4J^2\E U^2\). Likewise \(\mathcal M(\kappa_0t^2)/t^2\to2J\kappa_0\). Hence \(\mathcal M(\lambda)/W\to\kappa_0/(2J\,\E U^2)\), strictly inside \((7/5,13/5)\), so \eqref{eq:field} holds for small \(t\). Choosing \(K\) large enough puts \(t\le2^{-K}\) below \(t_1\) and all three thresholds (and keeps \(\lambda<1/2\)). The relative gap follows from Lemma~\ref{lem:block} and \(\A(t)^2\ge a_0^2t^2\ge\underline a_0^2t^2\). For the standard Gaussian, \(J=1/(2\sqrt\pi)\) and \(\E U^2=1\), and the allowed interval for \(\kappa_0\) contains \([0.79,1.46]\).
\end{proof}

We now set the precision parameters for Algorithm~\ref{alg:reduction} and complete the proof of Theorem~\ref{thm:main}. Fix a rational density bound \(L_g\ge\max(1,\|g\|_\infty)\) and choose
\begin{equation}\label{eq:trunc}
 s_R=\min\{s\in\mathbb Z_{\ge1}:\,C_Ls^{-\gamma}\le\beta/(256n)\},\qquad
 R=2^{s_R},\qquad
 \tau=\frac{\beta\,2^{-M}}{16L_gM}.
\end{equation}
By \eqref{eq:L}, \(q_R=T_\nu(R)\le\beta/(256n)\). Computing \(s_R\) by integer power comparisons (as for \(j_{\max}\)) gives \(s_R=O_\nu((n/\beta)^{1/\gamma})\), polynomial in \(N\), and \(\log_2(1/\tau)=M+O_\nu(\log M+\log(1/\beta))\), also polynomial in \(N\). Because the binary encoding of \(R=2^{s_R}\) has length \(s_R+1=O(s_R)\) (the radius itself is never enumerated) and the candidate field bits \eqref{eq:fieldbits} are polynomial, Lemma~\ref{lem:stability} produces the generic rational centres \(\mu_1,\ldots,\mu_n\) in polynomial time with polynomial bit length. The fixed rational constants \(C_L,C_A,K_\nu,c_1,L_g,d_\nu\) and \(\gamma\) are fixed in the reduction for the law \(\nu\), so the algorithm evaluates no density, probability or quantile.

\begin{proof}[Proof of Theorem~\ref{thm:main}]
Maximum independent set is NP-hard on \(2\)-connected planar cubic graphs~\cite[Theorem~4.1(a)]{mohar_face}. Run Algorithm~\ref{alg:reduction} on such a graph with one copy (\(r=1\)) of the candidate list from \eqref{eq:parameters}--\eqref{eq:fieldgrid}, the gap \(\beta\) from \eqref{eq:relativegap}, and the precision parameters \eqref{eq:trunc}. By \eqref{eq:trunc} and \(P_*\ge2^{-M}\), we have \(64nq_R\le\beta/4\) and \(2L_gM\tau\le\beta P_*/8\). For every complete labelled type \(T\) on which the decoder fails, Proposition~\ref{prop:failure} therefore gives
\[
 \bar p(T)\le\bigl(e^{\beta/4}(1-\beta)+\beta/8\bigr)P_*
 \le\Bigl(1-\frac{5\beta}{8}\Bigr)P_*.
\]
Indeed, \(e^{\beta/4}\le(1-\beta/4)^{-1}\) for \(\beta\le1/2\), which implies \(e^{\beta/4}(1-\beta)\le1-3\beta/4\). Meanwhile, the lower bound in Proposition~\ref{prop:failure} gives
\[
 \max_{T'}\bar p(T')\ge\Bigl(1-\frac{\beta}{64}-\frac{\beta}{8}\Bigr)P_*
 =\Bigl(1-\frac{9\beta}{64}\Bigr)P_*,
\]
strictly larger than \((1-5\beta/8)P_*\). Every global geometric mode of \(p_\mu\), including any tied mode and regardless of its hull, therefore decodes to a maximum independent set of \(G\). Both construction and decoding run in polynomial time and use a single modal oracle call.
\end{proof}

\section{Hardness of relative Delaunay approximation in the plane and its randomized consequence}\label{sec:amplification}
The same block construction also excludes constant-factor relative approximation. Analogous hardness is known for finding a most probable assignment to a specified set of variables in a belief network given the observed evidence~\cite{abdelbar}. Park and Darwiche multiply likelihood ratios across copies in that setting~\cite[Theorem~8 and Appendix~A]{park_darwiche}. Every suitable block incurs a relative probability loss \(\beta\) on every incorrect sign vector, with \(\beta^{-1}\) polynomial in the graph size. As in the product argument for discrete convex hulls~\cite[Lemma~7]{suri}, placing \(r\) independent copies in one configuration multiplies those losses across copies, and Proposition~\ref{prop:failure} retains this multiplication for complete geometric triangulations because \(k\) large sites can leave at most \(2k\) blocks undetermined. We derive both fixed and subexponentially small factors in the plane. Section~\ref{sec:line} treats Gaussian sites on the line.

\begin{theorem}\label{thm:factor}
Fix a bounded rotationally invariant planar density satisfying \eqref{eq:L}. For every fixed \(c\in(0,1)\), it is NP-hard to return a complete labelled type \(T\) satisfying
\[
 p_\mu(T)\ge c\max_{T'}p_\mu(T').
\]
The reduction uses one approximate-solver call and takes polynomial time. The sites of the constructed instance have rational generic centres \(\mu_i\) and positions \(X_i=\mu_i+Z_i\), where the displacements \(Z_i\) are independent draws from the fixed law. Consequently, unless \(P=NP\), no polynomial-time algorithm achieves any fixed factor \(c\in(0,1)\). In particular there is no polynomial-time relative approximation scheme. A randomized polynomial algorithm meeting this guarantee with probability at least \(2/3\) would imply \(NP=RP\).
\end{theorem}

Section~\ref{sec:amp-replication} proves Theorem~\ref{thm:factor}. Immediately after the proof, we show that the same replication argument extends to factors \(2^{-n^{1-\epsilon}}\) that decrease with the number of sites.

\subsection{The gap and the number of repetitions}\label{sec:amp-gap}
To amplify the relative gap of a single suitable block to an arbitrary target ratio \(c\in(0,1)\), we need both the block list size \(B\) and the reciprocal gap \(\beta^{-1}\) to grow polynomially in \(m\). For a general law under \eqref{eq:L}, the gap \(\beta=d_\nu c_1a^2/8\) from \eqref{eq:relativegap} already cancels the exponential factor \(2^{-m}\) when dividing the absolute loss of Lemma~\ref{lem:block} by the modal probability. With \(a=2^{-K_\nu\ell}\), we have
\[
 B=O_\nu(m^{3K_\nu/\gamma}),\qquad
 \beta^{-1}=O_\nu(m^{2K_\nu}),
\]
where both exponents depend only on \(\nu\). Repeating the entire candidate list \(r\) times lets the reduction amplify the gap without identifying which candidate block is suitable: whenever the decoder fails, every block that it reads is incorrect, so each copy contributes a factor \((1-\beta)\) unless a large site renders its suitable block undetermined. The standard Gaussian specialization, including repetition, is proved directly in Section~\ref{sec:socg-gaussian}.

\subsection{Amplification to a decaying factor and the randomized consequence}\label{sec:amp-replication}
For a general law satisfying \eqref{eq:L}, we choose enough copies of the candidate list to push the linear failed-output bound below \(c/8\) of \(P_*\), and then apply Proposition~\ref{prop:failure} with an inverse-polynomial tail bound per site. Only the sign-error tolerance \(\tau\) is exponentially small, affecting coordinate bit lengths by a polynomial number of bits while leaving the truncation radius \(R\) polynomial in bit length.

\begin{figure}[!htbp]
\centering
\includegraphics[width=\linewidth]{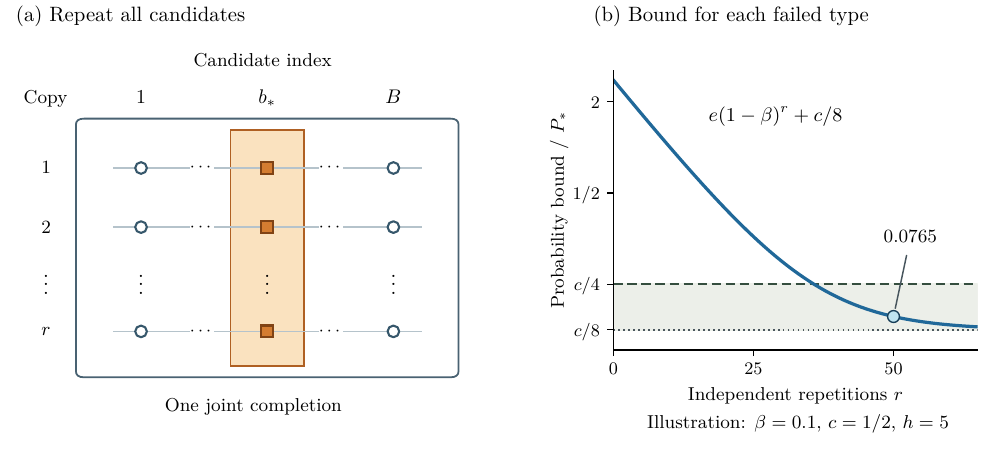}
\caption{Independent repetitions bound the probability of each failed type. In (a), every row repeats the full candidate list. The orange column marks a suitable candidate \(b_*\), and the outer frame denotes the joint completion. In (b), the curve bounds the probability of any one failed type divided by \(P_*\). For illustration, a hypothetical gap \(\beta=0.1\) and \(c=1/2\) give \(h=5\) and \(r=50\), at which the bound is \(e\,0.9^{50}+c/8\approx0.0765<c/4\).}\label{fig:amplification}
\end{figure}

\begin{proof}[Proof of Theorem~\ref{thm:factor}]
It suffices to treat rational \(c\in(0,1)\), because attaining a larger factor also attains \(c\). Build the candidate list of Section~\ref{sec:transfer-candidates}, set \(h=\lceil\log_2(1/c)\rceil+4\), and place
\[
 r=\lceil h/\beta\rceil
\]
independent copies of the entire list in separated boxes at mutual distance at least four (Fig.~\ref{fig:amplification}(a)). Complete their union once by Lemma~\ref{lem:completion}, obtaining \(M'=rBm\) cells and \(n'\le12M'+4\) sites. Each of the \(r\) copies contains a suitable block satisfying \eqref{eq:relativegap}. Apply Lemma~\ref{lem:stability} with
\[
 s_R=\min\{s\in\mathbb Z_{\ge1}:\,C_Ls^{-\gamma}\le1/(64n')\},\qquad
 R=2^{s_R},\qquad
 \tau=\frac{c\,2^{-M'}}{16L_gM'}.
\]
Then \(64n'q_R\le1\) by \eqref{eq:L}, and \(2L_gM'\tau\le cP_*/8\) because \(P_*\ge2^{-M'}\). Proposition~\ref{prop:failure} bounds the probability of every complete geometric type \(T\) on which the decoder fails by
\begin{equation}\label{eq:amplified}
 \bar p(T)\le\bigl(e(1-\beta)^r+c/8\bigr)P_*
 \le\bigl(e^{1-h}+c/8\bigr)P_*<\frac{c}{4}P_*,
\end{equation}
since \(h\ge\ln(1/c)+4\) implies \(e^{1-h}\le ce^{-3}<c/8\). Fig.~\ref{fig:amplification}(b) plots the first bound in \eqref{eq:amplified}. Meanwhile, the lower bound of Proposition~\ref{prop:failure} gives
\[
 \bar p_*=\max_{T'}\bar p(T')\ge\Bigl(1-\frac{1}{16}-\frac{c}{8}\Bigr)P_*\ge\frac{13}{16}P_*.
\]
Hence every complete labelled type \(T\) satisfying \(\bar p(T)\ge c\bar p_*\) has probability above \((c/4)P_*\) and therefore decodes to a maximum independent set, regardless of its hull or tie breaking.

All sizes and bit lengths remain polynomial. From Section~\ref{sec:amp-gap}, \(n'=O_\nu(hm^{1+3K_\nu/\gamma+2K_\nu})\) with \(m=O(N^3)\). Both \(s_R=O_\nu((n')^{1/\gamma})\) and the binary length of \(\tau\), which is \(M'+O_\nu(\log M'+\log(1/c))\), are polynomial in \(N\), so Lemma~\ref{lem:stability} produces the rational centres \(\mu_i\) in polynomial time with polynomial bit length. One call to the approximate solver would therefore solve maximum independent set on planar cubic graphs~\cite[Theorem~4.1(a)]{mohar_face}. In particular, a relative PTAS invoked at factor \(c=1/2\) would imply \(P=NP\).

For the randomized claim, given a decision instance \((G,k_0)\) of maximum independent set, run the solver once on the constructed instance and reject malformed outputs. Apply the decoder to any well-formed output (verifying that the returned set is independent) and accept \((G,k_0)\) exactly when the returned independent set has size at least \(k_0\). A no-instance is never accepted, while every yes-instance is accepted whenever the solver meets the approximation guarantee, which happens with probability at least \(2/3\). Thus the NP-complete decision problem lies in \(RP\subseteq NP\), giving \(NP=RP\).
\end{proof}

\medskip\noindent\textit{Factors that decrease with the number of sites.}
The same reduction remains polynomial whenever \(h\) is polynomial in the vertex count \(N\) and the rational factor \(c\) has polynomial bit length. Fix an integer \(k\ge1\), take \(c=2^{-N^k}\) and \(h=N^k+4\), and keep the exact same choices of \(R\) and \(\tau\). For a fixed integer \(a_\nu>3+9K_\nu/\gamma+6K_\nu\) depending only on \(\nu\), the completed instance has \(n'\le N^{k+a_\nu}\) sites for all sufficiently large \(N\) (smaller graphs are solved directly). Given any fixed \(\epsilon\in(0,1)\), choose \(k\ge a_\nu(1-\epsilon)/\epsilon\). Then
\[
 (n')^{1-\epsilon}\le N^{(k+a_\nu)(1-\epsilon)}\le N^k,
 \qquad 2^{-(n')^{1-\epsilon}}\ge c.
\]
Under the hypotheses of Theorem~\ref{thm:factor} and for every fixed \(\epsilon\in(0,1)\), it is therefore NP-hard to return a complete type of probability at least \(2^{-n^{1-\epsilon}}\) times the maximum, where \(n\) is the number of sites. The precision stays polynomial because \(s_R\) depends only on \(n'\), while the additional \(N^k\) bits of \(c\) enter only through \(\tau\).

\section{Hardness for elliptical laws with bounded projection densities}\label{sec:elliptical}
Because the Delaunay triangulation is not preserved by general linear maps, a fixed anisotropic noise law cannot be reduced to an isotropic one by a linear change of coordinates. Both planar hardness theorems nevertheless extend to every fixed invertible linear image of a rotation-invariant law satisfying \eqref{eq:L} with a bounded one-dimensional projection density. The key strategy is to keep the Euclidean base construction of Section~\ref{sec:geometry} unchanged and absorb the linear transformation into the coefficients of the linearized tests: dividing each cell's test by its positive directional scale at \(\mathbf D_q\) recovers a linear-test model driven by independent copies of the common symmetric unit projection \(U\) of \(Y\), with coefficients still lying in the fixed ranges \eqref{eq:ranges} used by Section~\ref{sec:heavy} and Lemma~\ref{lem:expansion}. The same proof also weakens the density hypothesis, showing that even certain singular planar laws are covered whenever their projections have a bounded density.

\begin{figure}[!htbp]
\centering
\resizebox{\linewidth}{!}{
\begin{tikzpicture}[x=1cm,y=1cm,font=\small,text=black,>=Stealth,line cap=round,line join=round]
\definecolor{ellblue}{HTML}{2875B7}
\definecolor{ellorange}{HTML}{DB7627}
\useasboundingbox (0,-.45) rectangle (15.5,4.65);
\node[anchor=west] at (0,4.4) {(a) Keep the Euclidean cell};
\node[anchor=west] at (5.35,4.4) {(b) Measure a projection};
\node[anchor=west] at (10.75,4.4) {(c) Normalize the test};
\begin{scope}[shift={(2.04,2.45)}]
\draw[black!55,line width=.7pt] (0,0) circle (1.12);
\fill[black] (0,0) circle (1.2pt);
\node[below left=2pt,font=\footnotesize] at (0,0) {$o_q$};
\draw[black!25,line width=.5pt] (0,0)--(1.12,0) (0,0)--(22.62:1.12);
\draw[ellblue,->,line width=.9pt] (1.12,0)--(1.93,0);
\draw[ellorange,->,line width=.9pt] (22.62:1.12)--(22.62:1.93);
\foreach \ang in {0,22.62,90,270}{\fill[black] (\ang:1.12) circle (1.8pt);}
\node[below left=2.5pt] at (1.12,0) {$A$};
\node[inner sep=0pt] at (.56,.66) {$D$};
\node[above=3pt] at (0,1.12) {$C$};
\node[below=3pt] at (0,-1.12) {$B$};
\node[below=3pt] at (1.93,0) {$n_A$};
\node[above=3pt] at (22.62:1.93) {$n_D$};
\draw[black!55,->,line width=.5pt] (.55,0) arc (0:22.62:.55);
\fill[ellorange!27] (0,0)--(.55,0) arc (0:22.62:.55)--cycle;
\end{scope}
\node[font=\footnotesize] at (2.28,.35) {$\angle(n_A,n_D)=2\arctan t$};
\begin{scope}[shift={(7.73,2.28)},scale=.95]
\fill[ellblue!10] plot[domain=0:360,samples=100,smooth,variable=\a] ({1.4*cos(\a)+.8*sin(\a)},{.2*cos(\a)+.7*sin(\a)})--cycle;
\draw[ellblue,line width=.9pt] plot[domain=0:360,samples=100,smooth,variable=\a] ({1.4*cos(\a)+.8*sin(\a)},{.2*cos(\a)+.7*sin(\a)});
\draw[black!55,dashed,line width=.55pt] (-1.61245,-1.12)--(-1.61245,1.37) (1.61245,-1.12)--(1.61245,1.37);
\draw[->,black!70,line width=.55pt] (-1.98,0)--(2.05,0) node[right,text=black] {$n$};
\fill[black] (0,0) circle (1.5pt);
\node[font=\footnotesize,inner sep=0pt] at (.04,-.23) {$0$};
\draw[ellblue,line width=1.0pt] (1.61245,0)--(1.61245,.52095);
\draw[ellblue,line width=.95pt] (1.61245,0)++(-.07,0)--++(0,.07)--++(.07,0);
\fill[black] (1.61245,.52095) circle (1.8pt);
\fill[ellblue] (-1.61245,-.52095) circle (1.8pt);
\draw[<->,black!70,line width=.6pt] (0,1.18)--(1.61245,1.18);
\node[above=3pt,font=\footnotesize] at (.806,1.18) {$s(n)$};
\draw[black!35,line width=.4pt] (0,.15)--(0,1.3);
\node[font=\footnotesize,anchor=east] at (-.09,.79) {$M(S^1)$};
\end{scope}
\node[font=\footnotesize] at (7.79,.35) {$s(n)=\|M^{\mathsf T}n\|=s(-n)$};
\node[align=center] at (13.02,3.36) {$\langle n,MY\rangle\overset{d}{=}s(n)U$};
\node[align=center,font=\footnotesize] at (13.02,2.60) {Divide the test by $s(n_D)>0$.};
\draw[black!45,line width=.5pt] (11.14,2.10)--(14.90,2.10);
\node[anchor=west,font=\footnotesize] at (11.20,1.71) {Site};
\node[anchor=east,font=\footnotesize] at (14.81,1.71) {Coefficient};
\draw[black!30,line width=.45pt] (11.14,1.43)--(14.90,1.43);
\node[anchor=west] at (11.20,1.02) {$A$};
\node[anchor=east] at (14.81,1.02) {$a_q\,s(n_A)/s(n_D)$};
\node[anchor=west] at (11.20,.37) {$D$};
\node[anchor=east] at (14.81,.37) {$-1$};
\end{tikzpicture}}
\caption{Directional scales in an elliptic displacement law. In (a), the Euclidean cell fixes the normals at \(A\) and \(D\), separated by \(2\arctan t\). In (b), the ellipse \(M(S^1)\) has projected half-width \(s(n)=\|M^{\mathsf T}n\|\). The same scale applies to \(-n\), including for noise with unbounded support. In (c), dividing the linear test by \(s(n_D)>0\) preserves its sign and gives the two displayed coefficients.}\label{fig:elliptic-projections}
\end{figure}
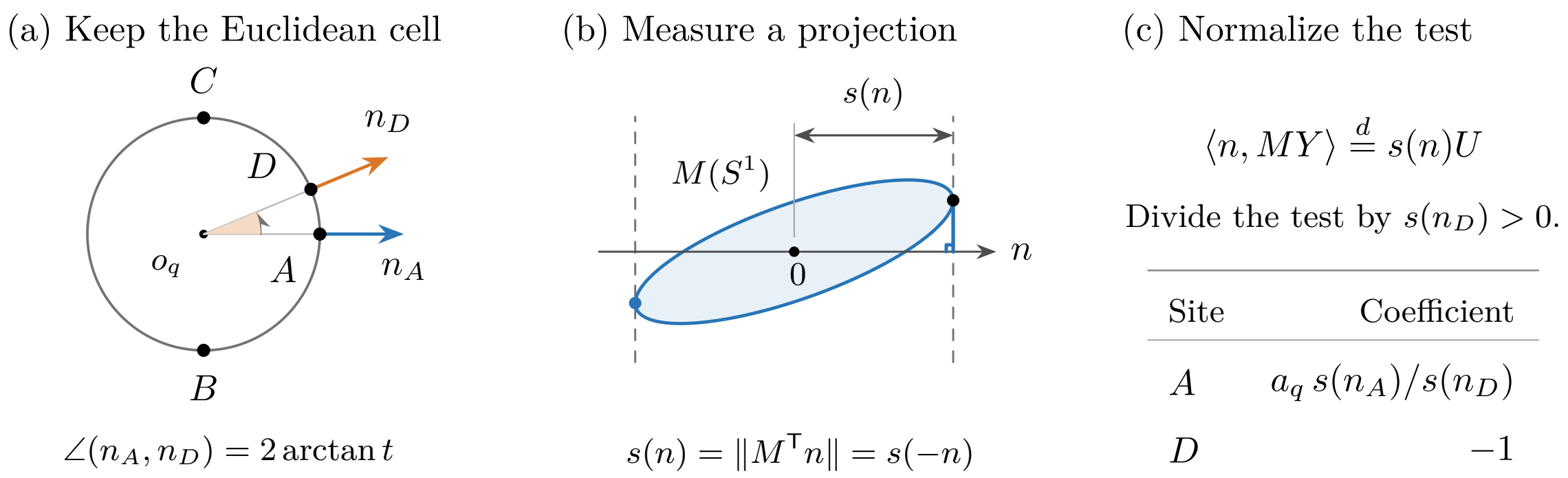

Let \(Y\) be a random vector in \(\R^2\) whose law is rotation-invariant, satisfies \eqref{eq:L}, and has a unit projection \(U\) with bounded density \(g\). Fix a real invertible \(2\times2\) matrix \(M\) with singular values \(0<s_{\min}\le s_{\max}\) and condition number \(\varkappa=s_{\max}/s_{\min}\), and let \(\nu\) be the law of \(MY\). For any unit vector \(n\in\R^2\),
\begin{equation}\label{eq:sn}
 \langle n,MY\rangle=\langle M^{\mathsf T}n,Y\rangle\overset{d}{=}s(n)\,U,
 \qquad s(n)=\|M^{\mathsf T}n\|,
\end{equation}
where the directional scale \(s(n)\) satisfies \(s_{\min}\le s(n)\le s_{\max}\), \(s(-n)=s(n)\), and \(|s(n)-s(n')|\le s_{\max}\|n-n'\|\). Fig.~\ref{fig:elliptic-projections} depicts this directional scaling while leaving the Euclidean cell untouched. Every Gaussian law \(N_2(0,\Sigma)\) with fixed positive definite \(\Sigma\) has this form, as do all bounded radial densities of Theorem~\ref{thm:main} (by \eqref{eq:greg}) and even laws with unbounded radial density such as \(\varphi(r)=\frac{3}{4\pi}r^{-1/2}\mathbf 1_{\{r\le1\}}\), whose projection density is bounded.

\begin{theorem}\label{thm:elliptical}
For every law \(\nu\) of this form, the conclusions of Theorems~\ref{thm:main} and~\ref{thm:factor} hold, including the factor \(2^{-n^{1-\epsilon}}\) of Section~\ref{sec:amplification}. The matrix \(M\) and the law of \(Y\) are fixed. Only the rational generic centres form the input.
\end{theorem}

\begin{proof}
We keep the Euclidean subdivision of Section~\ref{sec:geometry} unchanged. The argument proceeds in four steps: verifying almost-sure general position, checking lattice homogeneity under the symmetries preserved by \(\nu\), rescaling each cell's linear test to match \eqref{eq:ranges} with \(C_0=6\varkappa\), and translating the remaining constants through Table~\ref{tab:elliptic-map}.

\emph{General position.} Write \(Y=R(\cos\Theta,\sin\Theta)\) with \(\Theta\) uniform on \([0,2\pi)\) and independent of \(R\). A bounded projection density \(g\) excludes an atom of \(R\) at zero, and also at any \(r>0\): an atom of mass \(p>0\) at \(r>0\) would place mass of order \(p\sqrt{h/r}\) into the projection interval \([r-h,r]\) as \(h\to0^+\) (from the angular arc width near \(r\)), violating the upper bound \(h\|g\|_\infty\). Thus \(R\) is atomless. The preimage of any line or circle under an invertible affine map is a line or ellipse, which meets a circle of radius \(R\) in finitely many points except at most at one single radius where the curves coincide. Conditioning on \(R\) therefore gives probability zero in either case, and conditioning successively on the independent sites rules out coincidences, collinear triples and cocircular quadruples almost surely.

\emph{Homogeneity.} The elliptic law \(\nu\) is invariant under point reflection \(z\mapsto-z\), and translations permute its translates. By the lattice construction of Section~\ref{sec:geometry} and Section~\ref{app:contacts}, all bridges are translates of one another, all lower bridge endpoints are translates of one another, every upper bridge endpoint is a point reflection of a lower endpoint, and all designated cells are translates of one another.

\emph{Normalized linear tests.} The first-variation identity \eqref{eq:variation} is purely geometric and holds for arbitrary displacement vectors. At a site with outward unit normal \(n\), the displacement \(MY_i\) contributes \(\langle n,MY_i\rangle=s(n)\langle M^{\mathsf T}n/s(n),Y_i\rangle\), where the second factor is a unit projection of \(Y_i\) distributed as \(U\). Distinct sites yield independent projections. A site shared by two cells carries opposite outward normals in them, hence the same scale \(s(-n)=s(n)\) and opposite signs, preserving the sign couplings of Section~\ref{sec:geometry}. Writing \(s(\mathbf P)\) for \eqref{eq:sn} at the outward normal of vertex \(\mathbf P\) in cell \(q\), dividing \(\Lambda_q\) by \(s(\mathbf D_q)>0\) preserves \(\sgn\Lambda_q\) and produces linear tests of the exact form \eqref{eq:test}--\eqref{eq:bridgetest} in independent copies of \(U\), with normalized coefficients
\[
 \tilde a_q=a_q\frac{s(\mathbf A_q)}{s(\mathbf D_q)},\qquad
 \tilde b_q=b_q\frac{s(\mathbf B_q)}{s(\mathbf D_q)},\qquad
 \tilde c_q=c_q\frac{s(\mathbf C_q)}{s(\mathbf D_q)},
\]
analogous bridge coefficients \((\tilde a',\tilde b',\tilde c')\), and normalized field \(\theta_q/s(\mathbf D_q)\).

Because the angle between the normals at \(\mathbf A_q\) and \(\mathbf D_q\) is \(2\arctan t\le2t\), the Lipschitz bound on \(s\) gives \(|s(\mathbf D_q)/s(\mathbf A_q)-1|\le2\varkappa t\le1/4\) whenever \(t\le1/(8\varkappa)\). Combining this with the isotropic ranges \eqref{eq:ranges} (\(C_0=2\)) and \(1/\varkappa\le s(\mathbf P)/s(\mathbf D_q)\le\varkappa\) yields
\[
 |\tilde a_q-1|\le\frac{2t+2\varkappa t}{3/4}\le6\varkappa t,\qquad
 \frac{t}{6\varkappa}\le\tilde b_q\le6\varkappa t,\qquad
 0\le\tilde c_q\le6\varkappa t,
\]
with \(\tilde c_q\ge t/(6\varkappa)\) at bridge endpoints, and the same bounds with \(t'\) on every bridge. Thus the normalized tests satisfy \eqref{eq:ranges} with \(C_0=6\varkappa\ge6\). Because translations preserve all normals and point reflections reverse them while keeping \(s(-n)=s(n)\), homogeneity ensures that the normalized coefficients are identical across all bridges, across all bridge endpoints, and across all designated cells, with a single common scale \(s_{\mathrm d}=s(\mathbf D_d)\) at every designated cell \(d\).

\begin{table}[ht]
\caption{Correspondence between the isotropic reduction (Sections~\ref{sec:heavy}--\ref{sec:amplification}) and the elliptic reduction for \(\nu=\mathrm{law}(MY)\). Here \(s_{\min}\le s_{\max}\) are the singular values of \(M\), \(\varkappa=s_{\max}/s_{\min}\), \(g\) is the density of a unit projection of \(Y\), and \(s_{\mathrm d}=s(\mathbf D_d)\) is the common scale at designated cells.}\label{tab:elliptic-map}
\centering\small
\renewcommand{\arraystretch}{1.16}
\begin{tabular}{>{\raggedright\arraybackslash}p{0.26\linewidth}>{\raggedright\arraybackslash}p{0.30\linewidth}>{\raggedright\arraybackslash}p{0.35\linewidth}}
\toprule
Component & Isotropic case (\(M=I_2\)) & Elliptic case (\(MY\))\\
\midrule
Range constant and scale ceiling & \(C_0=2\), \(t\le t_1\) & \(C_0=6\varkappa\), \(t\le\min(t_1,1/(8\varkappa))\)\\
Block constants & \(t_1,c_\nu,C_\nu,c_1,d_\nu\) for \((g,C_0=2)\) & \(t_1,c_\nu,C_\nu,c_1,d_\nu\) for \((g,C_0=6\varkappa)\)\\
Field window \eqref{eq:field} & \(\tfrac75W\le\mathcal M(\lambda)\le\tfrac{13}{5}W\) & \(\tfrac75W\le\mathcal M(\lambda/s_{\mathrm d})\le\tfrac{13}{5}W\)\\
Scale functions and tail \eqref{eq:L} & \(\A,\widehat w,a_0\) of \(U\), constants \((C_L,\gamma)\) & \(\A,\widehat w,a_0\) of \(U\), same \(\gamma\), enlarged \(C_L\) via \(\|MY\|\le s_{\max}\|Y\|\)\\
Projection density bound \(L_g\) & \(L_g\ge\max(1,\|g\|_\infty)\) & \(L_g\ge\max(1,\|g\|_\infty/s_{\min})\)\\
Initial threshold for the field grid & \(2\|g\|_\infty t<4c_\nu^2a_0^2\) & \(2\|g\|_\infty t<4c_\nu^2a_0^2s_{\min}\)\\
\bottomrule
\end{tabular}
\end{table}

\emph{Transfer of block bounds and reductions.} Every remaining step of Sections~\ref{sec:heavy}--\ref{sec:amplification} transfers verbatim after updating the isotropic constants according to Table~\ref{tab:elliptic-map}. Two entries of the table deserve explicit verification. First, in Lemma~\ref{lem:stability}, the geometric comparison with \(\Gamma_q/\sigma\) acts on the unnormalized test \(\Lambda_q\), where the projection \(\langle n,MY_i\rangle\overset{d}{=}s(n)U\) at \(\mathbf D_q\) has density bounded by \(\|g\|_\infty/s_{\min}\le L_g\). Calling site \(i\) large when \(\|MY_i\|>R\) preserves \eqref{eq:L} with the same \(\gamma\) and a larger integer \(C_L\), because \(\|MY\|\le s_{\max}\|Y\|\), while \eqref{eq:Aupper} still holds for \(U\) because \(|U|\le\|Y\|\). Second, in the candidate field grid \eqref{eq:fieldgrid} of Section~\ref{sec:transfer-candidates}, lowering the initial threshold on \(t\) (via a larger fixed \(K_\nu\)) to \(2\|g\|_\infty t<4c_\nu^2a_0^2s_{\min}\) keeps the first grid bias below \(W\), since \(\mathcal M(\lambda_0/s_{\mathrm d})\le2\|g\|_\infty t^2t'/s_{\min}<4c_\nu^2a_0^2tt'\le W\). At the final grid value \(\lambda_{\text{last}}\ge1/4\), we have \(\mathcal M(\lambda_{\text{last}}/s_{\mathrm d})\ge\mathcal M(\min(y_1,1/(4s_{\max})))\), which exceeds \(13W/5\) for large \(K_\nu\). All other steps of Sections~\ref{sec:transfer} and~\ref{sec:amplification} use each block solely through its sign probabilities \(p_b(s)\), the relative gap \(\beta\), independence across sites, \(L_g\), and the trivial bound \(p_b(s)\le p_b^*\) on nonsuitable blocks. Theorems~\ref{thm:main} and~\ref{thm:factor}, including the decaying factor \(2^{-n^{1-\epsilon}}\), therefore follow unchanged.
\end{proof}

The assumption of a bounded projection density even admits singular planar laws. For instance, take \(R=1+X\), where \(X\) follows the uniform middle-thirds Cantor distribution, independently of the uniform angle \(\Theta\). The support of \(R\) has one-dimensional Lebesgue measure zero, so the union of circles of these radii has planar area zero by polar integration. Nevertheless, the unit projection has a bounded density. Indeed, for \(\alpha=\log2/\log3>1/2\), any interval of length \(h\le1\) has radial mass \(O(h^\alpha)\): picking \(j\) with \(3^{-j-1}<h\le3^{-j}\), the interval meets at most two level-\(j\) Cantor intervals of mass \(2^{-j}\). Conditional on \(R=r>|u|\), the projection density at abscissa \(u\) is \(1/(\pi\sqrt{(r-|u|)(r+|u|)})\). Because \(r\ge1\), the shells \(2^{-k-1}<r-|u|\le2^{-k}\) contribute \(O(2^{-k(\alpha-1/2)})\) uniformly in \(u\), which is summable over \(k\ge0\), while the kernel is bounded on \(r-|u|>1\). Compact support gives \eqref{eq:L}.

The reduction still embeds only finitely many rational constants, now depending on \(M\) and the law of \(Y\), without evaluating densities, probabilities or quantiles. Laws symmetric under point reflection that are not elliptically contoured have directional projection laws varying beyond a scalar factor, and we leave their complexity open.

Finally, Proposition~\ref{prop:oneblock} carries over whenever \(\E\|Y\|^2<\infty\). Writing \(e_1,e_2\) for the horizontal and vertical unit vectors, fix a rational \(\kappa_0\) satisfying
\[
 \frac{14}{5}J\E U^2\frac{s(e_2)^2}{s(e_1)}
 <\kappa_0<
 \frac{26}{5}J\E U^2\frac{s(e_2)^2}{s(e_1)},
\]
and set \(t=t'=2^{-K\ell}\), \(\lambda=\kappa_0t^2\). In the layout of Section~\ref{sec:geometry} and Section~\ref{app:contacts}, both bridge contacts lie on a vertical diameter (normal \(\pm e_2\)), whereas the normals at \(\mathbf A\) in bridges, bridge endpoints and designated cells are horizontal (\(\pm e_1\)). Dominated convergence as in Proposition~\ref{prop:oneblock} therefore yields
\[
 \frac{w_B}{t^2},\frac{w_C}{t^2}\longrightarrow
 4J^2\E U^2\left(\frac{s(e_2)}{s(e_1)}\right)^2,
 \qquad
 \frac{\mathcal M(\lambda/s_{\mathrm d})}{t^2}
 \longrightarrow\frac{2J\kappa_0}{s(e_1)},
\]
placing the normalized field inside \eqref{eq:field}. Because \(\widehat w(t,t)\le t^2\E U^2\) and \(\A(t)\le t\E|U|\), a sufficiently large fixed integer \(K\) gives one explicit block with the same inverse-polynomial relative gap, using constants fixed for \(M\) and \(Y\) without computing \(M\) from the input.

\section{Heavy tails with finite logarithmic moments and no positive radial moment}\label{sec:examples}
\begin{figure}[!htbp]
\centering
\includegraphics[width=\linewidth]{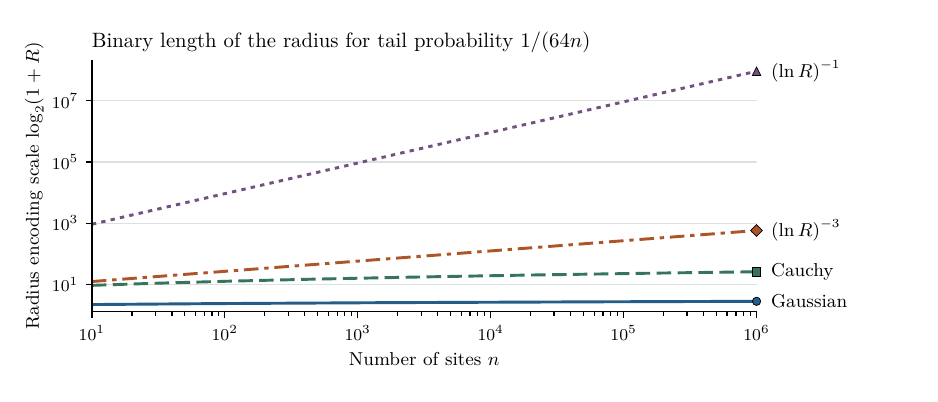}
\caption{Binary length of a truncation radius. Each curve shows \(\log_2(1+R)\) at the radius satisfying \(T_\nu(R)=1/(64n)\). Both axes are logarithmic. For \(T_\nu(R)=(\ln R)^{-\eta}\), the radius grows exponentially in \(n^{1/\eta}\), while its encoding length grows on the order of \(n^{1/\eta}\). The Gaussian, Cauchy and both logarithmic-tail laws shown satisfy (L).}\label{fig:tail-encoding}
\end{figure}

The tail assumption admits bounded noise, Gaussian and Cauchy noise, and laws with much larger displacements. We compare these examples through the number of bits needed to make a large displacement have probability at most \(1/(64n)\), the truncation level used in the product comparison. Fig.~\ref{fig:tail-encoding} compares the encoding lengths of the required radii.

Every finite positive moment of \(\|Z\|\) implies every positive moment of \(\log(1+\|Z\|)\), so Proposition~\ref{prop:equiv} gives (L). The standard planar Gaussian has \(T_\nu(R)=e^{-R^2/2}\), and a uniform law on a disk has compact support. The isotropic planar Cauchy law
\[
 \rho(z)=\frac{1}{2\pi(1+\|z\|^2)^{3/2}},\qquad
 T_\nu(R)=(1+R^2)^{-1/2},
\]
has finite moments of order \(0<a<1\). Its finite moments of order below one imply (L), so both hardness theorems apply. Scaling any of these fixed laws preserves their hypotheses.

More slowly decaying tails also qualify. For \(0<\beta<1\), the bounded radial density
\[
 \rho_\beta(z)=
 \begin{cases}
 (1-e^{-1})/\pi,&\|z\|\le1,\\[1mm]
 \displaystyle\frac{\beta}{2\pi\|z\|^2}
 (\ln\|z\|)^{\beta-1}e^{-(\ln\|z\|)^\beta},&\|z\|\ge e,\\
 0,&1<\|z\|<e
 \end{cases}
\]
integrates to one and has \(T_\nu(R)=e^{-(\ln R)^\beta}\) for \(R\ge e\). It satisfies (L) for every \(\gamma>0\). Every positive moment of \(\|Z\|\) diverges, since
\[
 \E\|Z\|^a\ge a\int_e^\infty R^{a-1}T_\nu(R)\,dR
 =a\int_1^\infty e^{au-u^\beta}\,du=\infty
 \qquad(a>0).
\]
The theorems thus cover laws for which no positive power of the displacement is integrable.

The bounded radial family
\begin{equation}\label{eq:logtail}
 \rho_\eta(z)=\frac{\eta\,\mathbf1_{\{\|z\|\ge e\}}}
 {2\pi\|z\|^2(\ln\|z\|)^{\eta+1}},\qquad
 T_\nu(R)=(\ln R)^{-\eta}\quad(R\ge e)
\end{equation}
shows the further reach of the product comparison. For every \(\eta>0\), choose a rational \(0<\gamma\le\eta\) and enlarge \(C_L\). Then (L) holds. A tail bound of \(1/(64n)\) needs a radius \(R=\exp((64n)^{1/\eta})\), whose binary length is polynomial in \(n\). Requiring tail \(2^{-s}\) instead needs \(\ln R\ge2^{s/\eta}\), so its bit length is exponential in \(s\). Keeping the probabilities of unaffected blocks is what permits the weaker truncation level in our reduction.

Condition (L) is exactly what allows all displacements to remain within a radius with polynomially many bits with a fixed positive probability (Proposition~\ref{prop:equiv}). Our transfer uses this confinement event to bound the modal probability from below. The complexity of modal Delaunay triangulation outside (L) remains open.

For a fixed uniform law on a disk, Theorem~\ref{thm:factor} gives worst-case hardness as the centres vary. A fixed generic configuration also has the stable regime described in Section~\ref{sec:intro-centres}: moving each centre by less than its structural tolerance preserves its Delaunay type. Multiplying all centres by a common positive factor multiplies this tolerance by that factor. Once the tolerance exceeds a bound on the noise norm, every realization has the type of the centres, with probability one.

\FloatBarrier

\section{Additive approximation for Gaussian sites by sampling or grid enumeration}\label{sec:approx}
Sampling realizations to approximate a geometric statistic goes back to L\"offler and Phillips~\cite{loffler_phillips}. J{\o}rgensen, L\"offler and Phillips show when finite per-site samples approximate the distribution of a geometric statistic on continuous inputs~\cite[Theorem~4]{jorgensen}. Here we count entire face lists and return one with largest empirical frequency. To implement it with finite bits, we round the coordinates of the standard Gaussian vectors that generate the displacements to rational values and bound the probability that rounding changes the type. This gives an additive guarantee. Lower bounds on the modal probability and on its gap from the competitors will then give relative approximation and exact recovery.

Throughout this section, site \(i\) is \(X_i=\mu_i+A_iZ_i\), where the centre \(\mu_i\in\Q^2\) and the invertible matrix \(A_i\in\Q^{2\times2}\) are part of the input. The vectors \(Z_1,\ldots,Z_n\) are independent, each with two independent \(N(0,1)\) coordinates. Thus \(X_i\sim N_2(\mu_i,A_iA_i^{\mathsf T})\), allowing each site its own, possibly anisotropic, covariance. The Gaussian reduction uses \(A_i=\sigma I_2\) with the same rational \(\sigma>0\) at every site. The input length \(\mathsf L\) includes the entries of these matrices and any supplied accuracy, confidence and probability bounds. Let \(\mathcal T_n\) be the finite set of complete labelled Delaunay types realizable by \(n\) planar points with labels \(1,\ldots,n\). For each \(T\in\mathcal T_n\), let \(p(T)\) be its probability under these Gaussian sites, and set \(p_*=\max_{T\in\mathcal T_n}p(T)\). We assume \(n\ge4\), since \(\mathcal T_n\) has one element for \(n\le3\).

\subsection{Rounding Gaussian coordinates}\label{sec:rounding}
To round the coordinates of the \(Z_i\), we divide the standard normal line into intervals of equal probability and choose one rational value per interval (Fig.~\ref{fig:quantile}). Rounding to these values can produce degeneracies. We test all \(\binom n3\) orientation and \(\binom n4\) incircle determinants. Coincident sites also make an orientation determinant vanish. If any test vanishes, we output the symbol \(\bot\). Otherwise we output the complete type. A \emph{valid} type is an element of \(\mathcal T_n\). The extra symbol \(\bot\) has probability zero under the original distribution \(p\). For any rounded output distribution \(\widetilde p\) on this common alphabet, the supremum over events defining total variation can equivalently be written as
\[
 \TV(p,\widetilde p)=\frac12\sum_{T\in\mathcal T_n\cup\{\bot\}}|p(T)-\widetilde p(T)|.
\]

\begin{samepage}
\begin{lemma}\label{lem:quantile}
For an integer \(q\ge2\), partition the standard normal line into \(q\) consecutive intervals of probability \(1/q\), choose a rational representative in the interior of each, and replace each coordinate of each \(Z_i\) by the representative of its interval. If \(\widetilde p_q\) is the resulting output distribution, then
\[
 \TV(p,\widetilde p_q)\le\frac{n^4}{q}.
\]
The representatives can be generated individually in time polynomial in \(\log q\), using polynomially many bits.
\end{lemma}
\end{samepage}

\begin{figure}[!htb]
\centering
\input{figures/figure_style}
\begin{tikzpicture}[ndt picture,x=2cm,y=2cm]
 \node[ndt part] at (-2.7,1.45) {(a) Move $D$ along the line};
 \definecolor{qcircle}{HTML}{367795}
 \draw[draw=qcircle,dash pattern=on 3pt off 1.7pt,line width=.72pt] (0,0) circle (1);
 \draw[ndt edge] (-1,0)--(0,1)--(1,0)--cycle;
 \draw[ndt construction] (-1,0)--(-1.5,-.5) (1,0)--(1.5,-.5);
 \draw[ndt arrow] (-2.2,-.5)--(2.2,-.5);
 \node[right] at (2.2,-.5) {$u$};
 \foreach \x in {-1.5,1.5} {\node[ndt orientation mark] at (\x,-.5) {};}
 \foreach \x in {-.8660254,.8660254} {\node[ndt incircle mark,draw=ndtBlue] at (\x,-.5) {};}
 \foreach \x/\lab in {-1.5/{-3/2},-.8660254/{-\sqrt3/2},.8660254/{\sqrt3/2},1.5/{3/2}} {
   \draw[black!45,line width=.4pt] (\x,-.55)--(\x,-.88);
   \node[below] at (\x,-.90) {$\lab$};
 }
 \node[ndt dot] at (-1,0) {}; \node[left] at (-1,0) {$A$};
 \node[ndt dot] at (1,0) {}; \node[right] at (1,0) {$B$};
 \node[ndt dot] at (0,1) {}; \node[above] at (0,1) {$C$};
 \node[ndt dot] at (.3,-.5) {};
 \node[above] at (.3,-.5) {$D(u)$};
\end{tikzpicture}

\vspace{-.7mm}
\begin{tikzpicture}[ndt picture,x=11cm,y=1cm]
 \node[ndt part] at (-.02,1.05) {(b)};
 \node[anchor=west] at (.06,1.05) {$v=\Phi(u)$: four crossings in eight equal-probability intervals};
 \foreach \a/\b in {0/.125,.125/.25,.75/.875,.875/1} {
  \fill[ndtOrange!23] (\a,0) rectangle (\b,.5);
 }
 \draw[ndt edge] (0,0) rectangle (1,.5);
 \foreach \j in {1,...,7} {\draw[black!45,line width=.5pt] (\j/8,0)--(\j/8,.5);}
 \foreach \x in {.0668072,.1932381,.8067619,.9331928} {
  \draw[black!58,line width=.68pt] (\x,.27)--(\x,.62);
 }
 \foreach \x in {0.04005916,0.15865525,0.30853754,0.45026178,0.54973822,0.69146246,0.84134475,0.95994084} {\fill[black] (\x,.18) circle (1.8pt);}
 \foreach \x in {.0668072,.9331928} {\node[ndt orientation mark] at (\x,.7) {};}
 \foreach \x in {.1932381,.8067619} {\node[ndt incircle mark,draw=ndtBlue] at (\x,.7) {};}
 \node[below,text height=1.8ex,text depth=.6ex] at (0,-.12) {$0$};
 \foreach \j in {1,...,7} {\node[below,text height=1.8ex,text depth=.6ex] at (\j/8,-.12) {$\j/8$};}
 \node[below,text height=1.8ex,text depth=.6ex] at (1,-.12) {$1$};
 \fill[black] (.04,-.97) circle (1.8pt);
 \node[anchor=west] at (.065,-.97) {representative image};
 \draw[black!58,line width=.68pt] (.39,-1.12)--(.39,-.82);
 \node[anchor=west] at (.415,-.97) {critical value};
 \fill[ndtOrange!23] (.76,-1.12) rectangle (.79,-.82);
 \node[anchor=west] at (.81,-.97) {bin with a crossing};
\end{tikzpicture}
\caption{Rounding one coordinate can change the type only in a bin with a crossing. In (a), \(A=(-1,0)\), \(B=(1,0)\), \(C=(0,1)\) and \(D=(u,-1/2)\). Orientation changes at \(u=\pm3/2\) (squares) and the incircle test changes at \(u=\pm\sqrt3/2\) (circles). In (b), \(\Phi\) maps these four values into bins \(1,2,7,8\). The black dots are \(\Phi(u)\) for the rational representatives \(u=\pm1/8,\pm1/2,\pm1,\pm7/4\), one per bin. Rounding within any unshaded bin preserves the complete type.}\label{fig:quantile}
\end{figure}

\begin{proof}
Round the \(2n\) coordinates of \(Z_1,\ldots,Z_n\) one at a time in a fixed order, using the same sample throughout. If the original and final triangulations differ, some rounding step changed the output. It therefore suffices to add up the probabilities of a change at each step. The coordinates start out independent, and each rounded value depends only on its own original coordinate. Conditional on the current values of all other coordinates, the coordinate about to be rounded still has the standard normal law. In particular, each of its \(q\) intervals still has probability \(1/q\).

With those other coordinates fixed, varying the remaining coordinate of \(Z_i\) moves \(X_i\) along a straight line in the direction of the corresponding column of \(A_i\) (Fig.~\ref{fig:quantile}(a)). Every determinant involving only the other sites is constant along the line. If any orientation or incircle determinant vanishes identically, the algorithm reports \(\bot\) both before and after this rounding step. Testing every determinant catches even degeneracies that a triangulation algorithm may never encounter. If no determinant vanishes identically, each pair of fixed sites defines a line that the moving site can meet at most once, and each triple defines a circle that it can meet at most twice. The corresponding determinants are polynomials of degree at most one and two in the moving coordinate. Thus at most \(\binom{n-1}{2}+2\binom{n-1}{3}\le n^3/2\) values are critical. On each interval between critical values, including the two outer intervals, every test keeps its sign and the triangulation stays the same.

The sampled coordinate and its representative lie in the interior of the same quantile interval almost surely. Rounding can change the output only if this interior contains a critical value (Fig.~\ref{fig:quantile}(b)). At most \(n^3/2\) intervals can do so. The conditional probability of a change is therefore at most \(n^3/(2q)\). Summing over \(2n\) steps bounds the probability that the coupled outputs differ, and hence their total variation distance, by \(n^4/q\), uniformly in the centres.

Given \(j\), we still need to construct a rational number inside its interval. Let \(\Phi\) be the cumulative distribution function of the standard normal. On the probability axis in Fig.~\ref{fig:quantile}(b), the \(j\)th interval runs from \((j-1)/q\) to \(j/q\), with midpoint \(v_j=(j-\tfrac12)/q\). The midpoint is \(1/(2q)\) from either endpoint on the probability axis. Thus any rational \(r_j\) with \(|\Phi(r_j)-v_j|<1/(4q)\) lies strictly inside its quantile interval on the real line, even for the unbounded first and last intervals.

We locate such a point by bisection on the real line. For \(K=\lceil\log_2 q\rceil+2\) the Gaussian Chernoff bound gives \(\Phi(-K)=1-\Phi(K)\le e^{-K^2/2}<1/(8q)\), so \([-K,K]\) contains the quantile \(\Phi^{-1}(v_j)\). At the midpoint \(r\) of the current interval, compute a rational \(\widetilde\Phi\) with \(|\widetilde\Phi-\Phi(r)|\le1/(16q)\). If \(|\widetilde\Phi-v_j|\le1/(8q)\), return \(r\). Indeed, \(|\Phi(r)-v_j|\le3/(16q)<1/(4q)\). Otherwise \(\Phi(r)-v_j\) has the sign of \(\widetilde\Phi-v_j\), which shows on which side of \(r\) the quantile lies, and we keep that half. The test must succeed once the search interval has length below \(1/(8q)\). Its midpoint is then within \(1/(16q)\) of the quantile, and \(\Phi'<1\) gives \(|\Phi(r)-v_j|<1/(16q)\), hence \(|\widetilde\Phi-v_j|<1/(8q)\). Starting from length \(2K\), this takes \(O(\log q)\) halvings, and every midpoint is a dyadic rational with \(O(\log q)\) bits.

To compute \(\Phi(r)\), we integrate the standard normal density from \(0\) to \(r\) and add \(1/2\). Expand \(e^{-w}\) at zero through degree \(k\) and substitute \(w=x^2/2\). The resulting polynomial in \(x\) has degree \(2k\). Every derivative of \(e^{-w}\) has magnitude at most one for \(w\ge0\), so the pointwise Taylor error is at most \(w^{k+1}/(k+1)!\). Between \(0\) and \(r\), whatever the sign of \(r\), we have \(w\le K^2/2\) along an interval of length \(|r|\le K\). Integrating the pointwise bound therefore gives truncation error at most \(K(K^2/2)^{k+1}/(k+1)!\). Since \(K=O(\log q)\), taking \(k\) a sufficiently large constant multiple of \(K^2\) and using a certified rational approximation to \(1/\sqrt{2\pi}\) gives the required accuracy \(1/(16q)\) with a number of bits polynomial in \(\log q\).
\end{proof}

\subsection{Additive approximation by sampling or enumeration}\label{sec:additive-algorithm}
The additive algorithm samples \(\widetilde p_q\) and returns a most frequent valid type. It stores only observed types. To compare that output with an unobserved mode, we need a probability bound that holds simultaneously for every type, regardless of the size of the support.

\Needspace{8\baselineskip}
\begin{theorem}\label{thm:additive}
Given independent Gaussian sites specified by rational centres and invertible rational \(2\times2\) covariance factors, and rational \(0<\varepsilon,\delta<1\), a randomized algorithm returns a valid type \(\widehat T\) such that
\[
 \Prob\{p(\widehat T)\ge p_*-\varepsilon\}\ge1-\delta.
\]
It uses \(O(\varepsilon^{-2}\log(2/\delta))\) independent discrete realizations and runs in time polynomial in \(n,\mathsf L,1/\varepsilon,\log(1/\delta)\).
\end{theorem}

\begin{algorithm}[ht]
\caption{Additive approximation for Gaussian input}\label{alg:additive}
\begin{algorithmic}[1]
\Require Rational centres, invertible rational covariance factors \(A_i\), and rational \(0<\varepsilon,\delta<1\)
\Ensure \(p(\widehat T)\ge p_*-\varepsilon\) with probability at least \(1-\delta\)
\State Choose the smallest power of two \(q\ge4n^4/\varepsilon\).
\State Set \(N=\lceil32\varepsilon^{-2}\rceil\lceil\log_2(2/\delta)\rceil\).
\For{\(j=1,\ldots,N\)}
 \State Sample \(2n\) uniform quantile indices, generate their rational representatives and form the rational sites.
 \State Evaluate all orientation and incircle predicates exactly. Record \(\bot\) if one vanishes, and the canonical type otherwise.
\EndFor
\State Return a most frequent valid type. If none was observed, return the Delaunay type of \((i,i^2)\), \(i=1,\ldots,n\).
\end{algorithmic}
\end{algorithm}

\begin{proof}
Algorithm~\ref{alg:additive} chooses \(q\) so that rounding changes the probability of any type by at most \(\varepsilon/4\) (Lemma~\ref{lem:quantile}). To control the sampling error, fix any ordering of the discrete outputs, including \(\bot\). By the Dvoretzky--Kiefer--Wolfowitz inequality~\cite{dkw} with Massart's sharp constant~\cite{massart}, with probability at least \(1-2\exp(-N\varepsilon^2/32)\) every empirical cumulative frequency is within \(\varepsilon/8\) of its population value. This bound holds for discrete distributions as well: applying their inverse distribution function to uniform draws gives the same cumulative frequencies at the corresponding probability thresholds. The algorithm takes \(N\ge32\varepsilon^{-2}\ln(2/\delta)\), since \(\log_2 x\ge\ln x\) for \(x\ge1\). The bound thus holds simultaneously at all thresholds with confidence at least \(1-\delta\).

Subtract the cumulative frequency just before a type from the cumulative frequency through it. Each of the two cumulative errors is at most \(\varepsilon/8\), so the resulting frequency \(\widehat p(T)\) differs from \(\widetilde p_q(T)\) by at most \(\varepsilon/4\). Its denominator counts all \(N\) draws, including \(\bot\). Adding the rounding error gives \(|p(T)-\widehat p(T)|\le\varepsilon/2\) for every valid type on this same event.

The algorithm returns a valid type \(\widehat T\) with a largest count. In particular, it occurs at least as often as any true mode \(T_*\), whose count is zero if it was never sampled. Applying the error bound first to \(T_*\) and then to \(\widehat T\) gives
\[
 p(T_*)\le\widehat p(T_*)+\varepsilon/2
 \le\widehat p(\widehat T)+\varepsilon/2
 \le p(\widehat T)+\varepsilon .
\]
If every draw is \(\bot\), all valid types have empirical frequency zero. The fallback is therefore also an empirical maximizer. It is valid because the points \((i,i^2)\) are generic, as shown in the proof of Proposition~\ref{prop:centre-rank}. Each draw takes \(2n\log_2q\) random bits. By Lemma~\ref{lem:quantile}, we generate its rational representatives in polynomial time with polynomial bit length. Exact determinant evaluation and sorted lists of labelled faces then let us compute and count the observed types in polynomial time.

Without sampling, enumerating the \(q^{2n}\) grid realizations for \(q=\lceil2n^4/\varepsilon\rceil\) gives \(\widetilde p_q\) exactly, and a most probable valid type of \(\widetilde p_q\), with the same fallback, loses at most \(2\TV(p,\widetilde p_q)\le\varepsilon\). This deterministic variant takes time \((n/\varepsilon)^{O(n)}\operatorname{poly}(\mathsf L)\).
\end{proof}

\section{Relative approximation and exact recovery of a unique modal triangulation}\label{sec:sampling-gaps}
When \(\varepsilon\ge p_*\), the additive guarantee allows every valid type. To obtain a useful relative guarantee, we supply a positive lower bound on \(p_*\) and choose a smaller additive tolerance. If a unique mode is also separated from its competitors by a known positive gap, a tolerance smaller than that gap forces exact recovery.

\begin{proposition}\label{prop:relative}
If a positive rational lower bound \(\underline p\le p_*\) is supplied, Gaussian input admits randomized relative search with
\[
 \Prob\{p(\widehat T)\ge(1-\varepsilon)p_*\}\ge1-\delta
\]
in time polynomial in \(n,\mathsf L,1/\varepsilon,1/\underline p,\log(1/\delta)\).
For arbitrary Gaussian input there is such an algorithm with bit cost
\[
 2^{O(n\log n)}
 \operatorname{poly}(\mathsf L,1/\varepsilon,\log(1/\delta)).
\]
\end{proposition}
\begin{proof}
With tolerance \(\varepsilon \underline p\), the algorithm of Theorem~\ref{thm:additive} returns a type satisfying \(p(\widehat T)\ge p_*-\varepsilon \underline p\ge(1-\varepsilon)p_*\) in time polynomial in \(1/(\varepsilon \underline p)\) and the other parameters. For the second assertion, the type count in the proof of Proposition~\ref{prop:centre-rank} gives \(p_*\ge 1/(n!C^n)\) for a fixed integer \(C>1\), since every realization is almost surely generic. Using this lower bound for \(\underline p\) gives the stated \(2^{O(n\log n)}\) factor.
\end{proof}

\begin{samepage}
\begin{proposition}\label{prop:histogram}
Suppose the mode \(T_*\) is unique, with probability \(p_*=p(T_*)\), gap
\(\Delta=p_*-\max_{T\ne T_*}p(T)>0\), and relative gap \(\gamma=\Delta/p_*\).
The empirical mode of \(N\) independent ideal Gaussian realizations equals \(T_*\) with probability at least \(1-\delta\) whenever
\[
 N\ge\frac{20}{p_*\gamma^2}
        \ln\frac{2|\mathcal T_n|}{\delta}.
\]
For Gaussian input with known inverse-polynomial lower bounds on \(p_*\) and \(\gamma\), exact modal recovery also has a randomized polynomial-time implementation in bits.
\end{proposition}
\end{samepage}
\begin{proof}
Each empirical frequency \(\widehat p(T)\) averages \(N\) independent indicators with variance at most \(p_*\). Bernstein's inequality~\cite[Section~2.7, inequality~(2.10)]{blm} at \(t=\Delta/3\) gives
\[
 \Prob\{|\widehat p(T)-p(T)|\ge\Delta/3\}
 \le2\exp\!\left(-\frac{N\Delta^2}{18p_*+2\Delta}\right)
 \le2\exp\!\left(-\frac{N\Delta^2}{20p_*}\right),
\]
since \(\Delta\le p_*\). As \(\Delta^2/p_*=p_*\gamma^2\), a union bound over \(\mathcal T_n\) shows that, for the stated \(N\), all these errors are below \(\Delta/3\) with probability at least \(1-\delta\). Then \(T_*\) has strictly largest empirical frequency. For the finite-bit implementation, let \(\underline p\) and \(\underline\gamma\) be the supplied positive lower bounds on the modal probability \(p_*\) and its relative gap \(\gamma\), respectively. Their product is at most \(p_*\gamma=\Delta\). Use half that product as the additive tolerance in Theorem~\ref{thm:additive}. Every other type has probability at most \(p_*-\Delta\), so only \(T_*\) can meet the guarantee.
\end{proof}

In Fig.~\ref{fig:four-centres}(b), the centres in label order are $(0,0)$, $(1,0)$, $(1,1.03)$ and $(0,1)$. The hull cycle $ABCD$ appeared in every draw at the two smallest scales and in 70.91\% at $\sigma=0.4$. CGAL~\cite{cgal} classified each sample, and exact integer predicates checked its face list independently on the same binary64 coordinates. The supplement gives marginal Wilson intervals~\cite[p.~210]{wilson} for all ten frequencies per scale. Under ideal independent sampling, Hoeffding's inequality~\cite[Theorem~2]{hoeffding} and a union bound give a simultaneous 95\% error bound of $0.097$ percentage points for all forty frequencies.

The coupling of Lemma~\ref{lem:quantile} uses independent Gaussian coordinates, and the algorithms need efficient finite sampling. The reductions for Delaunay triangulations require neither independence between coordinates nor an algorithm for evaluating the density or sampling efficiently. The sampling guarantees therefore do not automatically apply to every law covered by those theorems.

\providecommand{\leandecl}[1]{\texttt{#1}}
\section{Which results are verified in Lean, and under which hypotheses?}\label{sec:lean}
Every numbered statement of this paper and both algorithms have counterparts proved in Lean~4~\cite{lean4} with the Mathlib library~\cite{mathlib}, and so do the unnumbered claims listed in Table~\ref{tab:lean-claims}. A reader who doubts a step of the reductions can therefore consult a proof whose constructions, probability estimates and running-time bounds the Lean kernel has checked. The kernel certifies the formal statements, and the value of this certificate depends on how faithfully they express the printed ones and on what they assume. Tables~\ref{tab:lean} and~\ref{tab:lean-claims} form a dictionary between the results of this paper and their names in Lean, with the classical hypothesis that each one uses. Each result on the left links to its place in this paper. The names on the right belong to the module \texttt{Results} of the supplement, which restates every result under a name that describes its content, with the statement of the declaration that proves it. The file \texttt{README\_FORMALIZATION.md} in the folder \texttt{lean} of the supplement repeats this dictionary and gives, for every name, the proving declaration and, for every result, its premises and main hypotheses. The following paragraphs state the classical hypotheses and the conventions that fix the meaning of the formal statements, identify the proofs that Lean carries out by another route and the claims left outside, and explain how to repeat the check.

\begin{table}[!ht]
\caption{Dictionary between the numbered statements and the algorithms of this paper and their names in Lean, with the classical premise that each row uses: IS for \texttt{IndependentSetHard} and SS for \texttt{SubsetSumHard}. Each entry on the left links to the statement. The names belong to the namespace \texttt{MostLikelyDelaunay.Results}. Several names in one row prove the parts of the statement together, and some rows add more general forms that the text describes.}\label{tab:lean}
\centering\footnotesize\renewcommand{\arraystretch}{1.2}
\begin{tabular}{@{}>{\footnotesize\raggedright\arraybackslash}p{0.14\linewidth}>{\footnotesize\raggedright\arraybackslash}p{0.72\linewidth}>{\footnotesize\raggedright\arraybackslash}p{0.08\linewidth}@{}}
\toprule
Statement & Lean names & Premise\\
\midrule
\hyperref[prop:four-centres]{Prop.~\ref*{prop:four-centres}} & \leandecl{four\_site\_centre\_type}, \leandecl{quadrilateral\_types\_outrank\_centre\_type}, \leandecl{four\_site\_hull\_type\_probability} & none\\
\hyperref[prop:centre-rank]{Prop.~\ref*{prop:centre-rank}} & \leandecl{centre\_rank} & none\\
\hyperref[prop:line]{Prop.~\ref*{prop:line}} & \leandecl{sorted\_means\_half\_approximation}, \leandecl{sorted\_means\_half\_approximation\_sites}, \leandecl{sorted\_order\_half\_approximation}, \leandecl{sorted\_order\_optimal\_symmetric\_means}, \leandecl{sorted\_order\_optimal\_four\_sites} & none\\
\hyperref[thm:socg-gaussian]{Thm.~\ref*{thm:socg-gaussian}} & \leandecl{gaussian\_mode\_search\_np\_hard}, \leandecl{gaussian\_decaying\_factor\_search\_np\_hard} & IS\\
\hyperref[lem:geometry]{Lem.~\ref*{lem:geometry}} & \leandecl{contact\_layout}, \leandecl{contact\_layout\_machine} & none\\
\hyperref[lem:completion]{Lem.~\ref*{lem:completion}} & \leandecl{cell\_completion} & none\\
\hyperref[lem:stability]{Lem.~\ref*{lem:stability}} & \leandecl{local\_stability}, \leandecl{local\_stability\_machine\_output}, \leandecl{local\_stability\_machine\_time} & none\\
\hyperref[thm:gaussian-hull]{Thm.~\ref*{thm:gaussian-hull}} & \leandecl{hull\_mode\_search\_np\_hard}, \leandecl{hull\_label\_search\_np\_hard}, \leandecl{hull\_cyclic\_order\_search\_np\_hard}, \leandecl{hull\_one\_call\_reduction} & SS\\
\hyperref[def:L]{Def.~\ref*{def:L}} & \leandecl{LogarithmicTailCondition} & none\\
\hyperref[thm:main]{Thm.~\ref*{thm:main}} & \leandecl{radial\_mode\_search\_np\_hard} & IS\\
\hyperref[prop:equiv]{Prop.~\ref*{prop:equiv}} & \leandecl{confinement\_criterion}, \leandecl{confinement\_independent\_copies} & none\\
\hyperref[lem:near]{Lem.~\ref*{lem:near}} & \leandecl{near\_density\_symmetric\_continuous}, \leandecl{near\_density\_uniform\_limit}, \leandecl{near\_density\_window} & none\\
\hyperref[lem:weights]{Lem.~\ref*{lem:weights}} & \leandecl{weight\_bounds} & none\\
\hyperref[lem:expansion]{Lem.~\ref*{lem:expansion}} & \leandecl{block\_expansion}, \leandecl{zero\_field\_cycle\_expansion}, \leandecl{zero\_field\_cycle\_reversal}, \leandecl{field\_change\_bound}, \leandecl{field\_phase\_gap} & none\\
\hyperref[lem:block]{Lem.~\ref*{lem:block}} & \leandecl{block\_gap} & none\\
\hyperref[alg:reduction]{Alg.~\ref*{alg:reduction}} & \leandecl{radial\_one\_call\_reduction} & none\\
\hyperref[lem:localdisks]{Lem.~\ref*{lem:localdisks}} & \leandecl{local\_disks}, \leandecl{local\_disks\_dyadic}, \leandecl{local\_disks\_repeated} & none\\
\hyperref[lem:block-product]{Lem.~\ref*{lem:block-product}} & \leandecl{block\_product\_bound}, \leandecl{block\_product\_bound\_dyadic}, \leandecl{block\_product\_bound\_repeated}, \leandecl{block\_product\_bound\_layout} & none\\
\hyperref[prop:failure]{Prop.~\ref*{prop:failure}} & \leandecl{failure\_bound}, \leandecl{repeated\_failure\_bound}, \leandecl{failure\_bound\_any\_copies}, \leandecl{failure\_bound\_general}, \leandecl{failure\_bound\_general\_copies}, \leandecl{failure\_upper\_bound\_layout}, \leandecl{failure\_lower\_bound\_layout} & none\\
\hyperref[prop:oneblock]{Prop.~\ref*{prop:oneblock}} & \leandecl{one\_block\_certified}, \leandecl{gaussian\_one\_block\_certified}, \leandecl{gaussian\_squared\_density\_integral}, \leandecl{gaussian\_second\_moment} & none\\
\hyperref[thm:factor]{Thm.~\ref*{thm:factor}} & \leandecl{radial\_constant\_factor\_search\_np\_hard}, \leandecl{radial\_inapproximable\_unless\_p\_eq\_np}, \leandecl{radial\_scheme\_implies\_p\_eq\_np}, \leandecl{radial\_randomized\_factor\_implies\_np\_eq\_rp} & IS\\
\hyperref[thm:elliptical]{Thm.~\ref*{thm:elliptical}} & \leandecl{elliptic\_mode\_and\_factor\_search\_np\_hard}, \leandecl{elliptic\_inapproximable\_unless\_p\_eq\_np}, \leandecl{elliptic\_randomized\_factor\_implies\_np\_eq\_rp} & IS\\
\hyperref[lem:quantile]{Lem.~\ref*{lem:quantile}} & \leandecl{quantile\_coupling} & none\\
\hyperref[thm:additive]{Thm.~\ref*{thm:additive}} & \leandecl{additive\_sampling\_guarantee}, \leandecl{additive\_sampling\_realizations} & none\\
\hyperref[alg:additive]{Alg.~\ref*{alg:additive}} & \leandecl{additiveSamplingMachine}, \leandecl{additive\_sampling\_guarantee} & none\\
\hyperref[prop:relative]{Prop.~\ref*{prop:relative}} & \leandecl{relative\_search\_with\_lower\_bound}, \leandecl{relative\_search\_general} & none\\
\hyperref[prop:histogram]{Prop.~\ref*{prop:histogram}} & \leandecl{ideal\_histogram\_recovery}, \leandecl{histogram\_recovery\_machine} & none\\
\bottomrule
\end{tabular}
\end{table}

\begin{table}[!ht]
\caption{Dictionary between unnumbered claims of the text, with the places where each claim appears, and their names in Lean, with the classical premise that each row uses. The names belong to the namespace \texttt{MostLikelyDelaunay.Results}.}\label{tab:lean-claims}
\centering\footnotesize\renewcommand{\arraystretch}{1.2}
\begin{tabular}{@{}>{\footnotesize\raggedright\arraybackslash}p{0.24\linewidth}>{\footnotesize\raggedright\arraybackslash}p{0.62\linewidth}>{\footnotesize\raggedright\arraybackslash}p{0.08\linewidth}@{}}
\toprule
Claim & Lean names & Premise\\
\midrule
Hull edges are the pairs in exactly one face (Section~\ref{sec:introduction}) & \leandecl{hull\_edge\_iff\_single\_face}, \leandecl{hull\_edges\_determined\_by\_type} & none\\
General position almost surely (Secs.~\ref{sec:introduction} and~\ref{sec:elliptical}) & \leandecl{general\_position\_almost\_surely} & none\\
Factor \(1/2\) for Gaussian sites (Section~\ref{sec:introduction}) & \leandecl{gaussian\_half\_factor\_search\_np\_hard} & IS\\
Small noise and the stable regime (Secs.~\ref{sec:centres} and~\ref{sec:examples}) & \leandecl{small\_noise\_centre\_mode}, \leandecl{gaussian\_small\_noise\_centre\_mode}, \leandecl{stable\_regime\_centre\_type}, \leandecl{uniform\_disk\_stable\_regime} & none\\
Gaussian translation bound (Section~\ref{sec:centres}) & \leandecl{gaussian\_translation\_total\_variation}, \leandecl{gaussian\_translation\_chain}, \leandecl{gaussian\_translation\_type\_bound} & none\\
Moving the centres of Prop.~\ref{prop:four-centres} (Section~\ref{sec:centres}) & \leandecl{four\_site\_centre\_perturbation}, \leandecl{four\_site\_centre\_perturbation\_constant} & none\\
Centre or samples after Prop.~\ref{prop:centre-rank} (Section~\ref{sec:centres}) & \leandecl{centre\_or\_samples\_rule} & none\\
Cauchy law and slow tails (Sec.~\ref{sec:heavy}) & \leandecl{cauchy\_tail\_ratio\_tendsto\_zero}, \leandecl{log\_tail\_bridge\_weight\_comparable} & none\\
Negative remarks on projections and linear maps (Secs.~\ref{sec:heavy} and~\ref{sec:elliptical}) & \leandecl{asymmetric\_zero\_field\_marginal}, \leandecl{no\_common\_projection\_scale}, \leandecl{stretch\_flips\_incircle} & none\\
Factor \(2^{-n^{1-\epsilon}}\) for radial and elliptic laws (Secs.~\ref{sec:amp-replication} and~\ref{sec:elliptical}) & \leandecl{radial\_decaying\_factor\_search\_np\_hard}, \leandecl{elliptic\_mode\_and\_factor\_search\_np\_hard}, \leandecl{radial\_decaying\_factor\_implies\_p\_eq\_np}, \leandecl{radial\_randomized\_decaying\_factor\_implies\_np\_eq\_rp} & IS\\
Sorting with \(O(n\log n)\) comparisons and the unique oriented mode (Secs.~\ref{sec:introduction} and~\ref{sec:line}) & \leandecl{merge\_sort\_comparisons}, \leandecl{merge\_sort\_sorts}, \leandecl{unique\_oriented\_mode}, \leandecl{oriented\_mode\_iff\_sorted} & none\\
Example laws (Secs.~\ref{sec:elliptical} and~\ref{sec:examples}) & \leandecl{cauchy\_one\_call\_reduction}, \leandecl{cauchy\_factor\_reductions}, \leandecl{uniform\_disk\_factor\_reductions}, \leandecl{beta\_tail\_law\_admissible}, \leandecl{log\_tail\_law\_admissible}, \leandecl{gaussian\_covariance\_elliptic\_reductions}, \leandecl{radial\_laws\_elliptic\_hypotheses}, \leandecl{sqrt\_law\_elliptic\_hypotheses}, \leandecl{beta\_tail\_law\_no\_positive\_moment}, \leandecl{sqrt\_law\_density\_unbounded}, \leandecl{singular\_law\_mutually\_singular}, \leandecl{singular\_law\_elliptic\_reductions}, \leandecl{radial\_hypotheses\_scaling}, \leandecl{elliptic\_reductions\_scaling} & none\\
Deterministic enumeration (Sec.~\ref{sec:additive-algorithm}) & \leandecl{deterministic\_enumeration}, \leandecl{deterministic\_enumeration\_time} & none\\
\bottomrule
\end{tabular}
\end{table}

\textbf{Classical hypotheses.} The statements of this paper depend on two results from the literature, which the module \texttt{ClassicalHypotheses} of the supplement states as definitions. The premise \texttt{IndependentSetHard} asserts that \textsc{Maximum Independent Set} is NP-hard on simple cubic planar graphs, and Mohar proves this for the subclass of 2-connected graphs~\cite[Theorem~4.1(a)]{mohar_face}, which implies it for the whole class. The premise \texttt{SubsetSumHard} asserts that \textsc{Subset Sum} with positive items \(a_1,\ldots,a_m\) and a target \(1\le T\le A=\sum_i a_i\) is NP-hard~\cite[problem~18, pp.~95, 100]{karp}. Theorems~\ref{thm:socg-gaussian}, \ref{thm:main}, \ref{thm:factor} and~\ref{thm:elliptical} take \texttt{IndependentSetHard} as a premise, Theorem~\ref{thm:gaussian-hull} takes \texttt{SubsetSumHard}, and no other numbered statement uses either. The Lean statements of these four theorems, and those in the rows of Table~\ref{tab:lean-claims} with premise IS, write \texttt{IndependentSetHard} out as its definition, \texttt{NativeHard} applied to \texttt{MachineModel.basis} and \texttt{OneCallDecision.yes}, which reads \texttt{NativeHard basis yes} where these two namespaces are open. For the laws of Sections~\ref{sec:elliptical} and~\ref{sec:examples}, the declarations in the row of example laws of Table~\ref{tab:lean-claims} check the hypotheses of Theorems~\ref{thm:main}, \ref{thm:factor} and~\ref{thm:elliptical} and build the reductions, so their NP-hardness uses \texttt{IndependentSetHard} in the same way.

Neither premise is an axiom, so each one appears in the statement of every theorem that uses it. The reductions themselves are proved outright. For Theorem~\ref{thm:gaussian-hull}, for instance, \texttt{hull\_one\_call\_reduction} states the one-call reduction from \textsc{Subset Sum} with no premise at all. A premise enters only in the last step, which composes a reduction with the reductions from every problem in NP that the classical theorem provides. Deriving the two premises from the cited theorems requires writing their reductions as transition tables of the model described below. The formalization README explains this derivation informally, since a formal proof would need the Cook--Levin theorem and the cited reductions in this model. A third named premise, the classical duality between Delaunay triangulations and Voronoi diagrams, enters only declarations about labelled Voronoi diagrams, and no statement of this paper depends on it.

\textbf{Conventions.} Inputs are binary words, with rationals written in binary. Running time is measured on finite transition tables over three tapes with a built-in clock: each table halts within a number of steps bounded by a polynomial in its input length, and the degree of this polynomial is part of the table. The classes P, NP and RP are defined with these tables. A search problem is NP-hard when every language in NP reduces to it with one call to a solver. One table writes the query from the input, and a second table decides membership from the input and the reply. The second table must be correct for every acceptable reply, so a solver may break ties arbitrarily, and some acceptable reply of polynomial length must exist. Lemma~\ref{lem:geometry} and Algorithm~\ref{alg:reduction} also have versions that count bit operations, and the hardness theorems run them as transition tables. Planarity is combinatorial. A graph is planar when each connected component has a rotation system of genus zero, given by face labels on its darts that agree with a cyclic order at every vertex and satisfy Euler's relation. The faces of a plane drawing provide such labels, so this class contains the graphs of Mohar's theorem, and the equivalence with topological planarity is left outside the development. The Delaunay search problems are promise problems: their queries consist of rational centres in general position, and every reduction writes only such queries. The type of a configuration is formalized as its set of label triples whose open circumdisk contains no site. In general position these triples are the Delaunay triangles, and every displacement law considered here puts probability one on general position.

\textbf{Proofs that follow another route.} For some statements of Table~\ref{tab:lean}, the formal proof follows a different route from the body. Lean verifies Theorem~\ref{thm:socg-gaussian} by applying the reductions for radial laws to the standard Gaussian, as the last paragraph of Section~\ref{sec:model} notes. Theorem~\ref{thm:main} gives the exact mode, and the replication argument of Section~\ref{sec:amp-replication}, which also proves Theorem~\ref{thm:factor}, gives the factor \(2^{-n^{1-\epsilon}}\). The direct argument of Section~\ref{sec:socg-gaussian}, which differentiates Gaussian orthant probabilities~\cite{plackett}, is therefore not formalized, and the same holds for Sheppard's closed-form correlation recalled at the start of Section~\ref{sec:responses}. The radial reductions already contain the standard Gaussian, so this route proves Theorem~\ref{thm:socg-gaussian} with no formalization of Sheppard's formula or of the derivatives of Gaussian orthant probabilities. Theorem~\ref{thm:gaussian-hull} is verified for the construction of Section~\ref{sec:gaussian-hull} with a different calibration, defined in the supplement. The formal proof keeps \(U\), \(k\) and \(\lambda=2^{-k}\) from that section, with \(N=2m+4\), and changes \(\beta=\lambda^4\) to \(\beta=2^{-m}\lambda^2/(2^{16}N^2)\), the error unit for which \(2^{11}N^2\) units fill \(G/4\) exactly. With this unit the transition tables evaluate every quantity with \(m+2k+16+2\lceil\log_2(N+1)\rceil\) bits of precision, and the accounting below also covers the rounding of the centres that they compute. It then shows that the errors of the calibration, of the rounding and of the vertex perturbation keep every canonical hull probability within \(G/4=2^{11}N^2\beta\) of its ideal score, where \(G=2^{-m}\lambda^2/8\) is the score gap of Section~\ref{sec:gaussian-hull}. The calibration printed in that section, with \(\beta=\lambda^4\) and the bound \((5m+4)\beta<G/4\), is therefore not checked.

In a few other proofs, Lean replaces an intermediate estimate by another one with the same conclusion. For instance, it bounds the modal probability in Proposition~\ref{prop:centre-rank} from below by confining the sites to boxes of a grid and counting their possible neighbours, without Tutte's count of planar maps, which would need a formal enumeration of planar maps, and in Lemma~\ref{lem:stability} the bounds \(r_*\) and \(C_P\) are explicit formulas in the bit widths of the input, because the machine of that lemma prints them. The deterministic enumeration of Section~\ref{sec:additive-algorithm} uses a grid with a power of two below \(8n^4/\varepsilon\) cells per coordinate in place of \(\lceil2n^4/\varepsilon\rceil\), so that every grid point has an exact binary representation, and it keeps the running time stated there. Some formal statements are slightly stronger than the printed ones: Lemma~\ref{lem:completion} bounds the number of sites by \(8M+4\), Theorem~\ref{thm:factor} also allows \(c=1\), Theorem~\ref{thm:additive} and Lemma~\ref{lem:quantile} hold without invertible covariance factors, and Proposition~\ref{prop:equiv} holds for laws on any real normed space.

The formal versions of Lemmas~\ref{lem:localdisks} and~\ref{lem:block-product} and Proposition~\ref{prop:failure} cover the two configurations of Section~\ref{sec:transfer-local}, one copy of the candidate list and its \(r\) copies, for every radius \(R=2^{s_R}\) and every rational tolerance \(\tau\), and those of the first two lemmas for one copy allow any rational \(R\ge1\). For separated blocks with independent linear sign vectors and the site incidences of the construction, the formal version of \eqref{eq:productcomparison} restricts the event on the left to \(\mathcal N^c\), where \(\mathcal N\) is an exceptional event. Lemma~\ref{lem:localdisks} and part~3 of Lemma~\ref{lem:stability} give the two reading properties used here. In the upper bound of \eqref{eq:failure}, Lean adds \(\Prob(\mathcal N)\). The lower bound subtracts \(\Prob(\mathcal N)\) and uses the stability property that outside \(\mathcal N\) the linear signs determine the complete type whenever no site is large. The formal proofs of Theorems~\ref{thm:main} and~\ref{thm:factor} use the same constructions with other parameters. The machine-level proof of Theorem~\ref{thm:main} runs the repeated construction of Theorem~\ref{thm:factor} with \(c=1\), and the one-copy reduction of its printed proof is formalized as Algorithm~\ref{alg:reduction} and in Proposition~\ref{prop:failure}. Both formal proofs compute the number of copies, the tolerance \(\tau\) and the truncation index by their own formulas from an upper bound on \(m\), and for elliptic laws the formal proof of Theorem~\ref{thm:elliptical} carries out the transferred steps in its own modules. These choices serve the machine-level proofs. The transition tables compute every parameter by a closed formula in the size of the input before they build the instance, so the cell count enters through an explicit polynomial upper bound and the number of copies is an integer multiple of the denominator of \(\beta\), which avoids computing a rational ceiling. One verified construction with a target factor \(c\) then serves Theorems~\ref{thm:main} and~\ref{thm:factor} and the decreasing factors, and the exact mode is its case \(c=1\). The printed proofs use the smallest values that suffice, which keeps their arithmetic short, and the formal proofs check the inequalities that their own values need.

\textbf{Claims outside the development.} The sampling experiment of Fig.~\ref{fig:four-centres}(b) and Section~\ref{sec:sampling-gaps} is empirical, so the development does not reproduce it, while the certified intervals of Proposition~\ref{prop:four-centres} are formalized. The elliptic analogue of Proposition~\ref{prop:oneblock} at the end of Section~\ref{sec:elliptical} is also outside the development, and Theorem~\ref{thm:elliptical} gives NP-hardness for these laws through the candidate lists. Beyond these cases, unnumbered remarks are covered only where Table~\ref{tab:lean-claims} lists them.

\textbf{Repeating the check.} The data, figure generators, experimental code, certificates and formal proofs are available together at \url{https://github.com/tashimir/most-likely-delaunay-triangulations}. In the \texttt{lean} folder, \texttt{lean-toolchain} fixes Lean~4.34.1 and \texttt{lake-manifest.json} fixes each dependency of Mathlib at its own recorded revision and Mathlib itself at the revision
\begin{center}\small\texttt{d13f23b723b8a846827a245b89c10fc7d3f11612}.\end{center}
In that folder, \texttt{lake exe cache get} downloads the compiled Mathlib, and \texttt{lake build} compiles the 1372 modules, which the root file \texttt{MostLikelyDelaunay.lean} reaches through its imports. They include the module \texttt{Results}, so the build also checks every name of the dictionary. The command
\begin{center}\small\texttt{lake env python verification/check.py}\end{center}
recompiles the modules in dependency order in a fresh directory, with warnings treated as errors, and checks the axioms on which each of the 5929 public theorems depends. It prints their union, and with the option \texttt{-{}-report-dir} it also writes the list of each theorem to the file \texttt{axioms.txt}. Every list contains only \texttt{propext}, \texttt{Classical.choice} and \texttt{Quot.sound}, the standard axioms of Lean, so no proof uses \texttt{sorry} or an added axiom. The classical hypotheses are premises of theorems, so they appear in the statements and never in these lists. A clean build of the 1373 sources reported only these three axioms.

\section{Conclusion}\label{sec:conclusions}
Computing the Delaunay triangulation of observed sites is classical. Under uncertainty, a single displacement can affect several diagonals. Our reduction uses this dependence to encode a maximum independent set in the most probable face list. Even with independent standard Gaussian displacements, finding a triangulation at least half as probable as the mode is NP-hard. Sampling gives an additive guarantee (Theorem~\ref{thm:additive}). Its relative accuracy can be poor when the mode has factorially small probability (Proposition~\ref{prop:centre-rank}).

On the line, sorting Gaussian means under common variance gives a factor of \(1/2\) (Proposition~\ref{prop:line}). Can one find a most probable unoriented path in polynomial time for arbitrary means and \(n\)? The planar hull reduction in Section~\ref{sec:gaussian-hull} uses individual scales to encode item weights in cap probabilities. Is finding a most probable labelled hull still NP-hard when all sites have the same Gaussian covariance \(\sigma^2 I_2\)?

Which fixed displacement laws make finding a planar Delaunay mode NP-hard? Our reduction needs, for every \(n\), a disk about the origin with radius of polynomial bit length that contains all \(n\) displacements with probability at least \(1/2\). Such disks exist exactly under the logarithmic moment condition of Proposition~\ref{prop:equiv}. Proving hardness beyond this condition requires another way to control distant sites.

\let\usualbibliography\thebibliography
\renewcommand{\thebibliography}[1]{\usualbibliography{#1}\setlength{\itemsep}{2pt plus 0.4pt minus 0.2pt}}
\bibliographystyle{plainurl}
\bibliography{references}
\end{document}